\pdfoutput=1
\documentclass[preprint,12pt]{aastex}

\usepackage{amsmath}

\newcommand{\beq}{\begin{equation}}           
\newcommand{\eeq}{\end{equation}}             
\newcommand{\bfi}{\begin{figure}}           
\newcommand{\efi}{\end{figure}}             

\shortauthors{Dan\v{e}k \& Heyrovsk\'y}

\shorttitle{Triple-lens Critical Curves and Caustics}

\begin{document}

\title{Critical Curves and Caustics of Triple-lens Models}
	
\author{Kamil Dan\v{e}k and David Heyrovsk\'y}

\affil{Institute of Theoretical Physics, Faculty of Mathematics and Physics, \mbox{Charles University in Prague}, Czech Republic; \email{kamil.danek@utf.mff.cuni.cz}\email{heyrovsky@utf.mff.cuni.cz}}

\begin{abstract}

Among the 25 planetary systems detected up to now by gravitational microlensing, there are two cases of a star with two planets, and two cases of a binary star with a planet. Other, yet undetected types of triple lenses include triple stars or stars with a planet with a moon. The analysis and interpretation of such events is hindered by the lack of understanding of essential characteristics of triple lenses, such as their critical curves and caustics. We present here analytical and numerical methods for mapping the critical-curve topology and caustic cusp number in the parameter space of $n$-point-mass lenses. We apply the methods to the analysis of four symmetric triple-lens models, and obtain altogether 9 different critical-curve topologies and 32 caustic structures. While these results include various generic types, they represent just a subset of all possible triple-lens critical curves and caustics. Using the analyzed models, we demonstrate interesting features of triple lenses that do not occur in two-point-mass lenses. We show an example of a lens that cannot be described by the Chang-Refsdal model in the wide limit. In the close limit we demonstrate unusual structures of primary and secondary caustic loops, and explain the conditions for their occurrence. In the planetary limit we find that the presence of a planet may lead to a whole sequence of additional caustic metamorphoses. We show that a pair of planets may change the structure of the primary caustic even when placed far from their resonant position at the Einstein radius.

\end{abstract}

\keywords{gravitational lensing: micro --- methods: analytical --- planetary systems}

\section{Introduction}
\label{sec:intro}

In the past two decades gravitational microlensing surveys have been very successful, in particular as a tool for studying the stellar population toward the Galactic bulge. In a microlensing event, a star passing close to the line of sight to a background ``source" star is detected by its gravitational lens effect, which temporarily amplifies the flux from the source \citep[e.g.,][]{paczynski96}. The main advantage of the method is its sensitivity to low-mass objects, ranging from stellar down to planetary masses, with most of them too faint to be routinely detected by other means.

In addition to single-star microlensing events, right from their start of operations the surveys have detected events with binary-star lenses \citep{udalski_etal94,alard_etal95}. The frequency of binary events is lower than the frequency of binary stars, since binaries with too close or too far components often mimic single-lens events. The microlensing sensitivity to low mass ratios finally led in 2003 to the first detection of microlensing by a star with a planet \citep{bond_etal04}. By the time of writing, altogether 25 planetary systems had been detected by microlensing\footnote{http://exoplanetarchive.ipac.caltech.edu}. A majority involved a star with a single planet; nevertheless, four of them involved three-body systems. In two cases the lens was a star with two planets \citep{gaudi_etal08,han_etal13}, while the other two involved a binary with a planet \citep{gould_etal14,poleski_etal14}. Other possible triple-lens systems that had not been detected yet include triple stars, or even lenses formed by a star with a planet with a moon.

Turning to the underlying physics, in the case of a single lens the light from the source star is split into two images, which remain generally unresolved due to their small angular separation $\lesssim$ 1 mas. The accompanying temporary amplification of the flux from the source typically produces a simple symmetrically peaked light curve \citep[e.g.,][]{paczynski96}. Lenses with multiple components produce a higher number of images and lead to a greater diversity of light curves, which peak anytime the source crosses or approaches the caustic of the lens. Any simple caustic crossing leads to the appearance or disappearance of a pair of unresolved images at positions defining the critical curve of the lens. The caustic and the critical curve thus are key characteristics of the lens. They depend sensitively on the lens parameters: the masses and positions of the components. Understanding the range of possible critical-curve and caustic geometries is a prerequisite for successful analysis and interpretation of observed microlensing light curves.

Microlensing by a two-component lens, such as a binary star or a star with a planet, is well described by the two-point-mass lens model. Such a lens has only two relevant parameters: the mass ratio, and the projected component separation. The model has been analyzed in detail by \cite{schneider_weiss86}, \cite{erdl_schneider93}, and \cite{witt_petters93}; its limiting cases were studied by \cite{dominik99}. For any mass ratio the critical curve was shown to have three different topologies. In order of decreasing component separation these are: ``wide" with two separate loops, ``intermediate" (or resonant) with a single loop, and ``close" with an outer plus two inner loops. Each regime has a corresponding caustic geometry with the same number of separate non-overlapping loops: wide has two four-cusped loops, intermediate a single six-cusped loop, and close has one four-cusped plus two three-cusped loops. A source positioned outside the caustic has three images, while a source positioned inside the caustic has five images.

Microlensing by triple lenses can be described analogously by the three-point-mass lens model. There are five lens parameters: two relative masses, and three relative positions defining the two-dimensional configuration of the components in the plane of the sky. In comparison with the two-point-mass lens, the model has a number of qualitative differences. For example, varying the lens parameters may lead to a change in the cusp number of the caustic without any accompanying change in the topology of the critical curve. Such changes in the caustic structure occur via swallow-tail or butterfly metamorphoses \citep{schneider_etal92}. In addition, loops of the caustic may overlap, and individual loops may self-intersect. As a result, the caustic may have inner multiply nested regions. All caustics separate the outer four-image region from an inner six-image region. Only caustics with overlapping loops or self-intersections have additional eight-image regions, in case of double nesting even ten-image regions.

However, unlike in the two-point-mass lens case, the full range of critical-curve topologies and caustic structures of the three-point-mass lens has not been explored yet. Such a study would require a systematic mapping of the five-dimensional parameter space, detecting changes in the critical curve and caustic of the corresponding lens. Nevertheless, a range of published works have explored different specific regimes of triple lenses. The first studies explored binary lenses with an additional external shear \citep{grieger_etal89,witt_petters93}, with caustics already displaying swallow tails and butterflies. Most numerous are the studies involving a stellar lens with two (or more) planets. Some of them demonstrate effects on the light curve, others for example place constraints on the presence of a second planet in observed single-planet events. Without claiming completeness, we refer here to the works of \cite{gaudi_etal98}, \cite{bozza99}, \cite{han_etal01}, \cite{han_park02}, \cite{rattenbury_etal02}, \cite{han05}, \cite{kubas_etal08}, \cite{asada09}, \cite{ryu_etal11}, \cite{song_etal14}, and \cite{zhu_etal14}.

Lensing by a binary star with a planet has been explored less frequently \citep{bennett_etal99,lee_etal08,han08a,chung_park10}. In view of the two recently detected systems, this is bound to change. The close and far limits of triple-star lensing were investigated by \cite{bozza00a,bozza00b}. Finally, lensing by a star with a planet with a moon was studied by \cite{han_han02}, \cite{gaudi_etal03}, \cite{han08b}, and \cite{liebig_wambsganss10}, with prospects for detecting such systems remaining open.

Here we set out to systematically study the critical curves and caustics of triple lenses. Based on the work of \cite{erdl_schneider93} and \cite{witt_petters93}, we develop methods for efficient mapping of critical-curve topologies and caustic geometries in the parameter space of $n$-point-mass lenses. We then apply the methods to the analysis of simple triple-lens models. Initial steps of this research appeared in \cite{danek10}, \cite{danek_heyrovsky11}, and \cite{danek_heyrovsky14}.

We start in \S~\ref{sec:n-point} by introducing the basic concepts of $n$-point-mass lensing. In particular, we concentrate on the Jacobian and its properties, such as the equivalence of its contours with critical curves of re-scaled lens configurations \citep{danek_heyrovsky15}. Analytical and numerical methods for mapping critical-curve topologies in the lens parameter space are introduced in \S~\ref{sec:mapping_topologies}. In \S~\ref{sec:caustic} we discuss the caustic and its metamorphoses, and show how to track changes in cusp number using the cusp and morph curves \citep{danek_heyrovsky15}. We apply the methods to triple lenses in \S~\ref{sec:triple}, starting with a brief overview of their properties in \S~\ref{sec:triple-general}. In \S~\ref{sec:linear_symmetric} -- \S~\ref{sec:isosceles} we include a full analysis of four symmetric two-parameter triple-lens models, with an overview of the found critical curves and caustics in \S~\ref{sec:topologies-summary}. We end by summarizing the main results and highlights in \S~\ref{sec:summary}.

\section{The $n$-point-mass lens and its Jacobian}
\label{sec:n-point}

Galactic gravitational microlensing events can be described using a simple $n$-point-mass lens model, consisting of $n$ components (stars, planets) in a single lens plane with no external shear and no convergence due to continuous matter. Following \cite{witt90} we describe positions in the plane of the sky as points in the complex plane, with separations measured in units of the Einstein radius corresponding to the total mass of the lens.

The relation between the position of a background source $\zeta$ and the position $z$ of its image formed by the lens is expressed by the lens equation
\beq
\zeta=z-\sum^{n}_{j=1}{\frac{\mu_j }{\bar{z}-\bar{z}_j}}\,,
\label{eq:lenseq-n_complex}
\eeq
where $z_j$ and $\mu_j$ are the positions and fractional masses, respectively, of the individual components of the lens, and bars over variables denote their complex conjugation. The complex plane of the image positions $z$ is called the image plane, and the complex plane of the source positions $\zeta$ is called the source plane. The fractional masses are normalized to the total mass of the lens, so that $\sum^{n}_{j=1}\mu _j=1 $.

The positions of individual components $z_j$ generally change with time as they orbit around the center of mass of the lens. In this work we study the lensing properties of a ``snapshot" $n$-body configuration at a given instant. The case when the change of lens parameters is non-negligible on the timescale of the microlensing event can be described by a corresponding sequence of such snapshot configurations.

The $n$-point-mass lens with $n>1$ components produces between $n+1$ and $5\,(n-1)$ images (in steps of two) of any point in the source plane \citep{rhie03,khavinson_neumann06}. In the image plane images appear and disappear in pairs along the critical curve of the lens. The critical curve can be expressed as the set of all points $z_{cc}$ for which the sum
\beq
\sum^{n}_{j=1}{\frac{\mu_j }{(z_{cc}-z_j)^2}}=e^{-2\,{\rm i}\,\phi}\, ,
\label{eq:critical-npoint}
\eeq
lies on the unit circle \citep{witt90}. Here $\phi$ is a phase parameter spanning the interval $\phi\in[0,\pi)$. In the source plane the image number changes when the source crosses the caustic $\zeta_c$ of the lens. The caustic is obtained by tracing critical-curve points back to the source plane using equation~(\ref{eq:lenseq-n_complex}), i.e.,
\beq
\zeta_c=z_{cc}-\sum^{n}_{j=1}{\frac{\mu_j }{\bar{z}_{cc}-\bar{z}_j}}\,.
\label{eq:caustic}
\eeq

In mathematical terms, the critical curve is the set of points in the image plane with zero Jacobian of the lens equation. The Jacobian
\beq
{\rm det}\,J\,(z)= 1-\left|\,\sum^{n}_{j=1}{\frac{\mu _j }{(z-z_j)^2}}\,\right|^{\,2}
\label{eq:Jacobian}
\eeq
is discussed in detail in \cite{danek_heyrovsky15}. Here we only summarize its properties important for the following analyses. As seen from equation~(\ref{eq:Jacobian}), the Jacobian is a real function defined over the complex plane, running from $-\infty$ at the positions of all components $z=z_j$ to 1 at complex infinity and at the positions of all Jacobian maxima. These can be found as the roots of the polynomial obtained from
\beq
\sum^{n}_{j=1}{\frac{\mu_j }{(z-z_j)^2}}=0\,.
\label{eq:maxima-npoint}
\eeq
The degree of the corresponding polynomial indicates the Jacobian has up to $2\,n-2$ different maxima. The number may be lower if there are any degenerate roots; these correspond to higher-order maxima. A doubly degenerate root corresponds to a double maximum, a root with degeneracy 3 corresponds to a triple maximum, etc.

The saddle points of the Jacobian can be found similarly among the roots of the polynomial obtained from
\beq
\sum^{n}_{j=1}{\frac{\mu_j }{(z-z_j)^3}}=0\,.
\label{eq:saddle-npoint}
\eeq
The corresponding polynomial has up to $3\,n-3$ different roots. First we need to sort out any common roots of equations~(\ref{eq:maxima-npoint}) and (\ref{eq:saddle-npoint}), those correspond to higher-order maxima instead of saddles. All remaining roots are Jacobian saddle points. The number of different saddles may be reduced further if there are any degenerate roots; these identify higher-order saddles. A non-degenerate root corresponds to a simple saddle, a doubly degenerate root corresponds to a monkey saddle, etc.

Studying the contours of the Jacobian, \cite{danek_heyrovsky15} pointed out a remarkable correspondence. While the zero-Jacobian contour is the critical curve $z_{cc}$ of the lens, any other Jacobian contour $z_\lambda$ with ${\rm det}\,J\,(z_{\lambda})=\lambda$ is a re-scaled critical curve of a lens with the same components in re-scaled positions. Denoting by $z_{cc}(\mu_j,z_j)$ and $z_\lambda(\mu_j,z_j)$ the critical curve and ${\rm det}\,J=\lambda$ contour, respectively, of a lens with masses $\mu_j$ and positions $z_j$, we can express this correspondence by
\beq
z_\lambda(\mu_j,z_j)=z_{cc}(\mu_j,z_j\sqrt[4]{1-\lambda})\,/\,\sqrt[4]{1-\lambda}\,.
\label{eq:correspondence}
\eeq
For $\lambda=0$ we get the critical curve of the original configuration. Close to the original positions, the $\lambda\to-\infty$ contours are shrunk versions of the wide-limit critical curves. At the other limit, the highest-Jacobian $\lambda\to 1$ contours are expanded versions of the close-limit critical curves. A single Jacobian contour plot thus yields the critical curves for all scalings of the given lens configuration, from the close to the wide limit.

\section{Critical-curve topology regions in parameter space and their boundaries}
\label{sec:mapping_topologies}

\subsection{Critical-curve topology and its changes}
\label{sec:topology}

Here we summarize general properties of $n$-point-mass lens critical curves following \cite{danek_heyrovsky15}. Varying the phase parameter $\phi$ in the critical curve equation~(\ref{eq:critical-npoint}), we obtain $2\,n$ continuous line segments, which may connect to form $1\leq N_{loops}\leq 2\,n$ closed loops\footnote{We note that at least for $n<3$ the sharp upper bound is $2\,n-1$, and even for triple lenses we have found no more than 5 loops so far.} of the critical curve \citep{witt90}. Individual loops may lie in separate regions of the image plane, or they may lie nested inside other loops. The total number and mutual position of loops define the topology of the critical curve.

In the wide limit, all lenses have a critical curve with $N_{loops}=n$ separate loops, corresponding to $n$ Einstein rings of the individual components. In the close limit, the critical curve has $N_{loops}=1+N_{max}$ loops, where $N_{max}$ is the number of different Jacobian maxima. One loop corresponds to the Einstein ring of the total mass, plus there is a small loop around each maximum of the Jacobian. In view of the discussion following equation~(\ref{eq:maxima-npoint}), $N_{max}\leq 2\,n-2$, so that in the close limit $N_{loops}\leq 2\,n-1$. The equality holds if the Jacobian has only non-degenerate (simple) maxima. Any degenerate (higher-order) maximum reduces the number of loops. For example, the two-point-mass lens always has two different maxima, thus its critical curve always has three loops in the close limit.

When varying the parameters of the lens such as its scale, the topology may undergo changes when individual loops merge or split. This occurs at Jacobian saddle points when the critical curve passes through them \citep[e.g.,][]{erdl_schneider93}. Two loops come into contact at a simple saddle, three at a monkey saddle, more loops at gradually higher-order saddles. As shown by equation~(\ref{eq:saddle-npoint}), the Jacobian may have up to $3\,n-3$ different saddle points, with the highest number occurring when there are no higher-order saddles and no higher-order maxima. The number of different Jacobian contours passing through the set of saddles identifies the total number of changes in critical-curve topology encountered when varying the scale of the lens from the wide to the close limit. This can be seen as a consequence of the Jacobian-contour / critical-curve correspondence expressed by equation~(\ref{eq:correspondence}). Therefore, the critical curve of an $n$-point-mass lens may undergo no more than $3\,n-3$ changes in topology between the wide and close limits.

In the case of the two-point-mass lens, there are always three simple saddles. One lies on the axis between the components, while an off-axis pair of saddles lies on a different Jacobian contour. The two distinct saddle contours imply that the two-point-mass lens critical curve always undergoes two changes when varying the component separation $s$, and thus has exactly three topologies. The wide topology has two separate loops, the intermediate topology has a single merged loop, and the close topology has an outer loop plus two small inner loops around the Jacobian maxima. The topology sequence is independent of the second lens parameter, the mass ratio of the lens components.

Proceeding to lenses with more than two components, we note that the shape of the critical curve depends on $3\,n-4$ lens parameters \citep[e.g.,][]{danek_heyrovsky15}. Following the preceding discussion, boundaries between regions in parameter space with different critical-curve topology can be found by identifying parameter combinations, for which the critical curve passes through a saddle point of the Jacobian \citep[e.g.,][]{erdl_schneider93}. The search for these boundaries is thus mathematically reduced to finding the conditions for the occurrence of a common solution of equations~(\ref{eq:critical-npoint}) and (\ref{eq:saddle-npoint}) .

The usual analytical approach described in the following \S~\ref{sec:resultant} is based on rewriting both equations in polynomial form and computing their resultant. This step is then followed by a second resultant constructed from the first. However, this approach often yields unwieldy expressions. In addition, one has to check the results for spurious solutions. These do not occur in the two-point-mass lens, but they do appear in all the triple-lens models studied further below.

As an alternative, we present in \S~\ref{sec:scaling-resultant} and \S~\ref{sec:scaling-Jacobian} two efficient numerical methods for mapping the boundaries. These can be used in models with at least one scale-defining parameter, such as the component separation $s$ in the two-point-mass lens. In the first method we find the roots of the first resultant condition, while in the second we utilize the Jacobian scaling properties described by \cite{danek_heyrovsky15}. Both methods are free of spurious solutions.

\subsection{Analytical boundaries computed by resultant method}
\label{sec:resultant}

The method described here was pioneered in the context of critical-curve topology mapping by \cite{erdl_schneider93} and \cite{witt_petters93}. Multiplying the saddle-point equation~(\ref{eq:saddle-npoint}) by the product of its denominators yields
\beq
p_{sadd}(z)=\sum^{n}_{j=1}{\mu_j \prod^{n}_{k=1,\,k\neq j}{(z-z_k)^3}}=0\,,
\label{eq:saddle-npoint-poly}
\eeq
a polynomial of degree $3\,n-3$. In a similar manner we convert the critical-curve equation~(\ref{eq:critical-npoint}) to
\beq
p_{crit}(z)=\prod^{n}_{k=1}{(z-z_k)^2}-e^{2\,{\rm i}\,\phi}\,\sum^{n}_{j=1}{\mu_j \prod^{n}_{k=1,\,k\neq j}{(z-z_k)^2}}=0\,,
\label{eq:critical-npoint-poly}
\eeq
a polynomial of degree $2\,n$ for any value of the parameter $\phi$. The analytical condition for the existence of a common root of $p_{sadd}(z)$ and $p_{crit}(z)$ is
\beq
{\rm Res}_z(p_{sadd},p_{crit})=0\,,
\label{eq:resultant-first}
\eeq
where the resultant  ${\rm Res}_z(f,g)$ of two polynomials $f,\,g$ is a function of their coefficients. It may be computed by evaluating the determinant of the Sylvester or B\'ezout matrices, as described in Appendix~\ref{sec:appendix-resultant}.

The expression obtained from equation~(\ref{eq:resultant-first}) is a polynomial in terms of $e^{2\,{\rm i}\,\phi}$. If we denote $w=e^{2\,{\rm i}\,\phi}$, we can write the result as
\beq
p_{res}(w)=\sum^{m}_{j=0}{a_j\,w^{\,j}}=0\,,
\label{eq:resultant-first-poly}
\eeq
where the degree of the polynomial $m\leq 3\,n-3$. The boundary condition is now equivalent to the condition for $p_{res}$ to have a root on the unit circle.

In order to obtain the condition purely in terms of the lens parameters, \cite{witt_petters93} suggested the following approach for the two-point-mass lens. For a root along the unit circle $\overline{w^j}=w^{-j}$, so that if we take the complex conjugate of equation~(\ref{eq:resultant-first-poly}) and multiply it by $w^m$, we get another polynomial equation
\beq
p_{conj}(w)=\sum^{m}_{j=0}{\bar{a}_{m-j}\,w^{\,j}}=0\,.
\label{eq:resultant-first-conj}
\eeq
Along the boundary in parameter space, polynomials $p_{res}$ and $p_{conj}$ must have a common root, thus
\beq
{\rm Res}_w(p_{res},p_{conj})=0\,
\label{eq:resultant-second}
\eeq
yields the sought condition in terms of lens parameters. We point out that equation~(\ref{eq:resultant-second}) presents a single constraint in parameter space. Hence for two-parameter models (such as the two-point-mass lens or the triple-lens models described in \S~\ref{sec:linear_symmetric} -- \S~\ref{sec:isosceles}) it generally describes a set of curves, for three-parameter models a set of surfaces, etc.

We illustrate the approach here on the case of the two-point-mass lens \citep{erdl_schneider93,witt_petters93}, parameterized by the fractional mass of one component $\mu\in(0,1)$ and the separation between the components $s>0$. If we align the lens with the real axis and place the components symmetrically about the origin, their positions and masses are $\{z_1,z_2\}=\{-s/2,s/2\}$, and $\{\mu_1,\mu_2\}=\{\mu,1-\mu\}$. The polynomial equations for $p_{sadd}$ and $p_{crit}$ are
\beq
p_{sadd}(z)=z^3+\frac{3}{2}s\,(1-2\mu)z^2+\frac{3}{4}s^2 z+\frac{1}{8}s^3(1-2\mu)=0
\label{eq:saddle-binary-poly}
\eeq
and
\beq
p_{crit}(z)=z^4-\frac{1}{2}(s^2+e^{2\,{\rm i}\,\phi})z^2-s\,(1-2\mu)\,e^{2\,{\rm i}\,\phi}z+\frac{1}{16}s^2(s^2-4\,e^{2\,{\rm i}\,\phi})=0\,.
\label{eq:critical_binary-poly}
\eeq
Using them in equation~(\ref{eq:resultant-first}) leads to
\beq
{\rm Res}_z(p_{sadd},p_{crit})=-s^6\mu^2(1-\mu)^2\left[e^{6\,{\rm i}\,\phi}-3\,s^2(1-9\mu+9\mu^2)\,e^{4\,{\rm i}\,\phi}+3\,s^4e^{2\,{\rm i}\,\phi}-s^6\right]=0\,.
\label{eq:resultant-first-b}
\eeq
The factors in front of the square brackets are non-zero for any genuine two-point-mass lens. If we set $e^{2\,{\rm i}\,\phi}=w$, the term in the brackets yields the polynomial equation
\beq
p_{res}(w)=w^3-3s^2(1-9\mu+9\mu^2)w^2+3s^4w-s^6=0\,.
\label{eq:resultant-first-binary}
\eeq
Following equation~(\ref{eq:resultant-first-conj}) we construct
\beq
p_{conj}(w)=-s^6w^3+3s^4w^2-3s^2(1-9\mu+9\mu^2)w+1=0\,.
\label{eq:resultant-first-conj-binary}
\eeq
The resultant obtained from equation~(\ref{eq:resultant-second}) can be factorized as follows:
\begin{eqnarray}
\left[1-3\,s^2(1-9\mu+9\mu^2)+3\,s^4-s^6\right]& &\hspace{-.7cm} \left[1+3\,s^2(1-9\mu+9\mu^2)+3\,s^4+s^6\right] \nonumber\\  & &\hspace{-1cm}\times\left[1-3\,s^4+3\,s^8(1-9\mu+9\mu^2)-s^{12}\right]^2=0\,.
\label{eq:resultant-second-binary}
\end{eqnarray}
At least one of these three square brackets thus has to be equal to zero.

The first bracket in equation~(\ref{eq:resultant-second-binary}) is equal to $p_{res}(1)$, with $w=1$ corresponding to $\phi=0$. Therefore, it must include any transitions occurring on the critical curve along the real axis, in this case the passage of the critical curve through the central saddle point. Taken as a polynomial in $s$, the first bracket has a single real positive root \citep{erdl_schneider93}
\beq
s_w=\left[\sqrt[3]{\mu}+\sqrt[3]{1-\mu}\right]^{3/2}\,,
\label{eq:binary-wide}
\eeq
which is the boundary between the wide and intermediate topologies.

The second bracket is equal to $-p_{res}(-1)$, with $w=-1$ corresponding to $\phi=\pi/2$. There is only a single zero point of this expression in parameter space, $[\mu,\,s]=[0.5,\,\sqrt{0.5}]$, which corresponds to the intermediate -- close splitting along the imaginary axis of the critical curve of an equal-mass lens.

The third bracket corresponds to topology transitions that occur at any other values of $\phi$ on the critical curve. Here they describe the passage of the critical curve through the pair of saddle points off the real axis. Taken as a polynomial in $s$, the third bracket has a single real positive root \citep{erdl_schneider93,rhie_bennett99}
\beq
s_c=\left[\sqrt[3]{\mu}+\sqrt[3]{1-\mu}\right]^{-3/4}=s_w^{-1/2}\,,
\label{eq:binary-close}
\eeq
which is the boundary between the intermediate and close topologies. This boundary in fact passes through the single zero point of the second bracket in equation~(\ref{eq:resultant-second-binary}) at $\mu=0.5$ as well. The two curves given by equations~(\ref{eq:binary-wide}) and (\ref{eq:binary-close}) thus fully describe the division of the two-point-mass lens parameter space according to critical-curve topology.

Applying the same procedure to two-parameter models of triple lenses leads to the following problem. The method yields not only all the sought boundaries, but also curves corresponding to no change in critical-curve topology. These spurious results are additional solutions of equation~(\ref{eq:resultant-second}), which correspond to common roots of $p_{res}(w)$ and $p_{conj}(w)$ lying off the unit circle. In Appendix~\ref{sec:appendix-spurious} we describe how this occurs and why it does not occur in the two-point-mass lens. All curves corresponding to specific values of $w$ that can be factored out of the final equation~(\ref{eq:resultant-second}), such as $p_{res}(1)$ and $p_{res}(-1)$ in the case of equation~(\ref{eq:resultant-second-binary}), are genuine boundaries. Parameters satisfying the remaining parts of equation~(\ref{eq:resultant-second}) should be verified either by computing sample critical curves, or by confirming the existence of a root of $p_{res}(w)$ on the unit circle.

Alternatively, instead of computing the second resultant, one may solve the polynomial from equation~(\ref{eq:resultant-first-poly}) directly by re-substituting $w=\cos{2\,\phi}+{\rm i}\sin{2\,\phi}$. By taking the real and imaginary parts of $p_{res}$ separately, equation~(\ref{eq:resultant-first-poly}) may be treated as a set of two real equations. In case all the coefficients $a_j$ are real, such as in the two-point-mass lens or all the triple-lens models described in \S~\ref{sec:linear_symmetric} -- \S~\ref{sec:isosceles}, the imaginary part of equation~(\ref{eq:resultant-first-poly}) instantly yields the $\phi=0$ and $\phi=\pi/2$ solutions. The corresponding real parts then yield the $p_{res}(1)=0$ and $p_{res}(-1)=0$ boundaries, respectively. The other boundaries can be found by substituting $\cos{2\,\phi}=\xi$ and solving the two obtained real polynomial equations in $\xi$. Finding all roots $\xi_0\in[-1,\,1]$ of one of the polynomials and substituting them into the other polynomial yields the remaining boundary conditions.

\subsection{Numerical boundaries computed using the first resultant}
\label{sec:scaling-resultant}

For any particular $n-$point-mass lens model the lens masses $\mu_j$ and positions $z_j$ depend on a given set of parameters. Mapping the boundaries in parameter space numerically can be a tedious task. If we can compute analytically at least the first resultant and the polynomial $p_{res}$ given by equation~(\ref{eq:resultant-first-poly}), the task is reduced to keeping track of the roots of $p_{res}$ as one scans through parameter space. Positions at which the absolute value of any of the roots crosses unity correspond to points on the boundary. However, even this can be a very time-consuming exercise.

In case at least one of the parameters of the lens model defines an angular scale in the plane of the sky (such as the separation $s$ in the two-point-mass lens), it is possible to reduce the dimension of the space to be scanned by one. At the same time this approach directly yields all boundary points for each reduced parameter combination. In case there are several angular-scale parameters, we select one of them (further denoted by $s$) and instead of the others we use their ratio to $s$. We can now assume that all masses are independent of $s$ and all lens positions are directly proportional to $s$. All the models discussed in this paper satisfy these requirements.

If we divide equation~(\ref{eq:saddle-npoint-poly}) by $s^{3\,n-3}$, we obtain a polynomial equation of exactly the same form with $z'=z/s$ and re-scaled lens positions $z'_k=z_k/s$,
\beq
p'_{sadd}(z')=s^{-3\,n+3}\,\sum^{n}_{j=1}{\mu_j \prod^{n}_{k=1,\,k\neq j}{(z's-z'_ks)^3}}= \sum^{n}_{j=1}{\mu_j \prod^{n}_{k=1,\,k\neq j}{(z'-z'_k)^3}}=0\,.
\label{eq:saddle-npoint-poly-scalefree}
\eeq
The equation is thus explicitly independent of $s$. If we similarly divide equation~(\ref{eq:critical-npoint-poly}) by $s^{2\,n}$ we obtain
\beq
p'_{crit}(z')=\prod^{n}_{k=1}{(z'-z'_k)^2}-e^{2\,{\rm i}\,\phi}\,s^{-2}\,\sum^{n}_{j=1}{\mu_j \prod^{n}_{k=1,\,k\neq j}{(z'-z'_k)^2}}=0\,,
\label{eq:critical-npoint-poly-scalefree}
\eeq
a polynomial of exactly the same form as $p_{crit}(z)$ but with $e^{2\,{\rm i}\,\phi}\,s^{-2}$ instead of $e^{2\,{\rm i}\,\phi}$. Setting the resultant of $p'_{sadd}(z')$ and $p'_{crit}(z')$ equal to zero, we obtain a polynomial in terms of $w'=e^{2\,{\rm i}\,\phi}\,s^{-2}$,
\beq
p'_{res}(w')=\sum^{m}_{j=0}{a'_j\,w'^{\,j}}=0\,,
\label{eq:resultant-first-poly-scalefree}
\eeq
in which all coefficients $a'_j$ are independent of $s$, and the degree of the polynomial $m\leq3\,n-3$. Finding the boundaries now requires a scan of the reduced parameter space (skipping $s$), with the boundaries directly given by all roots of $p'_{res}(w')$. Any such root $w'_{root}$ yields the value of $s$ at the boundary, $s=|w'_{root}|^{-1/2}$. At the same time it also yields the position along the critical curve, $\phi=\arg(w'_{root})/2$, at which the critical curve splits.

Furthermore, if all coefficients $a'_j$ are real (such as in all the models discussed in this paper), all roots of $p'_{res}(w')$ either are real or they occur in complex-conjugate pairs. A complex-conjugate pair of roots yields the same boundary value of $s$, corresponding to a simultaneous split at complex-conjugate positions along the critical curve. It is thus useful to keep track of the discriminant of $p'_{res}(w')$ in the reduced parameter space. Its change of sign corresponds to a transition from a pair of real roots (two boundaries) to a pair of complex-conjugate roots (one boundary). Zero discriminant corresponds to a multiple root of $p'_{res}(w')$, describing a single boundary point and a single $\phi$ value along the critical curve.

For illustration, for the two-point-mass lens we obtain
\beq
p'_{res}(w')=w'^3-3(1-9\mu+9\mu^2)w'^2+3\,w'-1=0\,,
\label{eq:resultant-first-binary-scalefree}
\eeq
and its discriminant
\beq
\Delta(p'_{res})=-3^9\,\mu^2\,(1-\mu)^2\,(1-2\,\mu)^2\leq0\,.
\label{eq:discriminant-binary-scalefree}
\eeq
The only non-degenerate configuration with zero discriminant has $\mu=1/2$, for which there is a double real root plus another real root: $w'_{root}=\{-2,\,-2,\,1/4\}$. These correspond to boundary points $s=\{\sqrt{1/2},\,\sqrt{1/2},\,2\}$. For all other $\mu\in(0,\,1)$ the discriminant is negative. Thus, $p'_{res}(w')$ has one real root and a complex-conjugate pair of roots, resulting in two boundary values of $s$ for any $\mu$.

\subsection{Numerical boundaries computed using Jacobian-contour correspondence}
\label{sec:scaling-Jacobian}

We return here to the method mentioned briefly in \S~\ref{sec:topology}. Instead of computing the resultant, we can map the boundaries in parameter space utilizing the correspondence between Jacobian contours and critical curves of re-scaled lens configurations \citep{danek_heyrovsky15}, described here at the end of \S~\ref{sec:n-point}. Just like in the previous case, even this method requires having a single angular-scale defining parameter $s$, with all lens positions directly proportional to it, and all masses independent of it.

For a given set of the remaining parameters plus an arbitrary non-zero scale parameter $s_0$ we compute the positions of the Jacobian saddle points $z_{sadd}$ by finding all roots of $p_{sadd}(z)$ from equation~(\ref{eq:saddle-npoint-poly}). For each root we evaluate ${\rm det}\,J\,(z_{sadd})$ from equation~(\ref{eq:Jacobian}), which identifies the contour passing through the saddle point. Now the Jacobian-contour / critical-curve correspondence tells us, that in order for the critical curve to pass through the saddle point, we need to re-scale the lens positions by $\sqrt[4]{1-{\rm det}\,J\,(z_{sadd})}\,$, as shown in equation~(\ref{eq:correspondence}). Hence, the boundary value of the parameter $s$ is $s_0\sqrt[4]{1-{\rm det}\,J\,(z_{sadd})}\,$. Recalling the polynomial degree of $p_{sadd}(z)$, this approach yields up to $3\,n-3$ different boundaries for any given point of the reduced parameter space.

For illustration we refer to Figure~1 from \cite{danek_heyrovsky15}, which shows a Jacobian contour plot of a binary lens with $\mu=0.8$ and $s_0=1$. The real saddle point lies on the ${\rm det}\,J\approx-11.0$ contour, and the complex-conjugate saddle-point pair lies on the ${\rm det}\,J\approx0.711$ contour. The corresponding boundary separations are thus $s=\sqrt[4]{1+11}\approx1.861$ and $s=\sqrt[4]{1-0.711}\approx0.733$, respectively.

\section{Mapping the number of cusps of the caustic in parameter space}
\label{sec:caustic}

The caustic of the lens, defined here by equation~(\ref{eq:caustic}), consists of the same number of closed loops as the critical curve. We refer to \cite{danek_heyrovsky15} for an overall discussion of the properties of the $n$-point-mass lens caustic and its cusps. Here we are interested in tracking changes in the number of cusps in the parameter space of the lens.

Since individual loops or parts of the caustic may overlap in lenses with three or more components, we specify that we count the number of cusps encountered as we trace along each loop of the caustic. This means that a cusp superimposed over a fold is still a cusp, or that two or more cusps along different parts of the caustic that happen to lie at the same position in the source plane are still counted as two or more cusps.

The cusp number changes in caustic metamorphoses. The simplest of them is the beak-to-beak metamorphosis, in which two tangent folds reconnect and form two facing cusps \citep[see Figure 5 in][]{danek_heyrovsky15}. In the image plane this always corresponds to a change in critical-curve topology at a saddle point. The corresponding boundaries in parameter space are thus identical to those studied in \S~\ref{sec:mapping_topologies}.

While the beak-to-beak metamorphosis already occurs in two-point-mass lenses, two additional types occur in triple lenses: the swallow-tail and butterfly metamorphoses. They are discussed in more detail and illustrated in \cite{danek_heyrovsky15}. Both occur on the caustic without any change in topology or other significant effect on the critical curve. The lens-parameter conditions for the occurrence of these additional metamorphoses form new boundaries in parameter space, which subdivide the critical-curve topology regions discussed in \S~\ref{sec:mapping_topologies}.

For the following analysis we will use the cusp curve and morph curve introduced by \cite{danek_heyrovsky15}. The parametric form of the cusp curve is
\beq
\left[\sum^{n}_{j=1}{\frac{\mu_j }{(z-z_j)^3}}\right]^2=\Lambda\,\left[\sum^{n}_{j=1}{\frac{\mu_j }{(z-z_j)^2}}\right]^3\,,
\label{eq:cusp_curve_par1}
\eeq
where $\Lambda\geq0$ is a real parameter. The intersections of the cusp curve with the critical curve in the image plane identify the positions of cusp images along the critical curve. Since the curve is scale-invariant, its intersections with other Jacobian contours identify the positions of cusp images for all arbitrarily shrunk or expanded lens configurations.

Another scale-invariant curve in the image plane is the morph curve, defined in parametric form by
\beq
\left[\,\sum^{n}_{j=1}{\frac{\mu_j}{(z-z_j)^2}}\right] \left[\sum^{n}_{j=1}{\frac{\mu_j}{(z-z_j)^4}}\right]= (1+{\rm i}\,\Gamma\,)\, \left[\,\sum^{n}_{j=1}{\frac{\mu_j}{(z-z_j)^3}}\right]^2\,,
\label{eq:morph_curve_par1}
\eeq
where $\Gamma$ is a real parameter. Its intersections with the cusp curve away from the lens positions and higher-order maxima identify the positions of metamorphosis-point images. Intersections at saddles correspond to beak-to-beaks, simple intersections away from special points correspond to swallow tails, and intersections at two-branch self-intersection points of the cusp-curve correspond to butterflies if the critical curve is tangent to either of the branches \citep{danek_heyrovsky15}.

In order to find conditions for the occurrence of swallow-tail and butterfly metamorphoses, we first locate all intersections of the cusp curve and morph curve, and leave out the lens positions, saddle points, and higher-order maxima. We use the parametric polynomial forms of the curves obtained by multiplying equations~(\ref{eq:cusp_curve_par1}) and (\ref{eq:morph_curve_par1}) by all their denominators. We then employ a method analogous to the one used in \S~\ref{sec:resultant} for finding the topology boundaries in parameter space. We compute the resultant of the two polynomials and express the result as a polynomial in the morph-curve parameter $\Gamma$. Since we seek solutions with $\Gamma$ real, we compute a second resultant of the obtained polynomial with its complex conjugate.

The result can be factorized to obtain several conditions involving the non-scaling parameters of the lens, each as a polynomial in the cusp-curve parameter $\Lambda$. For any combination of the non-scaling parameters we look for real non-negative roots $\Lambda$. Equation~(\ref{eq:cusp_curve_par1}) then yields a set of $z$'s on the cusp curve: those that satisfy equation~(\ref{eq:morph_curve_par1}) and lie on the morph curve are the sought intersections. The final check is important, since the second resultant condition may introduce additional spurious solutions. In case the second resultant cannot be simply factorized or generally is too unwieldy, one can find the intersections numerically by gradually varying $\Lambda$ and tracing the branches of the cusp curve, checking each point $z$ with the morph-curve equation~(\ref{eq:morph_curve_par1}).

We then compute the Jacobian at the metamorphosis points to identify the critical curve passing through each of them. The corresponding scaling factor $\sqrt[4]{1-{\rm det}\,J}$ determines the separation of the lens components, just like in the saddle-point case in \S~\ref{sec:scaling-Jacobian}. The obtained parameter combinations form additional curves in parameter space, leading to its final division into subregions with different combinations of cusps on the loops of the caustic.

\section{Triple-lens models}
\label{sec:triple}

\subsection{General properties}
\label{sec:triple-general}

The general triple lens is described by five parameters: typically two relative masses and three spatial-configuration-defining parameters. Among these three it is advantageous to select a single scale parameter, such as the separation of two of the components, or the perimeter of the triangle connecting the components. The remaining two parameters can be for example angles or relative lengths. With such a choice one may utilize the methods from \S~\ref{sec:mapping_topologies} and \S~\ref{sec:caustic} for studying critical curves and caustics based on the scaling properties of the Jacobian.

The Jacobian surface has three poles at the lens positions, hence the critical curve has three loops in the wide limit. There are up to four different Jacobian maxima, therefore in the close limit the critical curve has up to five loops: an Einstein ring corresponding to the total mass, plus up to four small loops around the maxima. The surface has up to six different saddle points. As a consequence, the critical curve of any triple-lens configuration may undergo up to six topology changes when varying the scale parameter from the close to the wide limit.

The critical-curve polynomial $p_{crit}(z)$ from equation~(\ref{eq:critical-npoint-poly}) is of sixth degree, implying that the critical curve and caustic may each have no more than six loops. We point out here in advance that the models discussed further in \S~\ref{sec:linear_symmetric} -- \S~\ref{sec:isosceles} have only up to five loops.

A compact expression for $p_{crit}(z)$ can be obtained in terms of symmetric moments of the lens configuration. In general, five such moments are sufficient:
\beq
c_j=\frac{1}{j}\sum_{k=1}^3{z_k^j}
\label{eq:moments-pos}
\eeq
for $j=1\ldots 3$, and the mass-weighted
\beq
d_j=\sum_{k=1}^3{\mu_k\,z_k^j}
\label{eq:moments-mass}
\eeq
for $j=1,2$. With these definitions $d_1$ is the center of mass of the lens, and $c_1/3$ is its geometric center. In equal-mass models ($\mu_k=1/3$) the two centers coincide, so that $d_1=c_1/3$, and $d_2=2\,c_2/3$.

If we set the origin of the complex plane at the geometric center ($c_1=0$), we arrive at a simpler form of equation~(\ref{eq:critical-npoint-poly}) than if we used the center of mass. The critical-curve polynomial equation can then be written using the four other complex moments:
\begin{eqnarray}
p_{crit}(z)=z^6-(2\,c_2+e^{2\,{\rm i}\,\phi} &&\hspace{-7mm})\,z^4-2\,(c_3+d_1 e^{2\,{\rm i}\,\phi})\,z^3+ [\,c_2^2+(2\,c_2-3\,d_2)\,e^{2\,{\rm i}\,\phi}\,]\,z^2\nonumber \\ &&+\,2\,c_3(c_2-e^{2\,{\rm i}\,\phi})\,z+ c_3^2-(c_2^2+c_3\,d_1-c_2\,d_2)\,e^{2\,{\rm i}\,\phi}=0\,.
\label{eq:critical-triple-poly}
\end{eqnarray}

In a similar manner we may re-write the Jacobian saddle-point polynomial equation~(\ref{eq:saddle-npoint-poly}) in terms of the same four moments:
\begin{eqnarray}
p_{sadd}(z)=z^6+3\,d_1z^5-3\,(c_2&&\hspace{-6mm}-\,2\,d_2)\,z^4 +(7\,c_3+c_2\,d_1)\,z^3+3\,(c_2^2+2\,c_3\, d_1-c_2\,d_2)\,z^2\nonumber \\ && -\,3\,c_3(c_2-d_2)\,z-c_2^2(c_2-d_2)+c_3(c_3-c_2\,d_1)=0\,.
\label{eq:saddle-triple-poly}
\end{eqnarray}
For any specific triple-lens model it is sufficient to express the moments in terms of the model parameters and use them in equations~(\ref{eq:critical-triple-poly}) and (\ref{eq:saddle-triple-poly}). The transitions between different critical-curve topologies can then be mapped using the methods described in \S~\ref{sec:resultant} -- \S~\ref{sec:scaling-Jacobian}.

In the following \S~\ref{sec:linear_symmetric} -- \S~\ref{sec:isosceles} we illustrate the nature of triple lenses by performing the analysis described in \S~\ref{sec:mapping_topologies} and \S~\ref{sec:caustic} on four simple two-parameter models. We chose the most symmetric models with a single scale parameter, sketched in Figure~\ref{fig:model-sketches}. Two of the models are linear (symmetric with variable central mass, general asymmetric with equal masses), and two are triangular (equilateral with variable mass in one vertex, isosceles with variable vertex angle and equal masses). Each of the models, seen as 2D cuts through the full 5D triple-lens parameter space, intersects at least one other model. For example, the isosceles model includes equal-mass equilateral and equal-mass symmetric linear configurations. As a combination, these models form a reference set for further studies of the triple lens.

For all the models we present parameter-space maps of critical-curve topologies and cusp numbers, together with galleries of all different topologies. In the accompanying text we pay particular attention to interpreting the wealth of information contained in the figures. We point out any particularly interesting features of the critical curves and caustics. For better orientation, the text for each model is structured under the following headings: Model description, Jacobian surface character, Topology boundaries, Critical-curve topologies, Caustic structure, Jacobian contour plots, Planetary limits, Close limit, and Wide limit. For the first model, the linear symmetric configuration in \S~\ref{sec:linear_symmetric}, we additionally include the equations of the topology boundaries obtained from the resultant method, a gallery of all transition topologies with merging critical-curve loops, the polynomial expressions for the cusp and morph curves, and a gallery of caustics from all subregions of the parameter space subdivided by swallow-tail and butterfly boundaries.

\subsection{LS Model: Linear Symmetric Configuration}
\label{sec:linear_symmetric}

\noindent
{\em Model description ---} in the LS model, sketched in the first panel of Figure~\ref{fig:model-sketches}, the three lens components lie equidistantly along a line, spaced by separation $s$. The second parameter is the fractional mass $\mu$ of the central component. Placing the origin at the geometric center and the lenses along the real axis, the positions and masses of the components are $\{z_1,z_2,z_3\}= \{-s,0,s\}$ and $\{\mu_1,\mu_2,\mu_3\}= \{(1-\mu)/2,\mu,(1-\mu)/2\}$, with parameter ranges $\mu\in(0,1)$ and $s\in(0,\infty)$.

The $\mu=0$ limit describes an equal-mass binary with components 1 and 3 separated by $2\,s$. The $\mu=1$ case corresponds to a single central lens represented by component 2. The three-equal-masses case with $\mu=1/3$ is identical to special cases of two other models discussed further below: the $p=1/2$ configurations of the LA model in \S~\ref{sec:linear_equal-mass}, and the $\theta=\pi$ configurations of the TI model in \S~\ref{sec:isosceles}.

\noindent
{\em Jacobian surface character ---} for $\mu\neq1/9,1/5\,$: four simple maxima + six simple saddles; for $\mu=1/9\,$: two double maxima + four simple saddles; for $\mu=1/5\,$: four simple maxima + two simple saddles + two monkey saddles.

For $\mu\leq 1/9$ all maxima lie in complex-conjugate pairs along the imaginary axis; for $\mu>1/9$ there are two complex-conjugate pairs of simple maxima symmetrically displaced from the imaginary axis.

Two simple saddles lie along the real axis between neighboring lens components. The remaining saddles for $\mu\leq1/5$ lie in complex-conjugate pairs along the imaginary axis; for $\mu>1/5$ they lie in complex-conjugate pairs symmetrically displaced from the imaginary axis.

\noindent
{\em Topology boundaries ---} due to the high degree of symmetry of the LS model, two of the moments defined by equations~(\ref{eq:moments-pos}) and (\ref{eq:moments-mass}) are zero, $c_3=d_1=0$. This leaves only two non-zero moments, $c_2=s^2$ and $d_2=(1-\mu)\,s^2$. The critical-curve polynomial is reduced to
\beq
p_{crit}(z)=z^6-(2\,s^2+e^{2\,{\rm i}\,\phi})\,z^4+ [\,s^2+(3\mu-1)\,e^{2\,{\rm i}\,\phi}\,]\,s^2z^2-\mu\,s^4\,e^{2\,{\rm i}\,\phi}=0\,,
\label{eq:critical-LS-poly}
\eeq
and the saddle-point polynomial to
\beq
p_{sadd}(z)=z^6+3\,(1-2\mu)\,s^2 z^4+3\,\mu\,s^4 z^2-\mu s^6=0\,,
\label{eq:saddle-LS-poly}
\eeq
both in the form of cubic polynomials in $z^2$. Hence they can be solved analytically, yielding bulky explicit formulae for the critical curve and Jacobian saddle points. We proceed to identify the boundaries of parameter-space regions with different critical-curve topologies using the resultant method described in \S~\ref{sec:resultant}, applied here to the two cubic polynomials.

The first resultant yields the polynomial equation
\beq
p_{res}(w)=(1-9\mu)\,w^3+6\,(1-15\mu+18\mu^2)\,s^2 w^2-3\,(5+3\mu)\,s^4 w+8\,s^6=0\,,
\label{eq:resultant-first-LS}
\eeq
where $w=e^{2\,{\rm i}\,\phi}$. The second resultant can be factorized to yield three independent conditions,
\begin{eqnarray}
8\,s^6-3\,(5+3\mu)\,s^4+6\,(1-15\mu+18\mu^2)\,s^2+1-9\mu=0& &
\label{eq:topo-boundary-LS-w+1}\\
8\,s^6+3\,(5+3\mu)\,s^4+6\,(1-15\mu+18\mu^2)\,s^2-1+9\mu=0& &
\label{eq:topo-boundary-LS-w-1}\\
64\,s^{12}-48\,(1-15\mu+18\mu^2)\,s^8-3\,(5+3\mu)(1-9\mu)\,s^4-(1-9\mu)^2=0& &,
\label{eq:topo-boundary-LS-w-other}
\end{eqnarray}
each defining a curve in parameter space. The left-hand side of equation~(\ref{eq:topo-boundary-LS-w+1}) is equal to $p_{res}(1)$, hence it describes the splitting of the critical curve at $\phi=0$. Similarly, the left-hand side of equation~(\ref{eq:topo-boundary-LS-w-1}) is equal to $p_{res}(-1)$, hence it describes the splitting of the critical curve at $\phi=\pi/2$. These parameter-space curves thus are genuine boundaries between regions with different critical-curve topology. Due to the form of equation~(\ref{eq:critical-LS-poly}) any splitting points of the critical curve along the real or imaginary axes correspond to parameter combinations lying on one of these two boundaries.

The third equation~(\ref{eq:topo-boundary-LS-w-other}) describes the splitting of the critical curve at off-axis points with any other value of $\phi$. Checking the third curve following Appendix~\ref{sec:appendix-spurious}, we find that for $\mu<1/5$ it is a spurious solution of the resultant method. Only the $\mu\geq1/5$ part of the curve defines a genuine parameter-space boundary between different topologies.

A closer inspection of equations~(\ref{eq:topo-boundary-LS-w+1}) and (\ref{eq:topo-boundary-LS-w-1}) reveals that both are of third degree in $s^2$, while equation~(\ref{eq:topo-boundary-LS-w-other}) is of third degree in $s^4$. Hence, the scale parameter $s$ of all the boundaries can be expressed analytically as a function of $\mu$. Instead, we present the results graphically in the parameter-space plot in the left panel of Figure~\ref{fig:LS-parspace}. The boundaries described by equations~(\ref{eq:topo-boundary-LS-w+1}), (\ref{eq:topo-boundary-LS-w-1}), and (\ref{eq:topo-boundary-LS-w-other}) are illustrated by the black, orange, and cyan curves, respectively. The boundaries reach the limits of the plot at $[\mu,s]=[0,8^{-1/2}]$, $[0,1]$, $[1/9,0]$, and $[1,1]$. At the $[1/5,5^{-1/2}]$ triple point the critical curve passes through the monkey saddles.

\noindent
{\em Critical-curve topologies ---} the plot of the parameter space in the left panel of Figure~\ref{fig:LS-parspace} identifies regions with different critical-curve topology. The right panel and its blown-up detail in Figure~\ref{fig:LS-detail} include further subdivision by total cusp number, as discussed further below. For better orientation we mark the six topology regions in the left panel by letters A -- F from top to bottom and left to right. Starting from the widest regime we proceed gradually to lower values of the separation $s$ and mark regions in order of appearance. In case several regions appear at the same value of $s$, we mark them in order from lowest to highest central mass $\mu$.

The critical-curve topologies corresponding to all regions are illustrated by the examples in Figure~\ref{fig:LS-topologies}, where they are shown together with their caustics. Figure~\ref{fig:LS-transitions} includes examples of critical curves and caustics for transitions along the topology boundaries, marked by the letters of adjacent topology regions. The last two examples correspond to the boundary-intersection points seen in Figure~\ref{fig:LS-parspace}.

As shown in Figure~\ref{fig:LS-topologies}, the critical curve of the LS model consists of one (region B), three (regions A, C, E), or five loops (regions D, F). Not all the regions have a unique critical-curve topology: regions C+E, as well as D+F have the same topology. The linear symmetric triple lens thus has four distinct topologies: a single loop (B), three independent loops (A), an outer loop with two inner loops (C, E), or an outer loop with four inner loops (D, F). The topologies are summarized in the LS row of Table~\ref{tab:topologies}.

We inspect the curves in Figure~\ref{fig:LS-topologies} together with the parameter-space map in the left panel of Figure~\ref{fig:LS-parspace}, in order to understand the occurrence of the different topologies and transitions between them. In the wide limit (region A) the three independent loops correspond to Einstein rings of the three components. Reducing the separation first leads to a simultaneous merger of all three loops at the real-axis saddles forming a single loop in region B, as shown by transition AB in Figure~\ref{fig:LS-transitions}.

Further development with decreasing $s$ depends on the value of $\mu$. For $\mu<0.0753$ the next transition BC leads to region C, splitting off inwards two loops around the inner pair of Jacobian maxima along the imaginary axis, followed by transition CF to region F, splitting from the outer loop another pair of loops around the outer pair of maxima along the imaginary axis. For $\mu\in(0.0753,1/9)$ and $\mu\in(1/9,1/5)$ the transition BE to region E splits off inwards two loops along the imaginary axis, each of which encloses two maxima. For $\mu\in(0.0753,1/9)$ the next transition EF to region F splits each of the inner loops into a vertical pair along the imaginary axis, while for $\mu\in(1/9,1/5)$ transition DE to region D splits the inner loops into horizontal pairs bracing the imaginary axis. Finally, for $\mu>1/9$ transition BD to region D splits off simultaneously four loops from four different points along the outer loop.

The three remaining special cases, $\mu\approx0.0753$, $\mu=1/9$, and $\mu=1/5$, are illustrated and discussed together with their caustic structure under the ``Jacobian contour plots'' heading further below.

\noindent
{\em Caustic structure ---} the number of cusps on the caustic changes in beak-to-beak, butterfly, and swallow-tail metamorphoses. These can be found by studying the intersections of the cusp and morph curves, as discussed in \S~\ref{sec:caustic}. The polynomial form of both curves for the LS model is provided in Appendix~\ref{sec:appendix-curves}.

The butterfly and swallow-tail metamorphoses introduce additional boundary curves that lead to a finer division of parameter space by total cusp number. The number is color-coded in the right panel of Figure~\ref{fig:LS-parspace} and in Figure~\ref{fig:LS-detail}. Topology regions A and D are divided into two, B into three, and E into four subregions, with regions C and F undivided. For better orientation we mark each subregion by the letter of the region and a subscript number assigned similarly within the region from top to bottom and left to right.

A gallery of caustics corresponding to all 13 subregions can be found in Figure~\ref{fig:LS-caustics} together with transition caustics BCEF and BDE. For each marked subregion there is a pair of panels showing the full caustic plus a blown-up detail. Already at first inspection we can identify a range of local features that do not occur in two-point-mass lens caustics, such as self-intersecting loops (e.g., A$_2$, E$_2$), nested (C$_1$) or overlapping (D$_2$, F$_1$) loops. The caustic structures generated in butterfly metamorphoses can be seen in A$_2$, B$_2$, D$_1$, E$_1$, E$_2$; swallow-tail-generated structures in B$_3$, E$_4$.

Inspecting the caustic changes along parameter-space boundaries, we find there are two types of beak-to-beak metamorphoses. The transition from two tangent folds to two facing cusps may lead to the formation of an additional loop. This is always the case in two-point-mass lenses, and it occurs here in transitions B-A (B$_1$-A$_1$, B$_2$-A$_2$), B-D (B$_2$-D$_1$), B-E (B$_2$-E$_1$, B$_1$-E$_3$, B$_3$-E$_4$) and C-F (C$_1$-F$_1$). However, we find the transition may just as well result in the opposite, i.e., the merger of two loops. This can be seen here in transitions C-B (C$_1$-B$_3$), D-E (D$_2$-E$_2$), and F-E (F$_1$-E$_4$), always starting from overlapping loops and ending in a self-intersecting loop. Hence, in the triple lens the formation of a cusp pair in the beak-to-beak metamorphosis may either increase or decrease the number of caustic loops.

Studying the new boundaries in the right panel of Figure~\ref{fig:LS-parspace}, we note they are formed by three curves, all ending at $[\mu,s]=[1/9,0]$. The curve leading from $[1,\sqrt{2}+1]$ through regions A, B, E, and the curve leading from $[1,\sqrt{2}-1]$ through regions D, E are both associated with butterfly metamorphoses. The specific transitions include A$_1$-A$_2$, B$_1$-B$_2$, E$_3$-E$_1$ along the first, and D$_2$-D$_1$, E$_1$-E$_2$ along the second curve. The third curve leads from $[0,1]$ and shadows the adjacent topology boundary from the right side through regions B and E, as seen better in Figure~\ref{fig:LS-detail}. This curve is associated with swallow-tail metamorphoses (B$_1$-B$_3$, E$_3$-E$_4$).

The total cusp number in the LS model varies from 8 (subregion B$_1$ marked gray in Figures~\ref{fig:LS-parspace} and \ref{fig:LS-detail}) via 12 (A$_1$, B$_2$, C$_1$, E$_3$ -- red) and 16 (A$_2$, B$_3$, D$_2$, E$_1$, F$_1$ -- cyan) to 20 (D$_1$, E$_2$, E$_4$ -- brown). Due to the high symmetry of the model, single metamorphoses never occur -- always at least two occur simultaneously. As a result, the total cusp number in this case always changes in multiples of four, with the largest number of cusps appearing in the local transition from B$_1$ (8 cusps) to E$_4$ (20 cusps) just to the right of boundary intersection BCEF. There is more diversity in cusp numbers of individual caustic loops, which have 3, 4, 6, 8, 12 (B$_2$) or 16 (B$_3$) cusps. The occurrence of different loop combinations is summarized in column LS of Table~\ref{tab:caustic_structures}.

\noindent
{\em Jacobian contour plots ---} following the examples given by \cite{danek_heyrovsky15}, we present in Figure~\ref{fig:LS-contours} Jacobian contour plots for three special cases: $\mu\approx0.0753$, $\mu=1/9$, and $\mu=1/5$. The panels in the upper row illustrate the critical-curve sequences along vertical $\mu=const.$ cuts in the parameter-space plots. In each panel we mark the $s=0.5$ critical curve in bold for orientation. We recall that for lower $s$ values the critical curves correspond to the outer contours plus the inner loops around the Jacobian maxima, while for higher $s$ values they correspond to the inner contours around the lens positions. Note that the central ($\mu=1/9$) and right ($\mu=1/5$) columns are the only cases of the LS model with four saddles instead of the generic six, as discussed above.

The panels in the lower row include the corresponding cusp (orange) and morph (green) curves, which identify the number and distribution of cusps on the corresponding caustics. In all cases the contours closest to the lenses have four intersections with the cusp curve around each lens, corresponding to three four-cusped caustic loops in the wide-limit subregion A$_1$. The outermost contours corresponding to the close-limit Einstein ring have four cusp-curve intersections too. Hence, the central caustic loop has four cusps as well, whether in subregion F$_1$ (left), E$_3$ (central), or D$_2$ (right column).

At $\mu\approx0.0753$ (left column) the critical-curve sequence is A-B-F, and the caustic sequence is A$_1$-B$_1$-B$_3$-F$_1$. The transition from B via the boundary intersection BCEF directly to region F splits off simultaneously all four inner loops along the imaginary axis. The tiny loops around the inner pair of maxima are barely discernible just interior of the inner pair of saddles along the imaginary axis (see also Figure~\ref{fig:LS-transitions}). Each of the small contours around the maxima has three cusp-curve intersections, implying a set of four three-cusped caustic loops in the corresponding region F$_1$. Inspecting the morph curve, note the somewhat hard-to-see four simultaneous B$_1$-B$_3$ swallow-tail metamorphoses. These are indicated by the four simple intersections of the small figure-eight parts of the morph curve with the small loops of the cusp curve.

In the $\mu=1/9$ case (central column), the critical-curve sequence is A-B-E, and the caustic sequence is A$_1$-B$_1$-E$_3$. Here there are only two double maxima, so that the critical curve stays in region E even in the limit $s\to 0$. For this value of $\mu$ the close-limit topology of the critical curve has an outer Einstein ring with only two small loops inside. In the lower panel we see that either of these loops has four intersections with the cusp curve. This means that the corresponding caustic loops have four instead of three cusps in the $s\to 0$ limit in subregion E$_3$. This configuration is an example of the ``multiple caustics" case studied by \cite{bozza00b}.

The $\mu=1/9$ case is generally the only case with neither butterfly nor swallow-tail metamorphoses, as seen also from the right panel of Figure~\ref{fig:LS-parspace}. The cusp and morph curves intersect only at the positions of the lenses, saddles, and higher-order maxima. The higher-order maxima are isolated solutions of the morph-curve equation~(\ref{eq:morph_curve_par1}). When perturbed to lower $\mu$ values, these points turn into the horizontal figure-eight loops of the morph curve, as seen in the left $\mu\approx0.0753$ column. Similarly, when perturbed to higher $\mu$ values, they turn into vertical figure-eight loops that eventually merge with the outer part of the morph curve at the $\mu=1/5$ monkey saddles.

At $\mu=1/5$ (right column) the critical-curve sequence is A-B-D, and the caustic sequence is A$_1$-B$_1$-B$_2$-D$_2$. Here the transition from B to D via the pair of monkey-saddle points at BDE splits off simultaneously all four inner loops around off-axis maxima, with the splitting occurring pairwise from the two monkey saddles (see also Figure~\ref{fig:LS-transitions}). The small contours around the maxima each have three cusp-curve intersections, hence they correspond to four three-cusped caustic loops in subregion D$_2$. Inspecting the morph curves in the lower panel, we first notice the characteristic structure at the monkey saddles. In each of them the cusp curve arrives from four perpendicular directions, while the morph curve arrives from six symmetric directions. The corresponding caustic has two beak-to-beak metamorphoses with each involving three beaks. These can be seen in the BDE case triple-loop contact in the last panel of Figure~\ref{fig:LS-caustics}. Note also the additional tangent intersections of the cusp and morph curves on the imaginary axis closer to the origin. These correspond to two simultaneous B$_1$-B$_2$ butterfly metamorphoses, as expected from Figure~\ref{fig:LS-parspace}.

Note that yet another LS-model Jacobian plot (for the $\mu=1/3$ equal-mass case) can be found further in the right column of Figure~\ref{fig:LA-contours}, where it corresponds to the LA model with position parameter $p=1/2$. Its LS-model sequence of caustic subregions is A$_1$-B$_1$-B$_2$-D$_1$-D$_2$.

\noindent
{\em Planetary limits ---} the LS model has two different planetary-mass limits. In the $\mu\to 0$ case we have an equal-mass binary with separation $2\,s$ and a planet of mass $\mu$ placed exactly between its components (at the L1 Lagrangian point). In order of decreasing $s$ the caustic sequence is A$_1$-B$_1$-B$_3$-C$_1$-F$_1$. At both limits, the planet merely adds tiny extra loops to the binary caustic: the central loop in A$_1$, and the pair of extra three-cusped loops in C$_1$ and F$_1$. The planet perturbs primarily the $s=1$ wide -- intermediate binary transition, which occurs at the position of the planet. Here, in the single-loop subregions B$_1$ and B$_3$ the planet adds extra cusps to the caustic perpendicularly to the binary axis.

In the $\mu \to 1$ case we have a star with two equal-mass planets at opposite sides at a separation $s$. Here the caustic sequence is A$_1$-A$_2$-B$_2$-D$_1$-D$_2$. In the limiting subregions A$_1$ and D$_2$ the planets perturb the point-like single-lens caustic to form the central four-cusped loop \citep{chang_refsdal84}, adding two (A$_1$) or four (D$_2$) extra tiny loops. However, in subregions A$_2$, B$_2$, D$_1$ covering the separation range $s\in[\sqrt{2}-1,\sqrt{2}+1]$ the planets generate extra cusps on the central loop at positions perpendicular to the orientation of the system. The planets affect the caustic most prominently when they lie in the vicinity of the $s=1$ Einstein radius, causing all loops to merge and form the single-loop caustic B$_2$.

\noindent
{\em Close limit ---} for $\mu\neq1/9$ (subregions F$_1$, D$_2$): critical curve = Einstein ring + four small loops around Jacobian maxima; caustic = central four-cusped loop + four three-cusped loops escaping to $\infty$. For $\mu=1/9$ (E$_3$): critical curve = Einstein ring + two small loops around Jacobian maxima; caustic = central four-cusped loop + two four-cusped loops escaping to $\infty$ \citep[see also][]{bozza00b}.

\noindent
{\em Wide limit ---} for any $\mu$ (subregion A$_1$): critical curve = three independent Einstein rings with radii $\{\sqrt{(1-\mu)/2},\sqrt{\mu},\sqrt{(1-\mu)/2}\}$; caustic = three four-cusped weak-shear Chang-Refsdal loops.

\subsection{LA Model: Linear Asymmetric Configuration}
\label{sec:linear_equal-mass}

\noindent
{\em Model description ---} in the LA model three equal-mass lens components lie along a line with a variable separation of the outer two components, marked $2\,s$ to conform with the previous LS model. Here the second parameter is the relative separation $p$ of the central component from the left component in units of the outer component separation (see second sketch in Figure~\ref{fig:model-sketches}). Placing the origin at the geometric center and the lenses along the real axis, the positions and masses of the components are $\{z_1,z_2,z_3\}= \{-p-1,2\,p-1,2-p\}\,2\,s/3$ and $\{\mu_1,\mu_2,\mu_3\}= \{1/3,1/3,1/3\}$, with parameter ranges $p\in(0,1)$ and $s\in(0,\infty)$.

The parametrization is symmetric about $p=1/2$, so that the configurations, critical curves, and caustics for $p$ and $1-p$ are mutual mirror images. In the $p\to 0$ limit components 1 and 2 merge, so that the lens is reduced to a binary lens with masses $\{2/3,1/3\}$. Similarly, in the $p\to 1$ limit components 2 and 3 merge. For $p=1/2$ the components are spaced symmetrically, hence these configurations are identical to the $\mu=1/3$ configurations of the LS model and with the $\theta=\pi$ configurations of the TI model discussed further in \S~\ref{sec:isosceles}.

\noindent
{\em Jacobian surface character ---} for any $p\,$: four simple maxima + six simple saddles.

\noindent
{\em Topology boundaries ---} due to the components having equal masses, we get the center of mass $d_1=0$, and the remaining three moments are $c_2=4\,s^2(1-p+p^2)/3$, $d_2=2\,c_2/3$, and $c_3=8\,s^3(-p-1)(2\,p-1)(2-p)/27$. The critical-curve polynomial
\beq
p_{crit}(z)=z^6-(2\,c_2+e^{2\,{\rm i}\,\phi})\,z^4-2\,c_3\,z^3+ c_2^2\,z^2+ 2\,c_3(c_2-e^{2\,{\rm i}\,\phi})\,z+ c_3^2-c_2^2\,e^{2\,{\rm i}\,\phi}/3=0
\label{eq:critical-LA-poly}
\eeq
and the saddle-point polynomial
\beq
p_{sadd}(z)=z^6+c_2\,z^4+7\,c_3\,z^3+c_2^2\,z^2-\,c_3\,c_2\,z+c_3^2-c_2^3/3=0
\label{eq:saddle-LA-poly}
\eeq
generally cannot be solved analytically. We proceed to identify the boundaries of parameter-space regions with different critical-curve topologies using the resultant method described in \S~\ref{sec:resultant} applied directly to polynomials $p_{crit}$ and $p_{sadd}$.

The first resultant yields the polynomial equation
\begin{eqnarray}
p_{res}(w)=(8-\nu)\,w^6+6\,(12-\nu)&&\hspace{-6mm}c_2\,w^5+216\,c_2^2\,w^4- (216-198\,\nu+11\,\nu^2)\,c_2^3\,w^3\nonumber\\ &&- 3\,\nu\,(45-14\,\nu)\,c_2^4\,w^2- 18\,\nu^2\,c_2^5\,w+\nu^3\,c_2^6=0\,,
\label{eq:resultant-first-LA}
\end{eqnarray}
where $w=e^{2\,{\rm i}\,\phi}$ and the parameter $\nu\equiv27\,p^2(1-p)^2/(1-p+p^2)^3$. The equation depends on $s$ through the moment $c_2$. The second resultant can be factorized to yield three independent conditions. The first, $p_{res}(1)=0$, describes the splitting of the critical curve at $\phi=0$ as a 6th degree polynomial in $s^2$. The second, $p_{res}(-1)=0$, would describe the splitting of the critical curve at $\phi=\pi/2$. However, this condition has no solution in our parameter space, hence such splitting does not occur in the LA model. The third condition describes the splitting of the critical curve at any other value of $\phi$. It has the form of a 15th degree polynomial in $s^4$, which we do not present here explicitly. An inspection following Appendix~\ref{sec:appendix-spurious} reveals an entire spurious branch of the corresponding curve in parameter space.

The $p_{res}(1)=0$ and the 15th degree polynomial boundaries are marked in the parameter-space plot in the left panel of Figure~\ref{fig:LA-parspace} by the black and cyan curves, respectively. The curves meet the $p=0,1$ limits at $s=(\sqrt[3]{2}-1)^{-1/2}/2\approx0.981$ and $s=(\sqrt[3]{2}-1)^{1/4}/2\approx0.357$, respectively.

\noindent
{\em Critical-curve topologies ---} the division of parameter space by critical-curve topology is shown in the left panel of Figure~\ref{fig:LA-parspace}. In view of the $p\leftrightarrow (1-p)$ symmetry, we use letters to mark the six different regions within the left half of the parameter space (from top to bottom and left to right), and mark their symmetric counterparts by the same letters.

The topologies corresponding to all regions are illustrated by the examples presented in Figure~\ref{fig:LA-topologies} together with their caustics. The LA model permits a greater range of critical-curve topologies than the LS model, with each region having a unique topology. In addition to all the topologies seen previously in the LS model (Figure~\ref{fig:LS-topologies}), there is a two-loop (region B) and a four-loop (region A) topology. In region B one of the two independent loops contains two lens components, while in region A there is a simple loop plus a close-binary set of loops (two small loops within an outer loop). Neither of these can be achieved in the symmetric LS model. Note also that the single-loop topology (region D) occurs only for $p\in(0.239,0.761)$. In other cases with decreasing $s$ the loop around the pair of components undergoes the intermediate -- close binary transition (B-A) before connecting with the third-component loop (A-E). All topologies are summarized in the LA row of Table~\ref{tab:topologies}.

\noindent
{\em Caustic structure ---} the further subdivision of parameter space according to total cusp number is shown in the right panel of Figure~\ref{fig:LA-parspace}. The number of cusps on the caustic changes primarily in beak-to-beak metamorphoses. Unlike in the case of the LS model, all are of the standard type with cusp formation accompanying the splitting of caustic loops. Other metamorphoses occur only in the limited range $p\in(0.292,0.708)$, as seen from the extent of the additional duck-foot-shaped boundary in Figure~\ref{fig:LA-parspace}. The curve corresponds to a pair of swallow-tail metamorphoses, with the exception of its two points on the $p=1/2$ midline. The higher symmetry at these points leads to a pair of butterfly metamorphoses at each. The new boundary divides regions D, E, and F into two subregions each, while the other regions remain undivided. In comparison with the other models, each of the nine subregions has a unique caustic structure, as seen from column LA of Table~\ref{tab:caustic_structures}.

The LA model has a lower degree of symmetry than the LS model, so that beak-to-beak metamorphoses along the real axis may occur singly and the total cusp number thus may change by 2. The total cusp number varies from 8 to 20 in steps of two, with one exception: there is no eighteen-cusped caustic in the LA model. The newly occurring totals are 10 cusps (subregion B$_1$ marked blue in Figure~\ref{fig:LA-parspace}) and 14 cusps (A$_1$ -- green). In comparison with the LS model, there are caustic loops with 10 cusps (B$_1$), but there are no loops with 16 cusps.

\noindent
{\em Jacobian contour plots ---} in Figure~\ref{fig:LA-contours} we present Jacobian contour plots for three cases: $p=0.18,0.45,1/2$. The latter two were used by \cite{danek_heyrovsky15} to illustrate the swallow-tail and butterfly metamorphoses, respectively. The $p=1/2$ caustic sequence from wide to close limit is C$_1$-D$_1$-D$_2$-F$_1$-F$_2$, with transitions D$_1$-D$_2$ and F$_2$-F$_1$ each involving a pair of simultaneous butterfly metamorphoses. The $p=0.45$ caustic sequence is C$_1$-B$_1$-D$_1$-D$_2$-E$_2$-E$_1$-F$_2$, with transitions D$_1$-D$_2$ and E$_1$-E$_2$ each involving a pair of simultaneous swallow-tail metamorphoses.

The $p=0.18$ case in the left column illustrates a hierarchical triple system, with components 1 and 2 forming a binary and component 3 as a distant companion. The structure of the contours and curves near the first two components resembles a simple equal-mass binary \citep[see][]{danek_heyrovsky15} with its wide, intermediate, and close regimes. If we ignore the local structure around components 1 and 2, on the large scale the contours and curves again resemble a binary system with its three regimes, here with a 2:1 mass ratio since components 1 \& 2 act as a single body. The caustic sequence C$_1$-B$_1$-A$_1$-E$_1$-F$_2$ undergoes no swallow-tail, no butterfly, and no single-loop topology.

\noindent
{\em Planetary limits ---} the LA model has no planetary limit.

\noindent
{\em Close limit ---} for any $p$ (subregion F$_2$): critical curve = Einstein ring + four small loops around Jacobian maxima; caustic = central four-cusped loop + four three-cusped loops escaping to $\infty$.

\noindent
{\em Wide limit ---} the wide limit is more interesting, with three different regimes for arbitrarily large separation $s\to\infty$. In this limit the model behaves as an equal-mass binary lens plus an independent single lens. The three regimes are simply the close (subregion A$_1$), intermediate (B$_1$), and wide (C$_1$) regimes of the binary lens. The wide regime, which eventually dominates, leads to three independent single lenses with Einstein radii $\sqrt{1/3}$.

This behavior can be proved by the $s\to\infty$ asymptotic form of the boundaries in Figure~\ref{fig:LA-parspace}: in the $p<1/2$ part the A-B boundary is given by $2\,p\,s=\sqrt{1/3}$, and the B-C boundary by $2\,p\,s=\sqrt{8/3}$. Noting that $2\,p\,s$ is the separation between lens components 1 and 2 (see Figure~\ref{fig:model-sketches}), if we divide the two values by the Einstein radius of the binary ($\sqrt{2/3}$), we get the close -- intermediate and intermediate -- wide boundaries of the equal-mass binary lens. To summarize: critical curve = binary critical curve + Einstein ring with radius $\sqrt{1/3}\,$; caustic = binary caustic + four-cusped weak-shear Chang-Refsdal loop.

It is worth noticing that the presence of the distant third body along the binary axis has no effect on the critical curve topology or caustic structure of the binary lens, it merely adds a distant loop. Similarly, the structure of the plots in Figure~\ref{fig:LA-parspace} near the left boundary indicates that a tiny separation of components 1 and 2 along the axis to component 3 has no effect on the critical curve topology or caustic structure of the 2:1 mass-ratio binary lens, it merely adds two tiny loops.

\subsection{TE Model: Triangular Equilateral Configuration}
\label{sec:equilateral_vertex}

\noindent
{\em Model description ---} in the TE model the lens components lie at the vertices of an equilateral triangle with side length $s$ (see third sketch in Figure~\ref{fig:model-sketches}). We use the fractional mass $\mu$ of component 1 as the second parameter. Placing the origin at the geometric center and aligning the real axis with the median passing through component 1, the positions and masses of the components are $\{z_1,z_2,z_3\}= \{-1,e^{-{\rm i}\,\pi/3},e^{{\rm i}\,\pi/3}\}\,s/\sqrt{3}$ and $\{\mu_1,\mu_2,\mu_3\}= \{\mu,{(1-\mu)/2},{(1-\mu)/2}\}$, with parameter ranges $\mu\in(0,1)$ and $s\in(0,\infty)$.

The $\mu=0$ limit corresponds to an equal-mass binary formed by components 2 and 3, while the $\mu=1$ limit corresponds to component 1 as a single lens. The $\mu=1/3$ case is the most symmetric triple lens: three equal masses in an equilateral configuration. It is identical with the $\theta=\pi/3$ case of the TI model discussed further in \S~\ref{sec:isosceles}.

\noindent
{\em Jacobian surface character ---} for $\mu\neq 0.768,8/9\,$: four simple maxima + six simple saddles; for $\mu\approx0.768\,$: two simple maxima + one double maximum + five simple saddles; for $\mu=8/9\,$: four simple maxima + four simple saddles + one monkey saddle. In addition, note that the $\mu=1/3$ equal-mass case is unique due to its global three-fold symmetry.

\noindent
{\em Topology boundaries ---} due to the equilateral configuration $c_2=0$, the center of mass $d_1=(1-3\,\mu)\,s/\sqrt{12}$, $c_3=-s^3/\sqrt{27}$, and $d_2=-(1-3\,\mu)\,s^2/6$. The critical-curve polynomial
\beq
p_{crit}(z)=z^6-e^{2\,{\rm i}\,\phi}\,z^4-2\,(c_3+d_1 e^{2\,{\rm i}\,\phi})\,z^3   -3\,d_2\,e^{2\,{\rm i}\,\phi}\,z^2-\,2\,c_3\,e^{2\,{\rm i}\,\phi}\,z+ c_3(c_3-d_1\,e^{2\,{\rm i}\,\phi})=0
\label{eq:critical-TE-poly}
\eeq
and the saddle-point polynomial
\begin{eqnarray}
p_{sadd}(z)=z^6+3\,d_1z^5+6\,d_2\,z^4 +7\,c_3\,z^3+6\,c_3\, d_1\,z^2+3\,c_3\,d_2\,z+c_3^2=0
\label{eq:saddle-TE-poly}
\end{eqnarray}
cannot be solved analytically for a general $\mu$. We use the resultant method described in \S~\ref{sec:resultant} applied to polynomials $p_{crit}$ and $p_{sadd}$ to identify the boundaries of regions with different critical-curve topologies.

The first resultant yields the polynomial equation
\begin{eqnarray}
p_{res}(w)&&\hspace{-6mm}=(\tau^3\!-\!3\tau^2\!+\!1)\,w^6\nonumber\\ &&- 3\,\tau\,(3\tau^4\!-\!12\tau^3\!+\!8\tau^2\!+\!3\tau\!-\!3)\,s^2 w^5- 3\,\tau\,(23\tau^3\!-\!20\tau^2\!-\!8\tau\!+\!7)\,s^4 w^4\nonumber\\ &&+ (6\tau^3\!+\!12\tau^2\!-\!11)\,s^6 w^3+6\,\tau\,(\tau\!-\!2)\,s^8 w^2+3\,\tau\,s^{10} w-s^{12}=0\,,
\label{eq:resultant-first-TE}
\end{eqnarray}
where $w=e^{2\,{\rm i}\,\phi}$ and the parameter $\tau\equiv(3\,\mu-1)/2$ is equal to zero for an equal-mass lens. The second resultant can be factorized to yield four independent conditions. The first two, $p_{res}(1)=0$ and  $p_{res}(-1)=0$, describe the splitting of the critical curve at $\phi=0$ and $\phi=\pi/2$, respectively, in the form of 6th degree polynomials in $s^2$. The other two conditions describe the splitting of the critical curve at other values of $\phi$. Both have the form of polynomials in $s^4$ (3rd and 12th degree, respectively), which we do not present here explicitly. An inspection following Appendix~\ref{sec:appendix-spurious} reveals that only a segment of the 3rd degree polynomial curve corresponds to topology transitions, the rest is spurious.

The $p_{res}(1)=0$, $p_{res}(-1)=0$, the 3rd, and 12th degree polynomial boundaries are marked in the parameter-space plot in the left panel of Figure~\ref{fig:TE-parspace} by the black, orange, cyan, and green curves, respectively. The boundaries reach the $\mu=0$ side of the plot at $s=\sqrt{1/2}$ and $s=2$, the $\mu=1$ side at $s=1$, and the $s=0$ side at $\mu=1-(4/3)\sin{(\pi/18)}\approx0.768$. At the $[\mu,s]=[8/9,\sqrt{1/2}]$ triple point the critical curve passes through a monkey saddle.

\noindent
{\em Critical-curve topologies ---} the division of parameter space by critical-curve topology is shown in the left panel of Figure~\ref{fig:TE-parspace}. The right panel and its blown-up details in Figure~\ref{fig:TE-details} include the further subdivision according to total cusp number. The four curves identified above carve the parameter space in the left panel into ten different topology regions, marked by letters from top to bottom and left to right.

For each region examples of critical curves with their corresponding caustics are presented in Figure~\ref{fig:TE-topologies}. We see that the critical curve may have anywhere from one to five loops. In comparison with the previous models, there are two new topologies: a small loop within an outer loop (region D), and three small loops within an outer loop (I, J). Only one of the previous models' topologies does not occur here: the simple loop plus a close-binary set of loops (model LA region A). As seen from the TE row of Table~\ref{tab:topologies}, this model permits altogether seven different critical topologies, since the pairs of regions E+H, F+G, and I+J each share the same topology.

The beak-to-beak metamorphoses along most of the topology boundaries increase the number of critical-curve loops, with a decrease occurring only for the D-C, I-E, G-J, and F-J transitions. In addition, the TE model has the special case of the D-B transition, in which two simultaneous beak-to-beaks convert an outer-plus-inner loop combination to two separate loops, thus preserving the total number of loops.

\noindent
{\em Caustic structure ---} the subdivision of parameter space by number of cusps on the caustic is shown in the right panel of Figure~\ref{fig:TE-parspace} and in Figure~\ref{fig:TE-details}. As seen in both figures, many additional metamorphoses change the cusp number. There are four curves corresponding to butterfly and one curve corresponding to swallow-tail metamorphoses.

Two of the butterfly curves start from $[0,\sqrt{1/2}]$: one leading down through G to $[1/3,0]$, returning back through G, J, and F to $s=-1+\sqrt{3}-\sqrt{2-\sqrt{3}}\approx0.214$ at the right edge; the second goes up through B and A to $s=1+\sqrt{3}+\sqrt{2+\sqrt{3}}\approx4.66$ at the right edge. The other two butterfly curves rise monotonically from $[0.768,0]$ toward the right side: the third through J and F to $s=1+\sqrt{3}-\sqrt{2+\sqrt{3}}\approx0.800$; the fourth through J, E, C, and A to $s=-1+\sqrt{3}+\sqrt{2-\sqrt{3}}\approx1.25$.

Finally, the swallow-tail curve starts from $[0,\sqrt{1/2}]$ and leads up through B with a small sharp peak at $\mu\approx0.084$ to A, where it peaks sharply when it meets the butterfly curve at $\mu=1/3$, dropping down through A, C, E, and J to $[0.768,0]$. The curve corresponds to a complex-conjugate pair of swallow-tail metamorphoses everywhere except at the two sharp peaks. At the $\mu=1/3$ peak three symmetric butterfly metamorphoses occur simultaneously, while at the $\mu\approx0.084$ peak there is a complex-conjugate pair of butterfly metamorphoses.

The additional set of curves in the right panel of Figure~\ref{fig:TE-parspace} (with details better seen in Figure~\ref{fig:TE-details}) divides regions C, E, F, and G into three subregions each, regions A and B into five subregions, and region J into eight subregions. Regions D, H, and I remain undivided. The parameter space of the TE model is thus divided into 33 subregions, corresponding to 24 different loop combinations. These are sorted by cusp number and listed in column TE of Table~\ref{tab:caustic_structures}. The total cusp number varies from 10 to 20 in steps of two; i.e., there is no eight-cusped caustic. Caustics with 18 cusps that appear newly in this model occur in seven different subregions (see Table~\ref{tab:caustic_structures}). Individual caustic loops have 3, 4, 6, 7, 8, 9, 10, 12, or 14 cusps. In comparison with the previous models there is no sixteen-cusped loop (B$_3$ in LS model), while seven-cusped (I$_1$ shown in Figure~\ref{fig:TE-topologies}) and nine-cusped loops (D$_1$ in Figure~\ref{fig:TE-topologies}) are new here.

\noindent
{\em Jacobian contour plots ---} Figure~\ref{fig:TE-contours} includes contour plots for three special cases: $\mu=1/3$, $\mu\approx0.768$, and $\mu=8/9$. In the symmetric $\mu=1/3$ case the caustic sequence from wide to close limit is A$_1$-A$_4$-D$_1$-G$_2$, with the A$_1$-A$_4$ three simultaneous butterfly metamorphoses at $s=2^{2/3}\,3^{1/2}\approx2.75$. The D$_1$-A$_4$ three simultaneous beak-to-beaks occur at $s={[(5\sqrt{5}+11)/2]^{1/6}}\approx1.49$, and the D$_1$-G$_2$ three simultaneous beak-to-beaks occur at $s={[(5\sqrt{5}-11)/2]^{1/6}}\approx0.670$. There is no single-loop topology, but with the slightest perturbation to a higher $\mu$ the sequence would visit the C$_2$ single-loop region with a fourteen-cusped caustic.

The double-maximum case in the central $\mu\approx0.768$ column has a caustic sequence A$_1$-A$_2$-C$_1$-E$_2$-J$_3$-J$_7$ with two butterfly metamorphoses along the real axis. The plot requires a larger scale due to the close-limit J$_7$-J$_3$ butterfly point near the right edge of the inset.

The same holds also for the $\mu=8/9$ monkey-saddle case in the right column. The caustic sequence A$_1$-A$_2$-C$_1$-C$_3$-E$_1$-F$_2$-F$_3$ starts out similarly, but differs once the components get closer. The three butterfly points all lie on the real axis.

\noindent
{\em Planetary limits ---} the TE model has two different planetary-mass limits, similar to those of the LS model. The $\mu\to 0$ limit describes an equal-mass binary separated by $s$ with a planet of mass $\mu$ offset perpendicularly at the vertex of an equilateral triangle (at the L4 Lagrangian point for a face-on coplanar orbit\footnote{Note that in this case L4 is unstable, since $\mu_2=\mu_3$.}). In order of decreasing $s$, the caustic undergoes a sequence of seven metamorphoses: A$_1$-B$_1$-B$_5$-B$_3$-D$_1$-H$_1$-G$_2$-G$_1$. The two widest regimes, A$_1$ and B$_1$, correspond to a wide and intermediate binary caustic, respectively, each with an additional four-cusped loop due to the planet. Similarly, G$_1$ corresponds to a close binary caustic with two additional three-cusped loops due to the planet.

All six metamorphoses between B$_1$ and G$_1$ occur close to the $s=1/\sqrt{2}$ intermediate -- close binary transition, since the planet lies exactly at the location at which one three-cusped loop would split off in the binary limit. The sequence begins with a butterfly on the planetary loop leading to B$_5$, followed by a pair of swallow tails on the binary-caustic loop to B$_3$. In the peculiar transition to D$_1$ both loops connect in a pair of reverse beak-to-beaks, which simultaneously split off a three-cusped loop. A second three-cusped loop is formed along the real axis in a beak-to-beak to H$_1$, and a third and fourth three-cusped loop appear simultaneously in a pair of beak-to-beaks to G$_2$. Finally, the main caustic loop loses two cusps in a reverse butterfly to G$_1$.

The $\mu \to 1$ limit describes a single lens with two equal-mass planets at the same distance $s$, an angle of $\pi/3$ apart. Even in this case the caustic undergoes seven metamorphoses with decreasing $s$: A$_1$-A$_2$-A$_5$-C$_3$-E$_1$-F$_1$-F$_2$-F$_3$. In the A$_1$ wide limit, the caustic consists of three four-cusped weak-shear Chang-Refsdal loops \citep{chang_refsdal84}. In the F$_3$ close limit, the caustic has one four-cusped weak-shear Chang-Refsdal loop for the star, plus two pairs of three-cusped strong-shear Chang-Refsdal loops for the planets.

The butterfly metamorphosis from A$_1$ to A$_2$, found on the caustic loop of the star at the cusp facing the planets, occurs already at a large distance of the planets -- approaching $s\approx4.66$ for $\mu \to 1$. The next butterfly to A$_5$ near $s\approx1.25$ occurs on the caustic loop of the star at the tip of the first butterfly feature. The three following beak-to-beak transitions all occur near $s=1$, the Einstein radius of the single lens. First, the two planetary four-cusped loops connect with the star's loop leading to C$_3$, one pair of three-cusped planetary loops then disconnects to E$_1$, and another pair disconnects to F$_1$. The last two metamorphoses occur on the star's loop: reverse butterflies near $s\approx0.800$ to F$_2$, and $s\approx0.214$ to F$_3$. We point out that even though the planetary loops of the caustic connect and disconnect when the planets lie close to the star's Einstein ring, the star's loop of the caustic is affected by having extra cusps for a much wider range of planetary distances, $s\in(0.214,\,4.66)$.

\noindent
{\em Close limit ---} for $\mu\neq1/3,0.768$ (subregions F$_3$, G$_1$, G$_3$): critical curve = Einstein ring + four small loops around Jacobian maxima; caustic = central four-cusped loop + four 3-cusped loops escaping to $\infty$. For $\mu=1/3$ (G$_2$): critical curve = Einstein ring + four small loops around Jacobian maxima; caustic = central six-cusped loop + three three-cusped loops escaping to $\infty$ + one three-cusped loop staying at center. For $\mu=0.768$ (J$_7$): critical curve = Einstein ring + three small loops around Jacobian maxima; caustic = central four-cusped loop + one four-cusped and two three-cusped loops escaping to $\infty$.

The $\mu=1/3$ equal-mass case exhibits a new close-limit behavior: the central caustic loop has six instead of four cusps, and one of the four three-cusped loops stays at the center instead of escaping to $\infty$ (see Figure~\ref{fig:TE-topologies}). The central caustic loop is self-intersecting with a three-fold symmetry, so that a pair of cusps points in the direction of each of the lens components. In the $s \to 0$ limit each pair rapidly converges to a point, so that the shape of the shrinking caustic loop approaches a simple symmetric three-cusped loop, which is traced twice as one follows the corresponding Einstein ring. The nature of the central loop can also be seen from the contour plot in the left column of Figure~\ref{fig:TE-contours}. The six branches of the cusp curve extending symmetrically to the outer edges have six intersections with the outer contours, which represent the perturbed Einstein-ring loop in the close limit.

\noindent
{\em Wide limit ---} for any $\mu$ (subregion A$_1$): critical curve = three independent Einstein rings with radii $\{\sqrt{\mu},\sqrt{(1-\mu)/2},\sqrt{(1-\mu)/2}\}$; caustic = three four-cusped weak-shear Chang-Refsdal loops.

\subsection{TI Model: Triangular Isosceles Configuration}
\label{sec:isosceles}

\noindent
{\em Model description ---} the final, topologically richest model has three equal-mass lens components lying at the vertices of an isosceles triangle, with legs of length $s$ spanning a vertex angle $\theta$ (see fourth sketch in Figure~\ref{fig:model-sketches}). If we place the origin at the geometric center, align the axis of symmetry with the real axis, and set component 1 at the vertex angle, the positions and masses of the components are $\{z_1,z_2,z_3\}= \{{-e^{{\rm i}\,\theta/2}-e^{-{\rm i}\,\theta/2}} , -e^{{\rm i}\,\theta/2}+2\,e^{-{\rm i}\,\theta/2} , 2\,e^{{\rm i}\,\theta/2}-e^{-{\rm i}\,\theta/2} \} \, s/3$ and $\{\mu_1,\mu_2,\mu_3\}= \{1/3,1/3,1/3\}$, with parameters ${\theta\in(0,\pi]}$ and ${s\in(0,\infty)}$.

In the $\theta=0$ limit components 2 and 3 coincide, reducing the system to a binary lens with masses $\{1/3,2/3\}$ and separation $s$. We recall that in the LA model both $p=0$ and $p=1$ edges correspond to a binary lens with the same mass ratio but with separation $2\,s$. The $\theta=\pi$ case corresponds to the collinear limit with component 1 at the center. Hence, the corresponding configurations are identical to the $\mu=1/3$ case of the LS model, as well as the $p=1/2$ case of the LA model. Similarly, configurations with vertex angle $\theta=\pi/3$ are identical to the $\mu=1/3$ configurations of the TE model.

\noindent
{\em Jacobian surface character ---} for $\theta\neq 0.124\,\pi,0.708\,\pi,0.847\,\pi\,$: four simple maxima + six simple saddles; for $\theta\approx0.708\,\pi\,$: two simple maxima + one double maximum + five simple saddles; for $\theta\approx0.124\,\pi$ and $\theta\approx0.847\,\pi\,$: four simple maxima + four simple saddles + one monkey saddle. In addition, the $\theta=\pi/3$ equilateral case has global three-fold symmetry, as discussed in \S~\ref{sec:equilateral_vertex}.

\noindent
{\em Topology boundaries ---} since the masses are equal, the center of mass $d_1=0$, $c_2=[1-4\,\sin^2{(\theta/2)}]\,s^2/3$, $d_2=2\,c_2/3$, and $c_3=-2\,\cos{(\theta/2)}[1+8\,\sin^2{(\theta/2)}]\,s^3/27$. The critical-curve polynomial
\beq
p_{crit}(z)=z^6-(2\,c_2+e^{2\,{\rm i}\,\phi})\,z^4-2\,c_3\,z^3+c_2^2\,z^2+\,2\,c_3(c_2-e^{2\,{\rm i}\,\phi})\,z+ c_3^2-c_2^2\,e^{2\,{\rm i}\,\phi}/3=0
\label{eq:critical-TI-poly}
\eeq
and the saddle-point polynomial
\beq
p_{sadd}(z)=z^6+c_2\,z^4 +7\,c_3\,z^3+c_2^2\,z^2-c_2 c_3\,z-c_2^3/3+c_3^2=0
\label{eq:saddle-TI-poly}
\eeq
cannot be solved analytically for a general $\theta$. We use the resultant method described in \S~\ref{sec:resultant} applied to polynomials $p_{crit}$ and $p_{sadd}$ to identify the boundaries of regions in parameter space with different critical-curve topologies.

The first resultant yields the polynomial equation
\begin{eqnarray}
p_{res}(w)&&\hspace{-6mm}=\left[8\,(1-\kappa)^3+27\,\kappa\right]w^6+ 6\,(1-\kappa)\left[4\,(1-\kappa)^3+9\,\kappa\right]s^2 w^5\nonumber\\ &&+ 24\,(1-\kappa)^5s^4 w^4-\left[8\,(1-\kappa)^6+198\,\kappa\,(1-\kappa)^3+297\,\kappa^2\right]s^6 w^3\nonumber\\ &&+9\,\kappa\,(1-\kappa)\left[5\,(1-\kappa)^3+42\,\kappa\right]s^8 w^2-54\,\kappa^2 (1-\kappa)^2 s^{10} w-27\,\kappa^3 s^{12}=0,
\label{eq:resultant-first-TI}
\end{eqnarray}
where $w=e^{2\,{\rm i}\,\phi}$ and the parameter $\kappa\equiv4\,\sin^2{(\theta/2)}$ is equal to unity for an equilateral configuration. The second resultant can be factorized to yield three independent conditions. The first two, $p_{res}(1)=0$ and  $p_{res}(-1)=0$, describe the splitting of the critical curve at $\phi=0$ and $\phi=\pi/2$, respectively, in the form of 6th degree polynomials in $s^2$. The third condition describes the splitting of the critical curve at other values of $\phi$. It has the form of a 15th degree polynomial in $s^4$, which we do not present here explicitly. An inspection following Appendix~\ref{sec:appendix-spurious} reveals that two segments of the corresponding curve have to be omitted, since they yield spurious solutions with no change in critical-curve topology.

The $p_{res}(1)=0$, $p_{res}(-1)=0$, and the 15th degree polynomial boundaries are marked in the parameter-space plot in the left panel of Figure~\ref{fig:TI-parspace} by the black, orange, and cyan curves, respectively. The boundaries reach the $\theta=0$ side of the plot at $s=(\sqrt[3]{2}-1)^{1/4}\approx0.714$ and $s=(\sqrt[3]{2}-1)^{-1/2}\approx1.96$, the $\theta=\pi$ side at $s\approx0.546$ and $s\approx1.68$, and the $s=0$ side at $\theta\approx0.708\,\pi$. At either of the $[\theta,s]\approx[0.124\,\pi,1.76]$ and $[0.847\,\pi,0.478]$ triple points the critical curve passes through a monkey saddle.

\noindent
{\em Critical-curve topologies ---} the division of parameter space by critical-curve topology is shown in the left panel of Figure~\ref{fig:TI-parspace}. The cusp-number map in the right panel and its blown-up details in Figure~\ref{fig:TI-details} include further subdivision by cusp number. The three curves obtained from the resultant analysis carve the parameter space into twelve different regions, marked by letters A -- L in the left panel, from top to bottom and left to right.

For each region, examples of critical curves with their corresponding caustics are shown in Figure~\ref{fig:TI-topologies}. The critical curve may have anywhere from one to five loops. The TI model has all the topologies found in the three previous models, plus one new topology: two separate loops, one of which contains a small third loop (region B). As seen from the TI row of Table~\ref{tab:topologies}, this model permits altogether nine different critical topologies, since the pairs of regions E+J, I+L, and H+K each share the same topology.

The beak-to-beak metamorphoses along most of the topology boundaries increase the number of critical-curve loops, with a decrease occurring only for the F-G, I-J, H-L, and K-L transitions. Similarly to the D-B transition of the TE model, the TI model has the special F-C transition, in which two simultaneous beak-to-beaks convert an outer-plus-inner loop combination to two separate loops, preserving the total number of loops.

\noindent
{\em Caustic structure ---} the cusp-number maps in the right panel of Figure~\ref{fig:TI-parspace} and in Figure~\ref{fig:TI-details} include subdivision due to additional caustic metamorphoses. Four of the five additional curves define conditions for the butterfly and one for the swallow-tail metamorphoses. Two of the butterfly curves asymptotically approach $\theta\to 0$ as $s\to\infty$: one leads down through A, E, and H to $[\theta,s]=[\pi/3,0]$, rising back through H, L, and K to $s\approx0.502$ at the right edge; the other leads down through A, B, and C, rising up through C and D to $s=\infty$ at $\theta=\pi/2$, dropping down through D and G to $s\approx1.20$ at the right edge. The second pair of butterfly curves rises from $[0.708\,\pi,0]$ to the right edge: one through L and K to $[\pi,0.502]$; the other through L, J, and G to $[\pi,1.20]$. All the butterfly metamorphoses defined by these four curves occur along the symmetry axis of the lens.

The swallow-tail curve starts asymptotically at $s\to\infty$ as $\theta\to 0$, dropping down through A, B, and C, rising up through C and D to a sharp peak at $\theta\approx0.220\,\pi$, bouncing back to another sharp peak reaching the second butterfly curve at $\theta=\pi/3$, dropping finally through D, G, J, and L to $[0.708\,\pi,0]$. Just as in the TE model, the curve corresponds to a complex-conjugate pair of swallow-tail metamorphoses everywhere except at the two sharp peaks. At the $\theta=\pi/3$ peak, three symmetric butterfly metamorphoses occur simultaneously, while at the $\theta\approx0.220\,\pi$ peak there is a complex-conjugate pair of butterfly metamorphoses. These are the only two cases within the TI model with butterflies occurring off the symmetry axis.

The additional set of curves divides region E into two subregions, regions B, C, H, J, and K into three subregions each, regions A and G into four, region D into six, and region L into eight subregions, with regions F and I remaining undivided. The parameter space of the TI model is thus divided into 41 subregions, corresponding to 28 different caustic-loop combinations. All are sorted by cusp number and listed in column TI of Table~\ref{tab:caustic_structures}. The total cusp number varies from 8 to 20 in steps of two. Individual caustic loops have 3, 4, 5, 6, 7, 8, 9, 10, 12 or 14 cusps. In comparison with the previous models there is no sixteen-cusped loop (B$_3$ in LS model), while five-cusped loops appear newly here (in subregion B$_1$).

\noindent
{\em Jacobian contour plots ---} Figure~\ref{fig:TI-contours} includes contour plots for three special vertical transects through the parameter space. Vertex angle $\theta\approx0.220\,\pi$ in the left column corresponds to the left sharp peak of the swallow-tail curve in Figure~\ref{fig:TI-parspace}. The caustic sequence from wide to close limit is D$_1$-D$_4$-C$_2$-C$_3$-F$_1$-E$_2$-H$_2$-H$_1$, with no single-loop topology and no swallow-tail metamorphosis. The inset includes the H$_1$-H$_2$ butterfly metamorphosis close to its right edge. Note the special F-C pair of beak-to-beaks at the ${\rm Re}[z]\approx-0.14$ saddle points, in which two nested loops reconnect to form two separate loops.

The D$_1$-D$_4$ simultaneous complex-conjugate pair of butterflies is indicated by the morph curve passing through self-intersection points of the cusp curve near ${\rm Re}[z]\approx0.15$. These self-intersections disappear for any perturbation of $\theta$. For lower values the cusp curve at the upper point splits along a SW-NE axis, for higher values along a NW-SE axis. In either case only one of the two separated branches intersects the morph curve, corresponding to a complex-conjugate pair of swallow tails.

The central $\theta=\pi/2$ column has the caustic sequence D$_2$-G$_2$-G$_3$-F$_1$-I$_1$-H$_2$-H$_1$. The G$_2$-G$_3$ pair of swallow tails occurs in the image plane on the central small figure-eight loop of the morph curve, and the H$_1$-H$_2$ butterfly can be seen at the left edge of the inset. The peculiar feature occurring at this angle is the six-cusped loop around the left component in the wide limit (D$_2$). In the double-maximum $\theta\approx0.708\,\pi$ case in the right column the caustic sequence D$_3$-D$_2$-G$_2$-J$_2$-L$_4$-L$_7$ ends with a four-cusped caustic loop corresponding to the double-maximum critical-curve loop.

Two further transects of the TI model can be found in contour plots of other models. The symmetric $\theta=\pi/3$ case appears as the $\mu=1/3$ left column of the TE-model Figure~\ref{fig:TE-contours} (TI-model caustic sequence: D$_1$-D$_6$-F$_1$-H$_2$). The linear $\theta=\pi$ case appears rotated by $\pi/2$ as the $p=0.5$ plot of the LA model in Figure~\ref{fig:LA-contours} (TI-model: D$_3$-G$_1$-G$_4$-K$_1$-K$_3$).

\noindent
{\em Planetary limits ---} the TI model has no planetary limit.

\noindent
{\em Close limit ---} for $\theta\neq\pi/3,0.708\,\pi$ (subregions H$_1$, H$_3$, K$_3$): critical curve = Einstein ring + four small loops around Jacobian maxima; caustic = central four-cusped loop + four three-cusped loops escaping to $\infty$. For $\theta=\pi/3$ (H$_2$): critical curve = Einstein ring + four small loops around Jacobian maxima; caustic = central six-cusped loop + three three-cusped loops escaping to $\infty$ + one three-cusped loop staying at center. For $\theta=0.708\,\pi$ (L$_7$): critical curve = Einstein ring + three small loops around Jacobian maxima; caustic = central four-cusped loop + one four-cusped and two three-cusped loops escaping to $\infty$.

\noindent
{\em Wide limit ---} it is the wide limit that is particularly interesting here. For any fixed $\theta$ the $s\to\infty$ limit corresponds to three single lenses with Einstein radii $\sqrt{1/3}$. However, for $\theta\sim s^{-1}$ the wide limit describes an equal-mass binary formed by components 2 and 3 with a distant object offset perpendicularly from the binary axis. This is analogous to the $p\sim s^{-1}$ wide limit in the LA model with a distant object lying along the binary axis (see \S~\ref{sec:linear_equal-mass}). The TI model has a total of nine different caustic regimes for arbitrarily large leg length $s\to\infty$. In order of increasing vertex angle $\theta$ these are: A$_1$, A$_2$, A$_3$, A$_4$, B$_1$, C$_1$, D$_1$, D$_2$, and D$_3$. The regimes can be grouped by the character of the binary lens formed by components 2 and 3: close (topology region A), intermediate (C), and wide (D). Topology B and the caustic subregion structure are the results of perturbation of the binary by the distant companion -- as well as perturbation of the companion by the binary.

The additional topology B (see Figure~\ref{fig:TI-topologies}) occurs at the close -- intermediate boundary. The influence of the companion causes the opposite inner loop to connect with the outer loop before the adjacent one, rather than both connecting simultaneously. In the asymptotic ${s\to\infty}$ regime B becomes negligible, as can be seen from the topology boundaries. The separation of binary components 2 and 3 (see Figure~\ref{fig:model-sketches}) along the A-B boundary is asymptotically $2\,s\,\sin{(\theta/2)}\simeq(1+s^{-2}/6-s^{-3}/6)/\sqrt{3}$, while along the B-C boundary we get $2\,s\,\sin{(\theta/2)}\simeq(1+s^{-2}/6+s^{-3}/6)/\sqrt{3}$. The two expressions differ only at the $s^{-3}$ order. For completeness, the asymptotic C-D boundary yields $2\,s\,\sin{(\theta/2)}\simeq\sqrt{8/3}$. Normalizing the asymptotic separations by the $\sqrt{2/3}$ Einstein radius of the binary, we recover the standard close -- intermediate A-(B)-C and intermediate -- wide C-D boundaries of an equal-mass binary.

The wide-limit caustic in regions A through C always has a four-cusped weak-shear Chang-Refsdal loop for the distant companion. The binary part of the caustic is dominated by the close-binary structure in A$_1$: three loops with 4+3+3 cusps. Increasing the separation causes the main loop to undergo two subsequent butterfly metamorphoses to A$_2$ (6+3+3) and A$_3$ (8+3+3), followed by a pair of reverse swallow tails to A$_4$ (4+3+3). One three-cusped loop merges with the main loop to B$_1$ (5+3), followed by the other to reach the intermediate regime C$_1$ (6), which then splits into two in the wide-binary regime D$_1$ (4+4). The structure of the binary part does not change further in the D region.

At larger vertex angles in region D, it is the companion loop of the caustic that undergoes an interesting transition. The butterfly entrance to D$_2$ turns the loop into a six-cusped self-intersecting curve. This structure disappears in a reverse butterfly when exiting to D$_3$. While for any other angle an increase in $s$ would eventually return the six-cusped loop to the regular four-cusped regime, for $\theta=\pi/2$ the peculiar shape persists for $s\to\infty$. In this limit each of the three pairs of cusps converge to a point, so that the shape approaches a three-cusped loop traced twice. The caustic loop is similar to the six-cusped loop in the close limit of the equal-mass TE model.

We stress that in this regime all three components lie at arbitrarily large separations. Still, the configuration alters the caustic loop of the component at the vertex fundamentally. This effect is caused by the gravitational fields of the other two components, which combine so that they exactly cancel the lensing shear term. Hence, the Chang-Refsdal limit is not valid in this special case. Even though this occurs in the TI model for $s\to\infty$ only for $\theta=\pi/2$, for any fixed not-too-large value of $s$ there is an interval of angles, for which the simple Change-Refsdal shape of the caustic loop cannot be assumed.

\subsection{Overview of critical-curve topologies and caustic structures}
\label{sec:topologies-summary}

In Table~\ref{tab:topologies} we summarize the occurrence of the nine different critical-curve topologies in the parameter spaces of the four studied triple-lens models. The topologies are listed first in order of increasing total number of loops, then by decreasing number of separate outer loops.

The total loop number ranges from one to five, with single 1-loop and 5-loop topologies, two different 2-loop and 4-loop topologies, and three different 3-loop topologies. The number of outer separate loops ranges from one to three, and the number of inner loops ranges from zero to four. Four of the topologies occur in all studied models: single loop, three separate loops (the generic wide limit), an outer loop with two inner loops, and an outer loop with four inner loops (the typical close limit). These four are the only topologies found in the LS model. The LA model has six, the TE model seven, and the TI model has all nine topologies.

The topologies found in the analyzed models are only examples of the possible topologies of triple-lens critical curves, so the list in Table~\ref{tab:topologies} is by no means exhaustive. All of the studied models are symmetric by their geometry and mass combination; all have at least one reflection symmetry. Breaking the symmetries of the models opens up the possibility of additional critical-curve topologies. However, finding the overall number of topologies of the triple lens is beyond the scope of this work. Such a survey would require a detailed investigation of the geometry of the Jacobian surface and the positions and types of its maxima, saddles, and poles. The correspondence with Jacobian contours may prove to be another useful tool, which may identify the possible sequences of topology changes when scaling the lens from the wide to the close limit.

Finally, we point out that the topologies discussed here are those that occur in two-dimensional areas of the parameter space of their models. One may proceed further and classify also the critical-curve topologies of all transitions between the different marked topology regions, such as those shown in Figure~\ref{fig:LS-transitions}. While this would be a straightforward extension, we do not include it here since the present work is extended enough already.

Table~\ref{tab:caustic_structures} presents an overview of all 32 different caustic structures in the parameter spaces of the four studied triple-lens models. Here we identify the caustic structure purely by the combination of cusp numbers on the individual loops. In the notation used in the first column of the table, for example 16/10+3+3 identifies a caustic with a total 16 cusps on three loops: one with 10, and two with 3 cusps each. The caustics are ordered in the table first by increasing total cusp number, then in decreasing order by the number of cusps on individual loops.

The total cusp number runs from 8 to 20 in increments of 2. The eight-cusped caustics have only a single-loop structure, the ten-cusped caustics have two, twenty-cusped caustics three, eighteen-cusped caustics five, twelve- and fourteen-cusped caustics six each, and sixteen-cusped caustics nine different structures. Individual loops have 3 -- 10, 12, 14, or 16 cusps. Least common are the five-cusped loop (the 12/5+4+3 caustic in the TI/B$_1$ subregion), and the sixteen-cusped loop (the 16/16 caustic in the LS/B$_3$ subregion). There are only four caustic structures that occur in all four studied models: 12/12, 12/4+4+4 (includes the generic wide limit), 16/4+3+3+3+3 (includes the generic close limit), and 20/8+3+3+3+3. The LS model has 10, the LA model has only 9, the TE model has 24, and the TI model has 28 different caustic structures, lacking only four of all the found possibilities.

The list in Table~\ref{tab:caustic_structures} is based purely on the cusp numbers on loops of the caustic. If we took into account the combination of critical-curve topology and caustic structure, the list would expand to 37 different cases. The reason for this is that a given caustic structure may occur for different mutual positions of critical-curve loops. For example, two four-cusped loops of the caustic may correspond to two separate loops of the critical curve, or to an outer and inner loop of the critical curve. An inspection of the topologies in Table~\ref{tab:topologies} shows that this is the case for the 12/4+4+4 caustic in the LS model from the A versus E regions; the 16/6+6+4 caustic in the TI model D region versus LS model E region; the 14/4+4+3+3 caustic in the LA and TI model A regions versus the TE model J region and TI model L region; the 16/6+4+3+3 caustic in the TI model A region versus the TE model J region and TI model L region; and the 18/8+4+3+3 caustic in the TI model A region versus the TE model J region and TI model L region.

The list in Table~\ref{tab:caustic_structures} does not include the caustics of lens configurations at metamorphosis points. The classification along subregion boundaries in parameter space would require additional counting of beak-to-beak, swallow-tail, and butterfly points. Even here we note that the caustic structures of the analyzed models are only examples of triple-lens caustics, and the list in Table~\ref{tab:caustic_structures} is not exhaustive. Analyzing less symmetric systems reveals that the frequency of metamorphosis types is different from most of the analyzed models, with swallow tails occurring more frequently than butterflies (as seen here in the LA model).

\section{Summary}
\label{sec:summary}

We presented an initial exploration of the character of critical curves and caustics of triple gravitational lenses. The topic is of particular interest for the analysis of observed planetary microlensing events. Already four published events involved triple-lens systems: in two cases the lens was a star with two planets \citep{gaudi_etal08,han_etal13}; in two cases a binary lens with a planet \citep{gould_etal14,poleski_etal14}. Moreover, even triple systems formed by a star with a planet with a moon might be detectable by microlensing \citep[e.g.,][]{liebig_wambsganss10}.

Our approach is based on analyzing the properties of the lens in the image plane, with the main relevant results from \cite{danek_heyrovsky15} introduced in \S~\ref{sec:n-point}. We describe in detail in \S~\ref{sec:mapping_topologies} and \S~\ref{sec:caustic} several analytical and numerical methods for parameter-space mapping of the topology of the critical curve and the number of cusps of the caustic. The methods are presented in a form applicable to any $n$-point-mass lens.

In \S~\ref{sec:triple-general} we present the main relevant polynomial equations for triple lenses in a compact form in terms of moments of the mass distribution. In \S~\ref{sec:linear_symmetric} -- \S~\ref{sec:isosceles} we apply the described methods to the analysis of four simple two-parameter models of triple lenses. For each of them in turn we discuss separately the Jacobian surface character, topology boundaries, critical-curve topologies, caustic structure, Jacobian contour plots, planetary limits, close limit, and wide limit. The combined results explain the properties of lenses on four 2D cuts through the full 5D parameter space of the general  triple lens. Since each of the studied models intersects at least one other along a 1D set of parameter combinations, the presented results may serve as a reference framework for further studies.

The results include the description of a range of triple-lens features, such as the occurrence of swallow-tail and butterfly metamorphoses, Jacobian monkey saddles, or double maxima. We demonstrated that the appearance of two new cusps in the beak-to-beak metamorphosis is not always accompanied by the splitting of critical-curve loops. It may lead to the opposite, the merger of loops, as discussed in the caustic structure paragraphs of \S~\ref{sec:linear_symmetric}. Finally, two simultaneous beak-to-beaks may lead to a reconnection preserving the number of loops. Within the studied models this occurs along the D-B transition in the TE model, and along the F-C transition in the TI model.

The planetary limits are particularly noteworthy in the equilateral TE model. The $\mu\to 0$ limit shows that a planet placed near the close -- intermediate beak-to-beak point of a binary lens may turn the single transition into a sequence of six caustic metamorphoses. The $\mu\to 1$ limit demonstrates that a pair of planets may form additional cusps on the Chang-Refsdal four-cusped caustic of the star at a very broad range of separations, in this case $s\in(0.214,\,4.66)$ Einstein radii.

Unusual behavior in the close limit occurs in the presence of double maxima (LS, TE, and TI models), and in the equal-mass equilateral lens (TE and TI models). Any double maximum causes the lens to have a four-cusped escaping caustic loop instead of a pair of three-cusped loops \citep{bozza00b}. The symmetry of the equal-mass equilateral lens causes it to have a self-intersecting six-cusped (asymptotically three-cusped) primary caustic loop, instead of the generic four-cusped loop. In addition, one of the weak three-cusped caustics stays at the origin instead of escaping away.

Very interesting wide-limit behavior was found in the TI model with vertex angle ${\theta=\pi/2}$. Here one of the caustic loops retains a self-intersecting six-cusped shape approaching a three-cusped shape for arbitrarily large separations of all components. Such a lens violates the generic Chang-Refsdal limit \citep{chang_refsdal84}. In this case the gravitational fields of the two other components exactly combine to cancel the lensing shear term. The shape is thus dominated by higher-order terms, leading to its different geometry.

For smaller angles, the wide limit of the TI model demonstrates the sensitivity of the binary caustic structure of the two close components to a distant companion. Here the presence of the third perpendicularly offset body perturbs primarily the caustic structure near the close -- intermediate transition. As a result, the caustic undergoes a sequence of five metamorphoses instead of a single one.

We compiled the results found for all four models in the tables presented in \S~\ref{sec:topologies-summary}. The critical curves have 9 different topologies, and the caustics have 32 different structures, when identified by the combination of numbers of cusps on individual loops. A joint classification of critical-curve topologies and caustic structures increases the number further to 37 different situations. The lists are just a sample of the full range provided by the general triple lens.

In Table~\ref{tab:caustic_structures} we provided for each model the parameters of examples of all different caustic structures. These can be used to guide more detailed studies of critical-curve and caustic metamorphoses, to generate examples for testing light-curve fitting codes and algorithms, or as a starting point for exploring the behavior of less symmetric triple lenses.

\acknowledgements

We thank the anonymous referee for suggestions that helped improve the manuscript. Work on this project was supported by Czech Science Foundation grant GACR P209-10-1318 and Charles University grant SVV-260089.

\appendix
\section{Computing the resultant of two polynomials}
\label{sec:appendix-resultant}

A polynomial $f(x)$ of degree $d\geq1$ with roots $\xi_1, \ldots \xi_d\,$ can be written as
\beq
f(x) =\sum_{j=0}^d a_j x^j=a_d\prod_{j=1}^d (x-\xi_j)\,,
\label{eq:polynomial_f}
\eeq
where $a_d\neq0$. Similarly, for another polynomial $g(x)$ of degree $e\geq1$ with roots $\eta_1, \ldots \eta_e\,$ we have
\beq
g(x) =\sum_{j=0}^e b_j x^j=b_e\prod_{j=1}^e (x-\eta_j)
\label{eq:polynomial_g}
\eeq
with $b_e\neq0$. The resultant of the two polynomials is a function of their coefficients that is proportional to the product of the differences of all combinations of individual roots, specifically \citep[e.g.,][]{escofier01},
\beq
{\rm Res}_x(f,g)=a_d^e\, b_e^d\,\prod_{j=1}^d\,\prod_{k=1}^e\,(\xi_j-\eta_k)\,.
\label{eq:resultant}
\eeq
Using equations~(\ref{eq:polynomial_f}) and (\ref{eq:polynomial_g}), we can also write the resultant in the form \citep[e.g.,][]{sturmfels02}
\beq
{\rm Res}_x(f,g)=a_d^e\,\prod_{j=1}^d\,g(\xi_j)=(-1)^{d\,e}\,b_e^d\,\prod_{k=1}^e\,f(\eta_k)\,.
\label{eq:resultant_1}
\eeq
It follows from equation~(\ref{eq:resultant}) that the condition for the two polynomials to have at least one common root is equivalent to the condition
\beq
{\rm Res}_x(f,g)=0\,,
\label{eq:resultant_0}
\eeq
an equation in terms of the coefficients of the polynomials.

A common way to compute the resultant is based on the Sylvester matrix $S(f,g)$. This $(d+e)\times(d+e)$ matrix is formed by arranging the coefficients of $f(x)$ from highest to lowest order, filling the sequence repeatedly in the first $e$ rows gradually staggered to the right, followed by $d$ rows of coefficients of $g(x)$ staggered in a similar manner, filling the rest with zeros \citep[e.g.,][]{petters_etal01}. Explicitly,
\beq
S(f,g)=  \left(
  \begin{array}{lllllllll}
      a_d & a_{d-1} & \multicolumn{2}{c}{\cdots} & a_0 & 0 & \cdots & 0 \\
        0 & a_d & \multicolumn{2}{c}{\cdots} & a_1 & a_0 & \cdots & 0 \\
        \multicolumn{2}{c}{\vdots} &  &  &  &  &  & \vdots \\
        0 & 0 & \multicolumn{5}{c}{\cdots} & a_0 \\
      b_e & b_{e-1} & \cdots & b_0 & 0 & \multicolumn{2}{c}{\cdots} & 0 \\
        0 & b_e & \cdots & b_1 & b_0 & \multicolumn{2}{c}{\cdots} & 0 \\
        \multicolumn{2}{c}{\vdots} &  &  &  &  &  & \vdots \\
        0 & 0 & \multicolumn{5}{c}{\cdots} & b_0 \\
  \end{array}
  \right)\,.
\label{eq:Sylvester}
\eeq
The resultant is then obtained simply by computing the determinant of the Sylvester matrix,
\beq
{\rm Res}_x(f,g)={\rm det}\,S(f,g)\,.
\label{eq:resultant-Sylvester}
\eeq
In some texts \citep[e.g.,][]{escofier01} the Sylvester matrix is defined as the transpose of the matrix in equation~(\ref{eq:Sylvester}), i.e., with coefficients in staggered columns rather than rows. However, this does not change the determinant, so that equation~(\ref{eq:resultant-Sylvester}) remains valid. In other texts the form of the matrix in equation~(\ref{eq:Sylvester}) remains the same, but the coefficients are filled in reverse, from lowest to highest order \citep[e.g.,][]{erdl_schneider93,sturmfels02}. The determinant of such a matrix is not always equal to the resultant defined by equation~(\ref{eq:resultant}), it differs by a factor $(-1)^{d\,e}$. However, this difference plays no role when seeking the null-resultant condition given by equation~(\ref{eq:resultant_0}), which is equivalent to ${\rm det}\,S(f,g)=0$.

An alternative way to compute the resultant utilizes the smaller $n\times n$ B\'ezout matrix $B(f,g)$, where $n={\rm Max}\,(d,e)$. The matrix is obtained by anti-symmetrizing the direct product of the polynomials, dividing the result by the difference of their variables, and ordering the terms according to the powers of the variables. The elements of the matrix are thus defined by \citep[e.g.,][]{sturmfels02}
\beq
\sum_{j,k=0}^{n-1} B_{jk}(f,g)\,x^j\,y^k =\frac{f(x)\,g(y)-f(y)\,g(x)}{x-y}\,.
\label{eq:Bezout}
\eeq
If we assume that the degree of the first polynomial $d\geq e$, substituting the expressions from equations~(\ref{eq:polynomial_f}) and (\ref{eq:polynomial_g}) in equation~(\ref{eq:Bezout}) then yields $B_{jk}(f,g)$ in terms of the coefficients:
\beq
\sum_{j,k=0}^{d-1} B_{jk}(f,g)\,x^j\,y^k =\sum_{j,k=0}^{d-1} x^j\,y^k\,\hspace{-3mm}\sum_{l={\rm Max}(j+1,k+1)}^{{\rm Min}(d,\,j+k+1)}\hspace{-5mm}a_l\,b_{j+k-l+1}\, -\sum_{j,k=0}^{e-1} x^j\,y^k\,\hspace{-3mm} \sum_{l={\rm Max}(j+1,k+1)}^{{\rm Min}(e,\,j+k+1)}\hspace{-5mm}a_{j+k-l+1}\,b_l\,.
\label{eq:Bezout-coefficients}
\eeq
For polynomials of the same degree ($d=e$) the result simplifies to directly yield
\beq
B_{jk}(f,g) =\sum_{l={\rm Max}(j+1,k+1)}^{{\rm Min}(d,\,j+k+1)}\hspace{-5mm}(a_l\,b_{j+k-l+1}\,-a_{j+k-l+1}\,b_l)\,.
\label{eq:Bezout-coefficients-special}
\eeq
Assuming the more general $d\geq e$ case, the resultant is computed as the determinant of the B\'ezout matrix multiplied by a non-zero factor:
\beq
{\rm Res}_x(f,g)=(-1)^{d(d-1)/2}\,a_d^{e-d}\,{\rm det}\,B(f,g)\,.
\label{eq:resultant-Bezout}
\eeq
Hence, the condition for a common root from equation~(\ref{eq:resultant_0}) is equivalent to ${\rm det}\,B(f,g)=0$.

The B\'ezout matrix has the advantage of being smaller, while the Sylvester matrix is usually sparser. If any particular symbolic-manipulation software package fails to compute the determinant of one of them, one can try using the other, before resorting to numerical computations.

\section{Spurious results of the resultant method}
\label{sec:appendix-spurious}

The polynomial $p_{res}(w)$ defined by equation~(\ref{eq:resultant-first-poly}) has a set of roots $w_1,\,\ldots\,w_m$. The polynomial $p_{conj}(w)$ defined by equation~(\ref{eq:resultant-first-conj}) then has roots $\bar{w}_1^{-1},\,\ldots\,\bar{w}_m^{-1}$. The search for a common unit root using equation~(\ref{eq:resultant-second}) assumes that for at least one of the roots $w_j=\bar{w}_j^{-1}$, and hence it lies on the unit circle. However, equation~(\ref{eq:resultant-second}) is satisfied even by a non-unit root $w_j$ if it is equal to the inverse conjugate of a different root, i.e., $w_j=\bar{w}_k^{-1}$ for some other $w_k\neq w_j$. This implies that $w_k$ is another non-unit root satisfying equation~(\ref{eq:resultant-second}). In case $p_{res}$ has only real coefficients (e.g., in the case of the two-point-mass lens or the triple-lens models from \S~\ref{sec:linear_symmetric} -- \S~\ref{sec:isosceles}) and $w_j$ lies off the real axis, $\bar{w}_j$ and $\bar{w}_k$ are two more non-unit roots of $p_{res}$. The total number of such spurious solutions depends on the degree of the polynomial $p_{res}$, which is $\leq 3\,n-3$, as discussed in \S~\ref{sec:resultant}.

In the case of the two-point-mass lens, the polynomial given by equation~(\ref{eq:resultant-first-binary}) is of third degree and has real coefficients. Spurious solutions off the real axis would require $p_{res}$ to have at least four different roots  $\{w_1,\,w_2=\bar{w}_1^{-1},\,\bar{w}_1,\,\bar{w}_2=w_1^{-1}\}$, hence they do not occur. A spurious solution $w_1$ along the real axis would require $p_{res}$ to have a set of real roots $\{w_1,\,w_2=w_1^{-1},\,w_3\}$ and $p_{conj}$ a corresponding set of roots $\{w_1^{-1},\,w_1,\,w_3^{-1}\}$. We could then write
\beq
p_{res}(w)=(w-w_1)(w-w_1^{-1})(w-w_3)\,.
\label{eq:spurious}
\eeq
Expanding this expression and comparing it term-by-term with equation~(\ref{eq:resultant-first-binary}), we find that the only possible real solution is $w_1=w_2=1$, which lies on the unit circle and thus is not spurious. Moreover, this double-root solution requires either $\mu=0$ or $\mu=1$, and thus it corresponds to the single-lens limit rather than a genuine binary. The nonexistence of real spurious solutions can also be seen from the discriminant of $p_{res}$ computed from equation~(\ref{eq:resultant-first-binary}). The result is always negative for $\mu\neq0,\,0.5,\,1$. Therefore, except for these values $p_{res}$ has one root on and two roots off the real axis. For the only other non-degenerate case, $\mu=0.5$, the roots are $\{-2\,s^2,\,-2\,s^2,\,s^2/4\}$. We see that even in this case there is no reciprocal pair of roots and hence no spurious solution. For the two-point-mass lens all solutions of equation~(\ref{eq:resultant-second}) thus correspond to topology changes of the critical curve.

In all the triple-lens cases analyzed in \S~\ref{sec:linear_symmetric} -- \S~\ref{sec:isosceles}, spurious solutions do occur. In the LS model the 6th degree saddle-point and critical-curve polynomials in fact can be studied as 3rd degree polynomials in $z^2$. The first resultant then leads to a 3rd degree $p_{res}(w)$, which can be shown to have a pair of real spurious roots $w_1=w_2^{-1}$ when the central mass $\mu<0.2$. In the three other models $p_{res}(w)$ is of 6th degree. Regions in parameter space with spurious solutions can be found for each of them.

\section{Cusp and morph curve polynomials of the LS model}
\label{sec:appendix-curves}

We present here the explicit form of the cusp and morph curve polynomials for the simplest triple-lens model studied in this work, the linear symmetric model from \S~\ref{sec:linear_symmetric}.

The cusp curve is generally defined by equation~(\ref{eq:cusp_curve_par1}), with $n=3$ for the triple lens. We first multiply the equation by $(z-z_1)^6(z-z_2)^6(z-z_3)^6$. Next we substitute the positions and masses of the LS-model components, $\{z_1,z_2,z_3\}=\{-s,0,s\}$ and $\{\mu_1,\mu_2,\mu_3\}= {\{(1-\mu)/2,\mu,(1-\mu)/2\}}$. We obtain the parametric polynomial form of the cusp curve,
\begin{eqnarray}
&&\hspace{-6mm}p_{cusp}(z)=\nonumber \\ &&\hspace{-3mm}(1-\Lambda)\,z^{12}+3\,[\,2-4\mu+\Lambda(3\mu-1)\,]\,s^2z^{10}+ 3\,[\,3-10\mu+12\mu^2-\Lambda(1-5\mu+9\mu^2)\,]\,s^4z^8\nonumber \\ &&\hspace{-3mm}+[\,4\mu(4-9\mu) -\Lambda(1-3\mu+9\mu^2-27 \mu^3)\,]\,s^6z^6-3\,[\,2-7\mu+\Lambda(1-5\mu+9\mu^2)\,]\,\mu\, s^8z^4\nonumber \\ &&\hspace{-3mm} -3\,[\,2+\Lambda(1-3\mu)\,]\,\mu^2s^{10}z^2+(1-\Lambda\mu)\,\mu^2s^{12}=0\,,
\label{eq:cuspcurve-LS-poly}
\end{eqnarray}
where $\Lambda>0$ is a real non-negative parameter.

In a similar manner we convert the morph-curve equation~(\ref{eq:morph_curve_par1}) for $n=3$, using exactly the same steps. After multiplication by $(z-z_1)^6(z-z_2)^6(z-z_3)^6$ and substitution of the LS model positions and masses, we obtain the parametric polynomial form of the morph curve,
\begin{eqnarray}
&&\hspace{-6mm}p_{morph}(z)=\nonumber \\ &&\hspace{-3mm}-{\rm i}\,\Gamma z^{12} +[\,1-\mu-6\,{\rm i}\,\Gamma(1-2\mu)\,]\,s^2z^{10}-[\,2-8\mu+6\mu^2+3\,{\rm i}\,\Gamma (3-10\mu+12\mu^2)\,]\,s^4z^8\nonumber \\ &&\hspace{-3mm}+[\,1-12\mu+11\mu^2-4\,{\rm i}\,\Gamma\mu (4-9\mu)\,]\,s^6z^6+\,[\,4-4\mu+3\,{\rm i}\,\Gamma(2-7\mu)\,]\,\mu\,s^8z^4\nonumber \\ &&\hspace{-3mm} +[\,1-\mu+6\,{\rm i}\,\Gamma\mu\,]\,\mu\,s^{10}z^2-{\rm i}\,\Gamma\mu^2s^{12}=0\,,
\label{eq:morphcurve-LS-poly}
\end{eqnarray}
where $\Gamma$ is a real parameter.

We see that in the LS model both polynomials are of sixth degree in $z^2$. Intersections of the cusp curve given by equation~(\ref{eq:cuspcurve-LS-poly}) and the critical curve given by equation~(\ref{eq:critical-LS-poly}) identify cusp images along the critical curve. Intersections of the morph curve given by equation~(\ref{eq:morphcurve-LS-poly}) and the cusp curve identify the images of caustic-metamorphosis points in the image plane, as discussed in \S~\ref{sec:caustic}.

\clearpage
\bfi
\includegraphics[scale=.3]{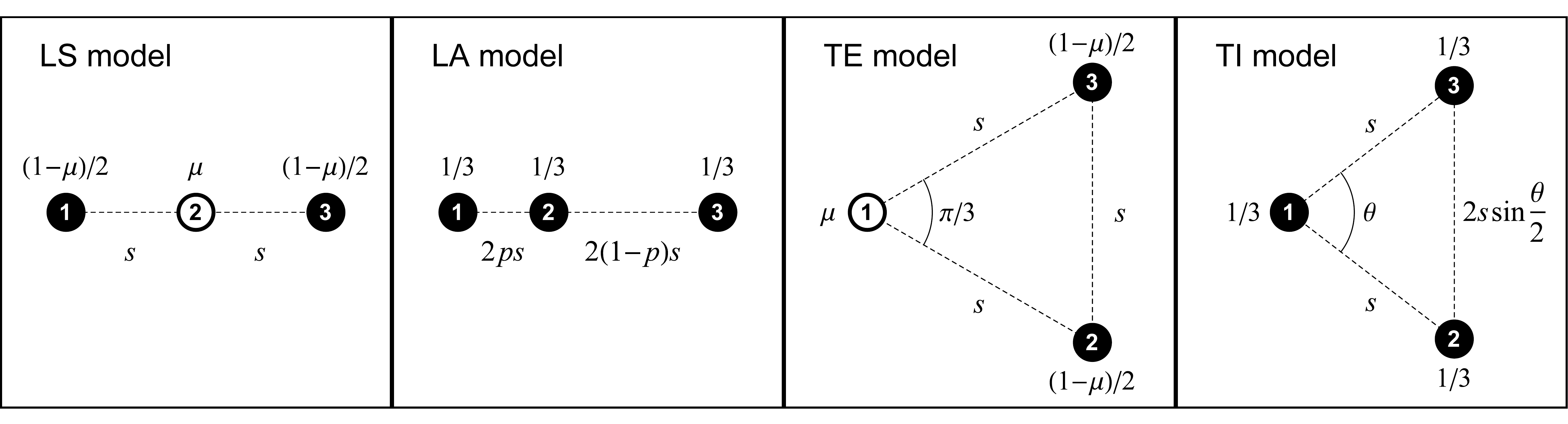}
\caption{Configurations of the analyzed triple-lens models. Lens components are marked by numbered circles and labelled by their masses; black circles indicate equal mass. Model parameters: Linear Symmetric -- central mass $\mu$, separation $s$ of neighbors; Linear Asymmetric -- relative position $p$ of central component, half-separation $s$ of outer components; Triangular Equilateral -- mass $\mu$ of component 1, side length $s\,$; Triangular Isosceles -- vertex angle $\theta$, leg length $s\,$.}
\label{fig:model-sketches}
\efi

\clearpage
\bfi
\hspace{-2mm}
\includegraphics[scale=.9]{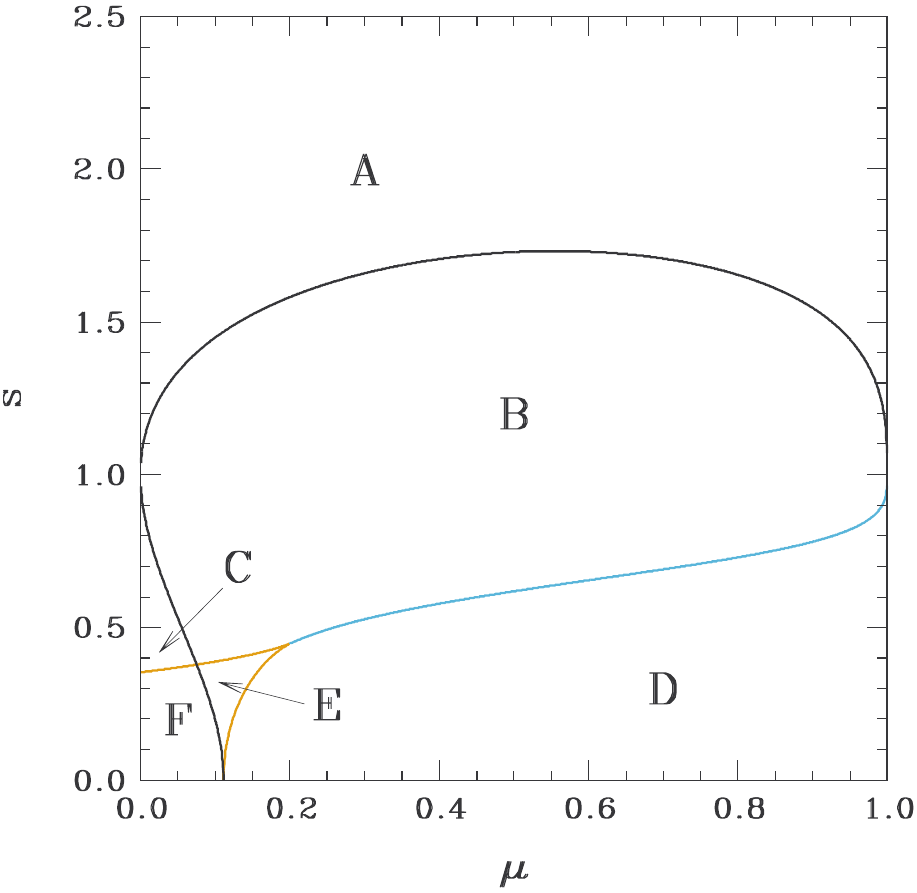}
\hspace{2mm}
\includegraphics[scale=.9]{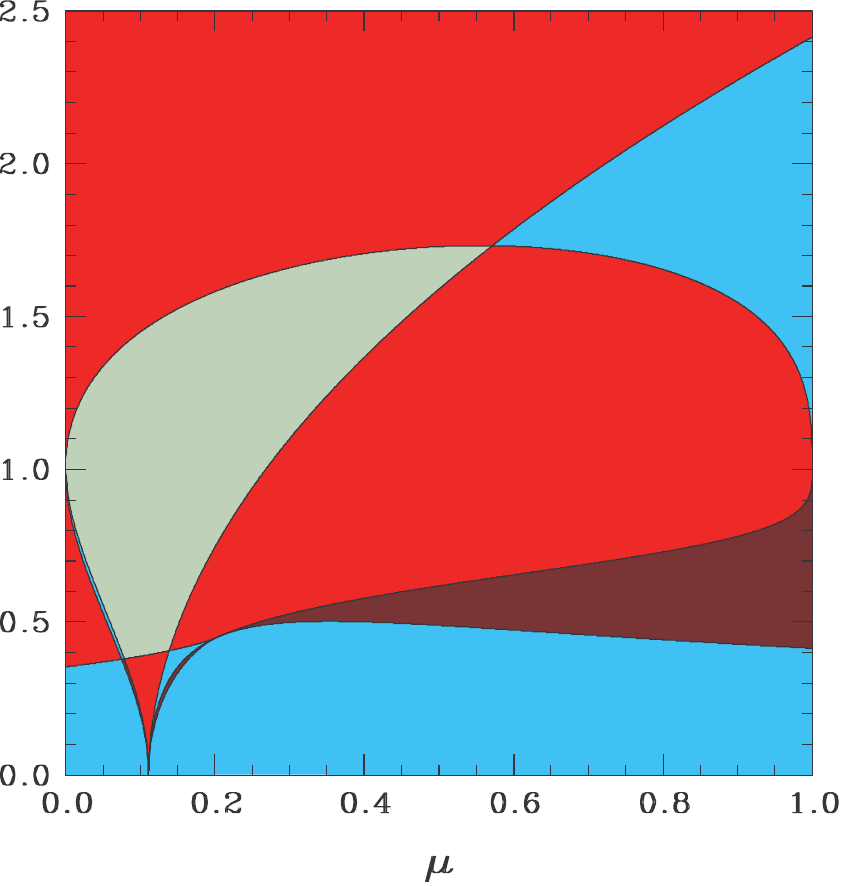}
\caption{LS model parameter-space division. Left panel: by critical-curve topology (marked by letters), with black boundary given by equation~(\ref{eq:topo-boundary-LS-w+1}), orange by equation~(\ref{eq:topo-boundary-LS-w-1}), cyan by equation~(\ref{eq:topo-boundary-LS-w-other}). Right panel: by total number of cusps on caustic [\,8 cusps -- gray; 12 -- red; 16 -- cyan; 20 -- brown]. Region near lower left corner of right panel is blown up in Figure~\ref{fig:LS-detail}.}
\label{fig:LS-parspace}
\efi

\clearpage
\bfi
\includegraphics[scale=2.2]{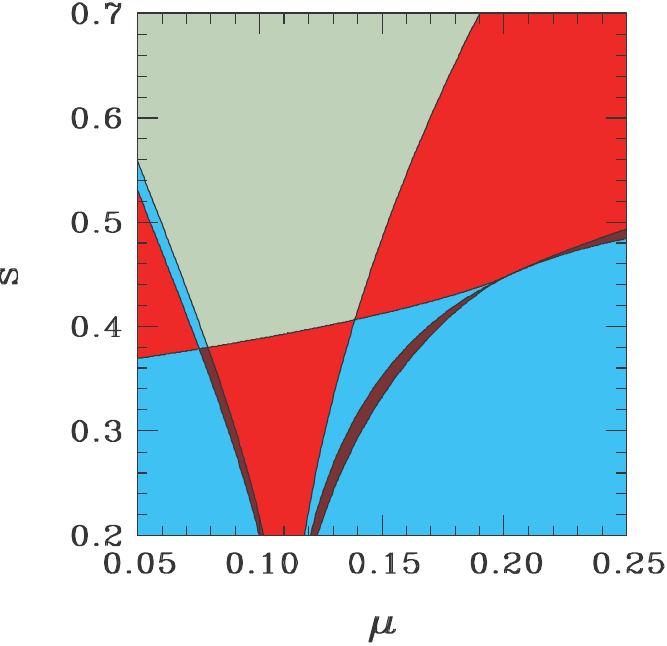}
\caption{LS model: detail of parameter-space division by total cusp number from Figure~\ref{fig:LS-parspace}.}
\label{fig:LS-detail}
\efi

\clearpage
\bfi
\includegraphics[width=16.5 cm]{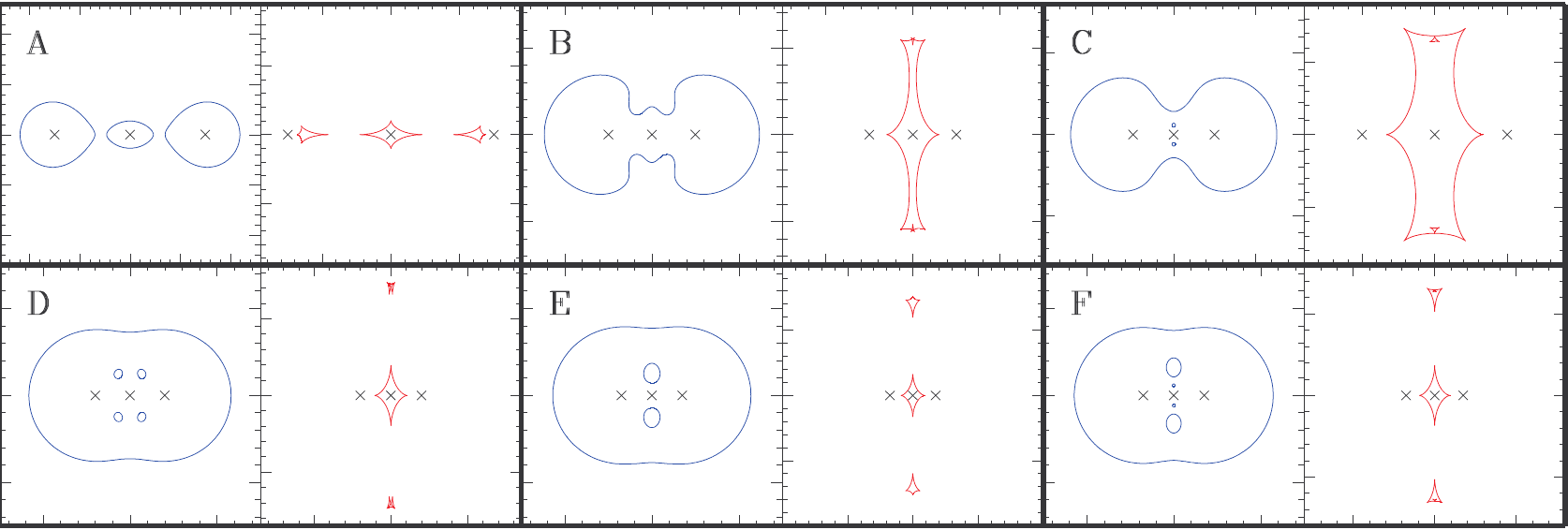}
\caption{LS model: gallery of topologies of critical curves (blue) and corresponding caustics (red) with lens positions marked by black crosses. Letters correspond to regions marked in Figure~\ref{fig:LS-parspace}. Caustic subregions and lens parameters $[\mu, s]$ of examples: A$_1$ $[0.1,1.5]$, B$_2$ $[0.2,0.5]$, C$_1$ $[0.04,0.5]$, D$_2$ $[0.2,0.4]$, E$_3$ $[1/9,0.35]$, F$_1$ $[0.07,0.35]$.}
\label{fig:LS-topologies}
\efi

\clearpage
\bfi
\includegraphics[width=16.5 cm]{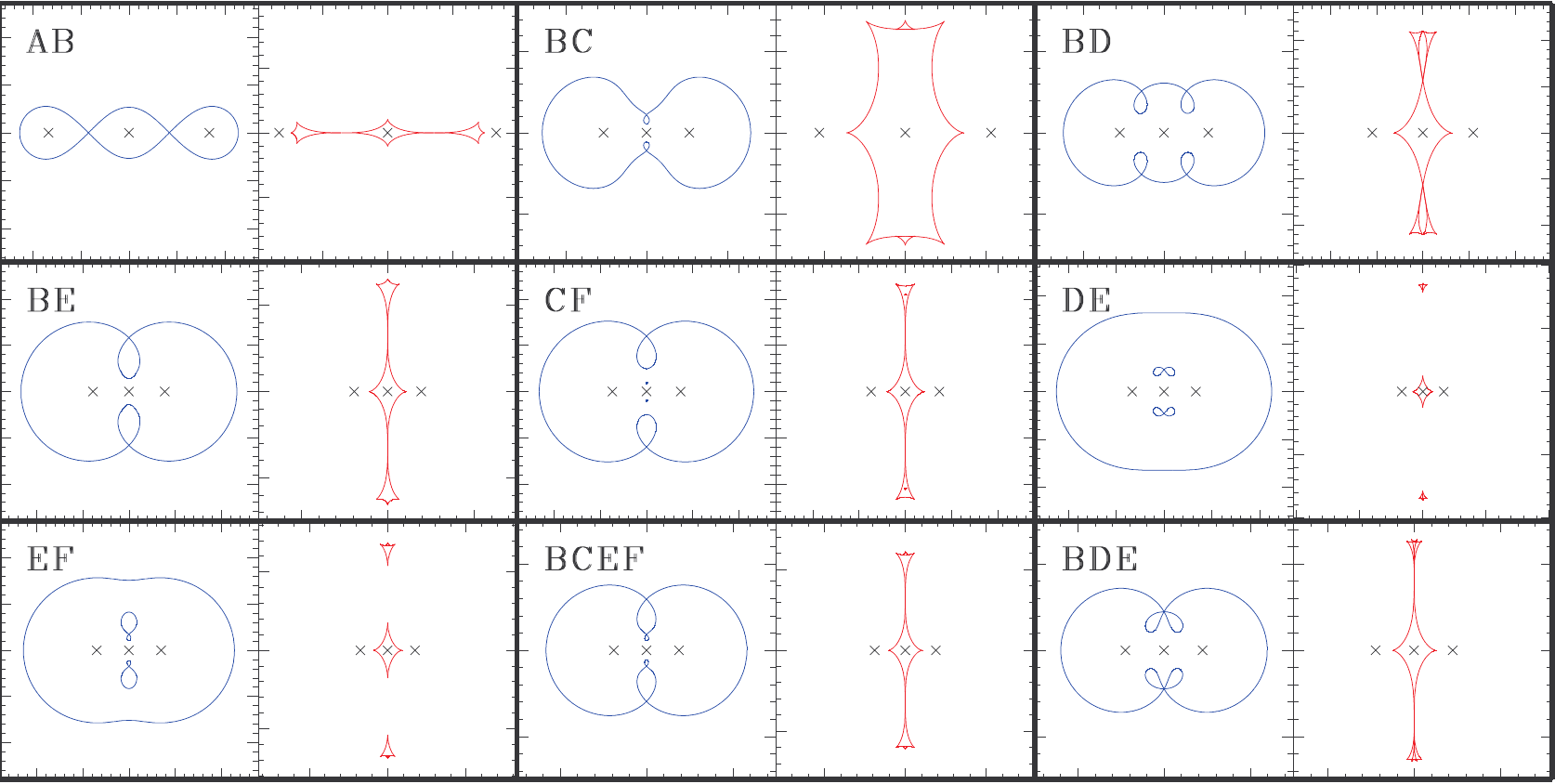}
\caption{LS model: gallery of critical curves (blue) and corresponding caustics (red) in transitions between the regions marked in Figure~\ref{fig:LS-parspace}. Letters denote all adjacent regions. Lens parameters $[\mu, s]$ of examples: AB $[1/3,1.6777]$, BC $[0.05,0.53075]$, BD $[1/3,0.54593]$, BE $[0.1,0.38852]$, CF $[0.05,0.36951]$, DE $[0.15,0.33]$, EF $[0.08,0.34866]$, BCEF $[0.07530,0.37866]$, BDE $[1/5,5^{-1/2}]$.}
\label{fig:LS-transitions}
\efi

\clearpage
\bfi
\includegraphics[width=16.5 cm]{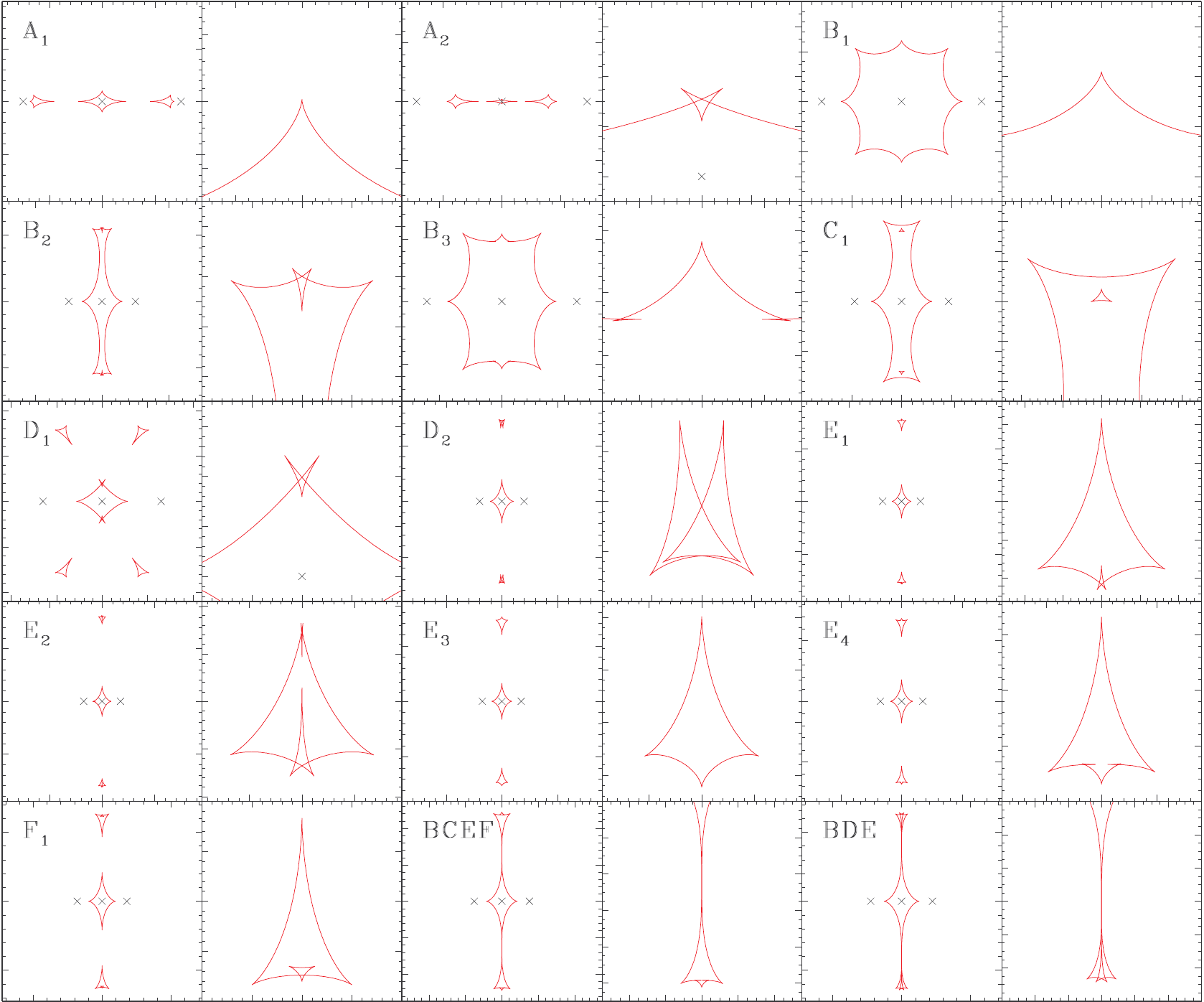}
\caption{LS model: gallery of caustics with blown-up details from 13 subregions of right panel of Figure~\ref{fig:LS-parspace}, plus two transition caustics. Labels mark cusp-number subregions, with numbers assigned within given region from top to bottom and left to right in Figure~\ref{fig:LS-parspace}. Lens parameters $[\mu, s]$ of examples: A$_1$ $[0.1,1.5]$, A$_2$ $[0.95,1.45]$, B$_1$ $[0.2,0.8]$, B$_2$ $[0.2,0.5]$, B$_3$ $[0.04,0.6]$, C$_1$ $[0.04,0.47]$, D$_1$ $[0.7,0.65]$, D$_2$ $[0.2,0.4]$, E$_1$ $[0.15,0.36]$, E$_2$ $[0.155,0.35]$, E$_3$ $[1/9,0.35]$, E$_4$ $[0.08,0.35]$, F$_1$ $[0.07,0.36]$; BCEF $[0.07530,0.37866]$, BDE $[1/5,5^{-1/2}]$.}
\label{fig:LS-caustics}
\efi

\clearpage
\bfi
\hspace*{-2mm}
\includegraphics[scale=.38]{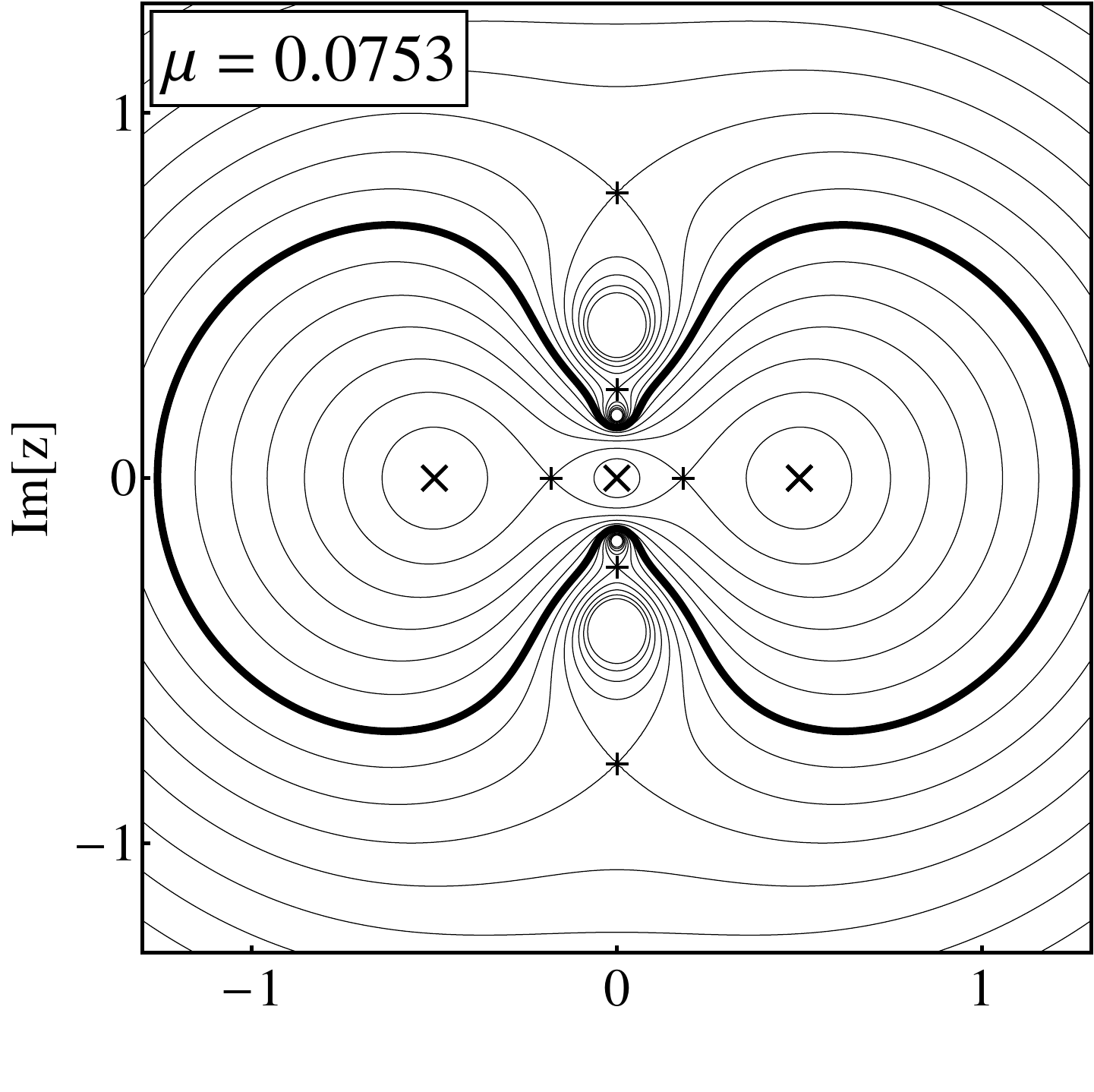}
\hspace{-5mm}
\includegraphics[scale=.38]{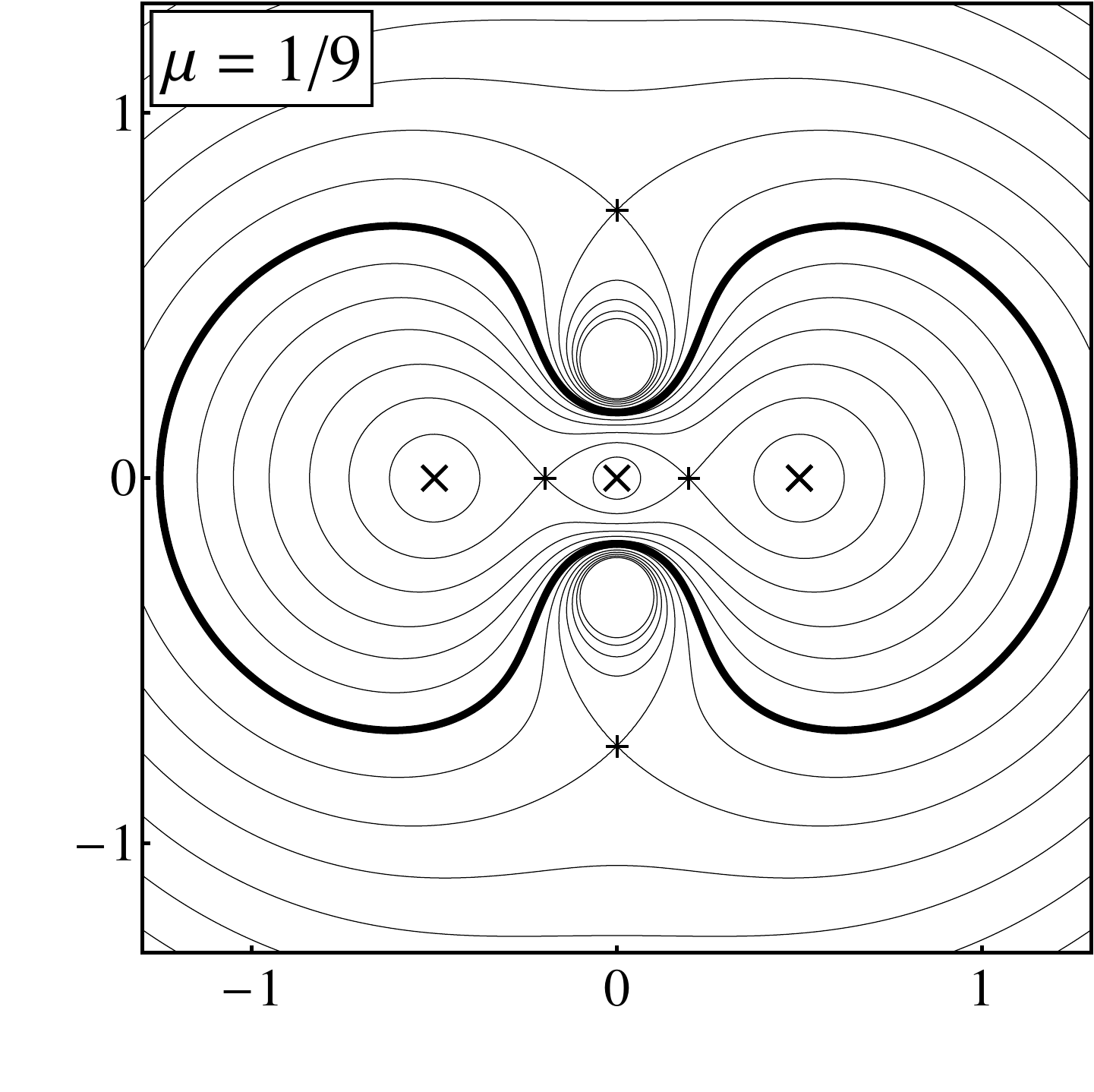}
\hspace{-5mm}
\includegraphics[scale=.38]{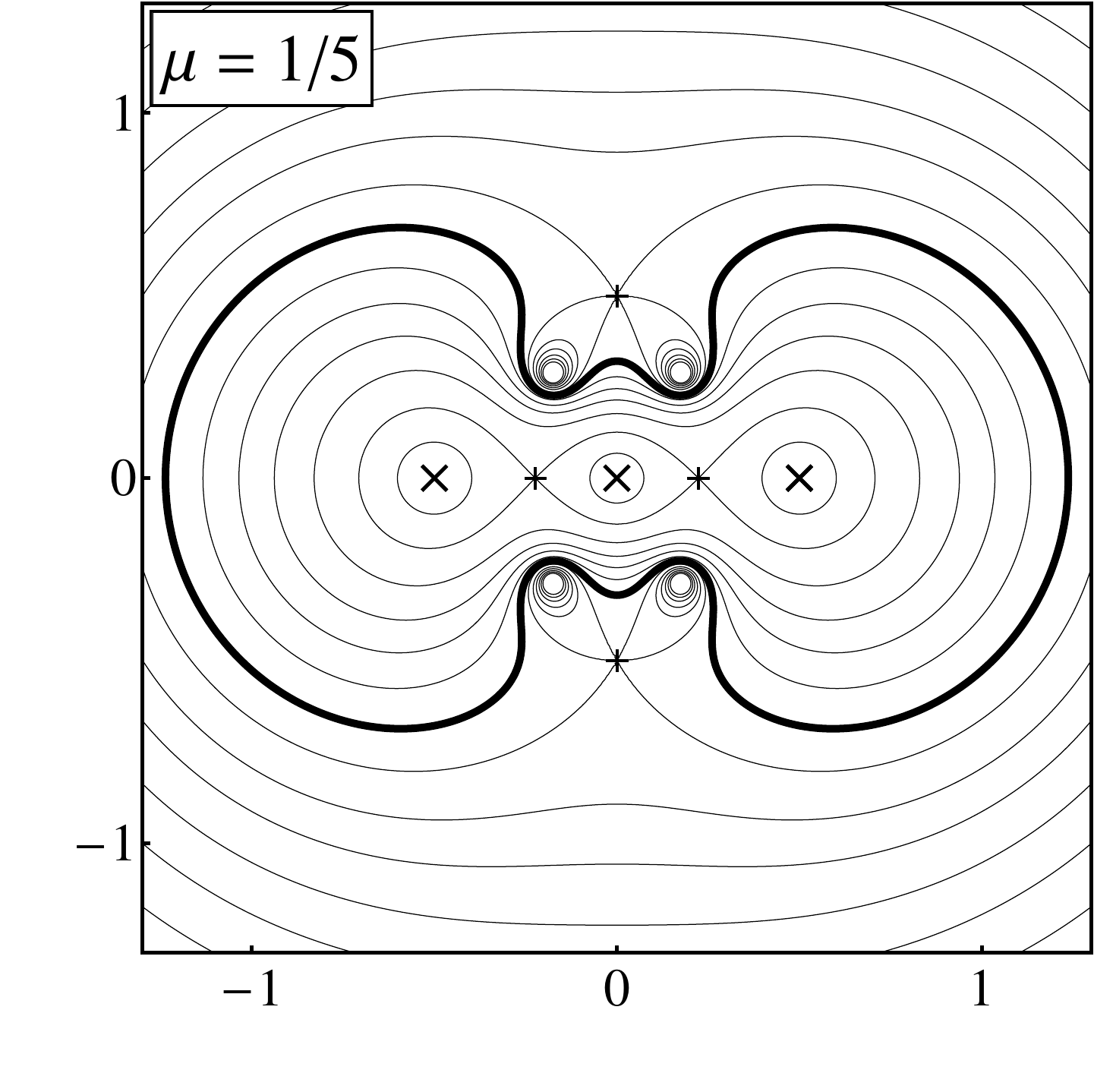}
\vspace*{-3mm}
\\
\hspace*{-1.6mm}
\includegraphics[scale=.38]{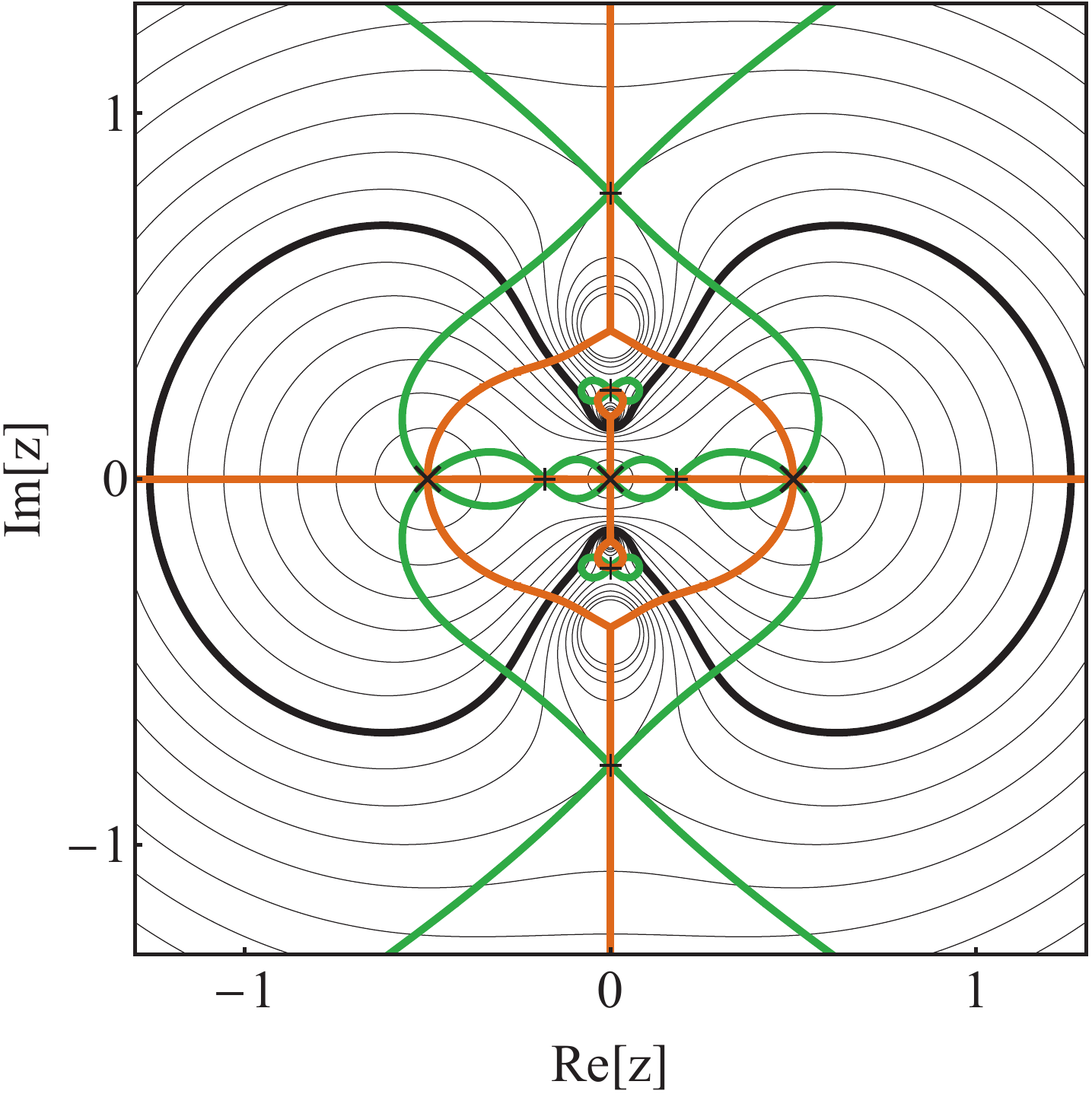}
\hspace{-1.4mm}
\includegraphics[scale=.38]{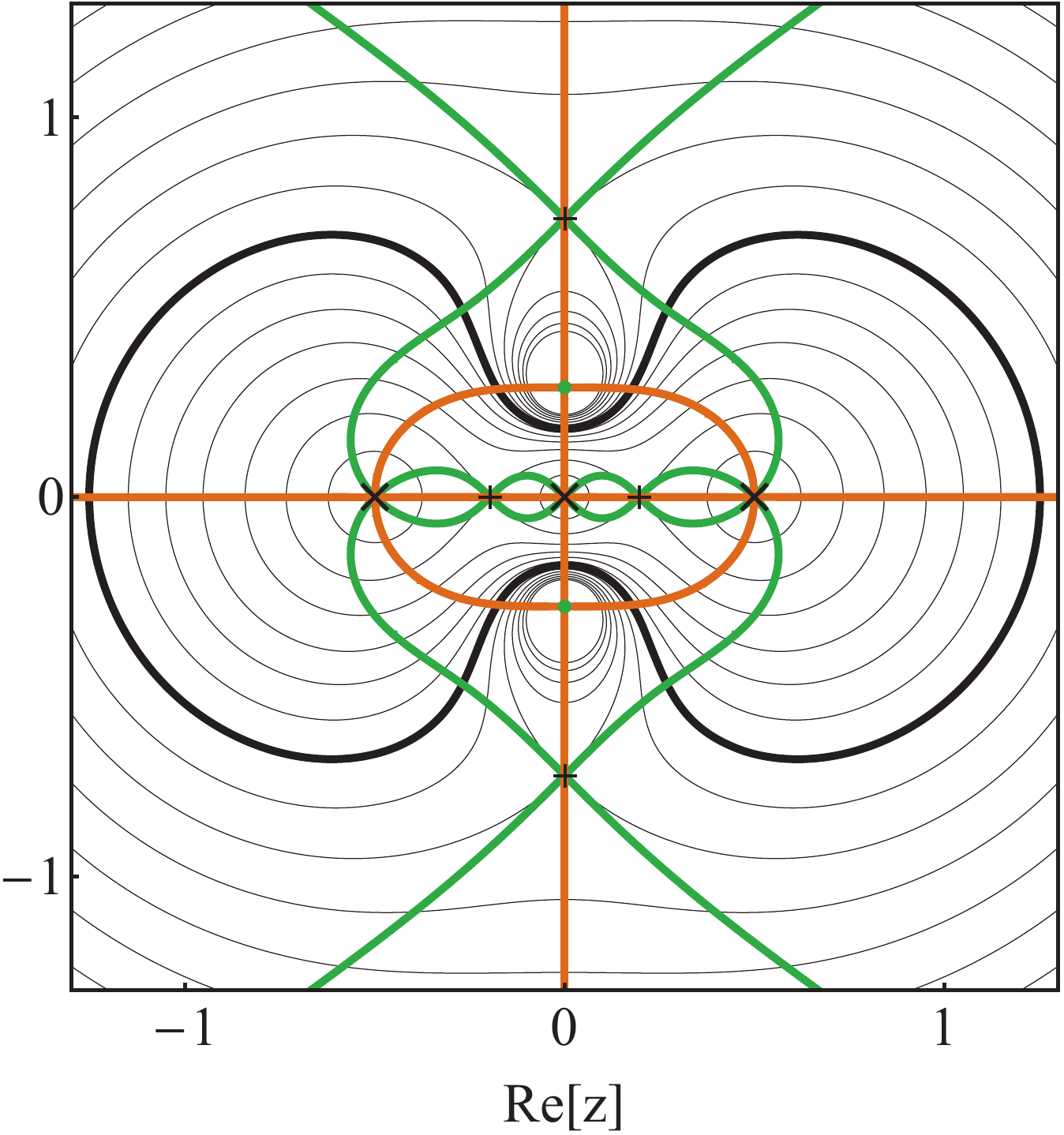}
\hspace{-1.4mm}
\includegraphics[scale=.38]{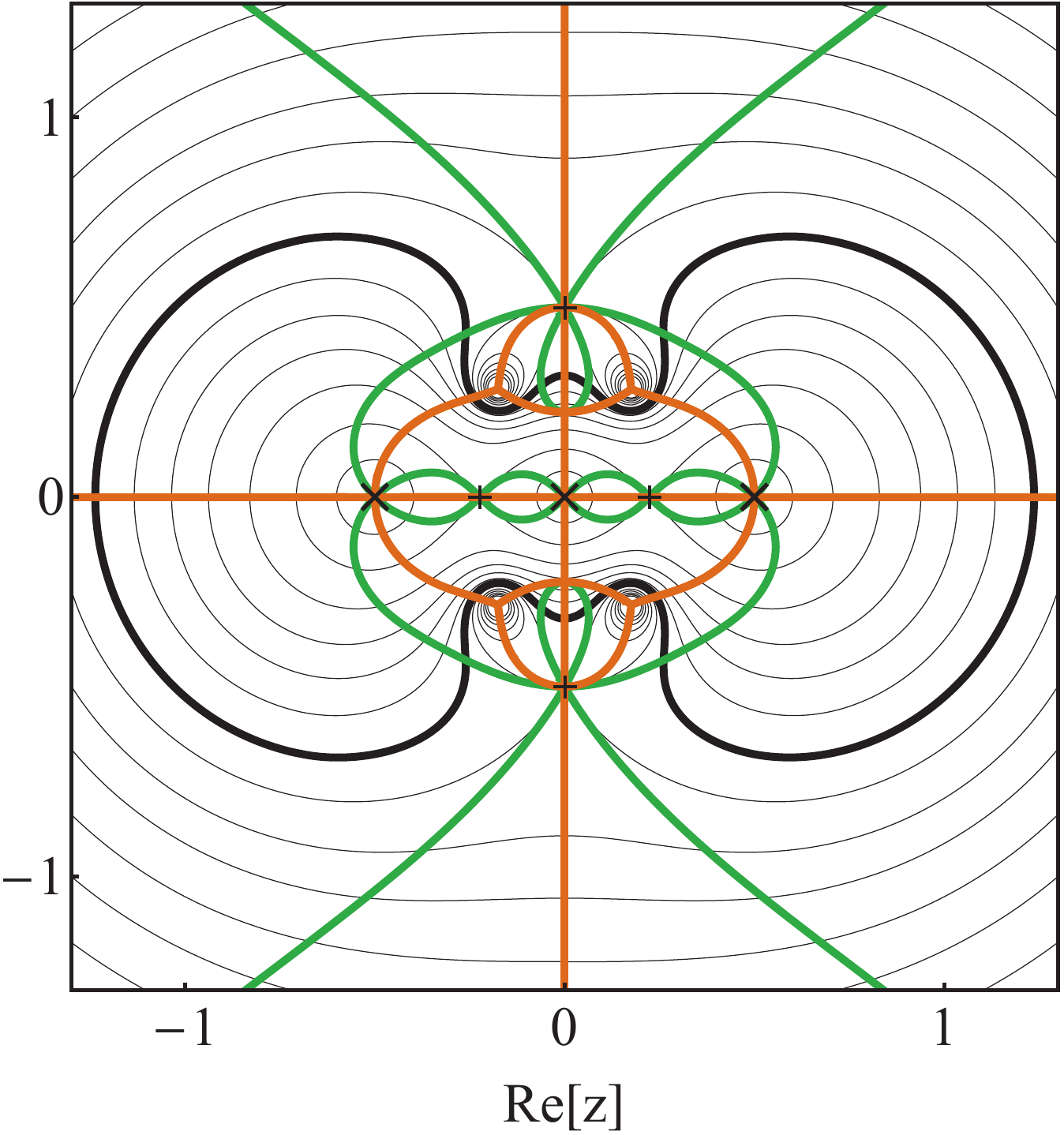}
\caption{LS model: Jacobian contour plots for BCEF boundary intersection at $\mu\approx0.0753$ (left column), for double maxima at $\mu=1/9$ (central column), and for BDE monkey saddles at $\mu=1/5$ (right column). Lower row includes cusp curves (orange) and morph curves (green). Contour values differ from column to column. Bold black contour: critical curve for $s=0.5\,$; crosses: lens components; pluses: Jacobian saddle points.}
\label{fig:LS-contours}
\efi

\clearpage
\bfi
\hspace{-2mm}
\includegraphics[scale=.9]{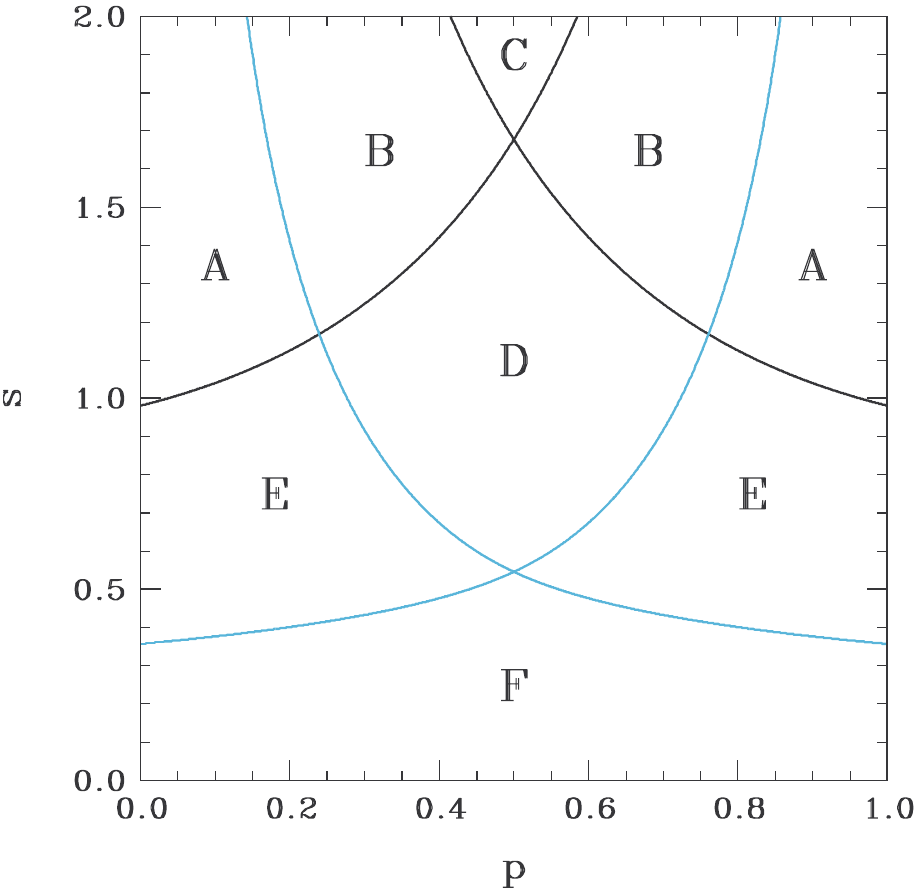}
\hspace{2mm}
\includegraphics[scale=.9]{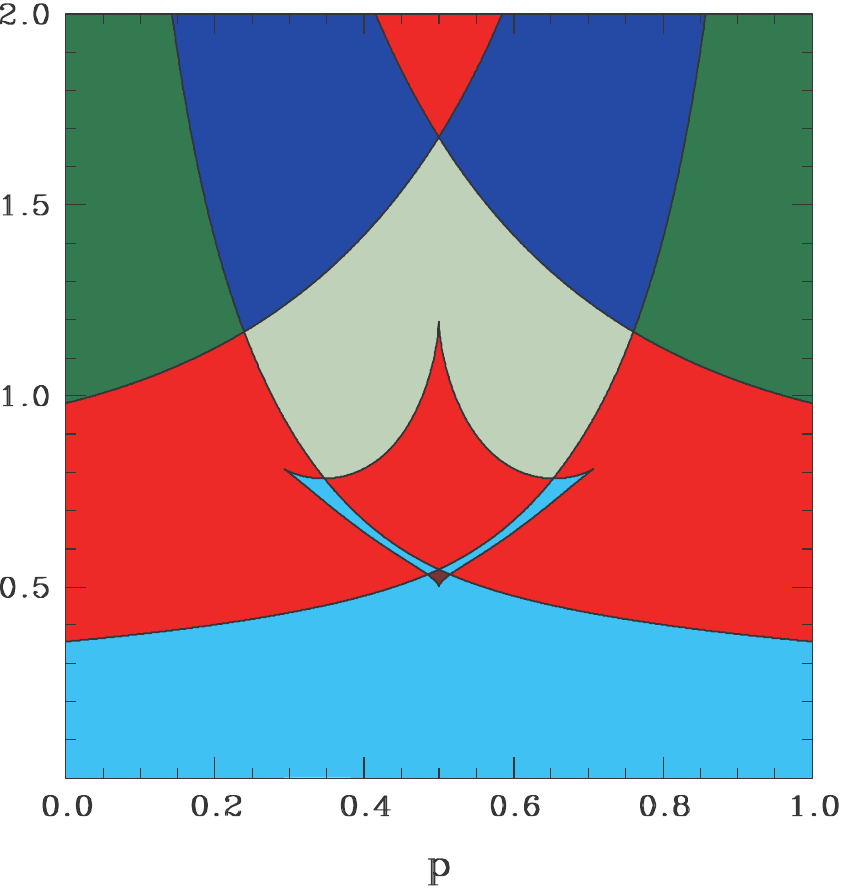}
\caption{LA model parameter-space division. Left panel: by critical-curve topology (marked by letters), with black boundary given by $p_{res}(1)=0$, cyan by 15th degree polynomial in $s^4$ (see \S~\ref{sec:linear_equal-mass}). Right panel: by total number of cusps on caustic [\,8 cusps -- gray; 10 -- blue; 12 -- red; 14 -- green; 16 -- cyan; 20 -- brown].}
\label{fig:LA-parspace}
\efi

\clearpage
\bfi
\includegraphics[width=16.5 cm]{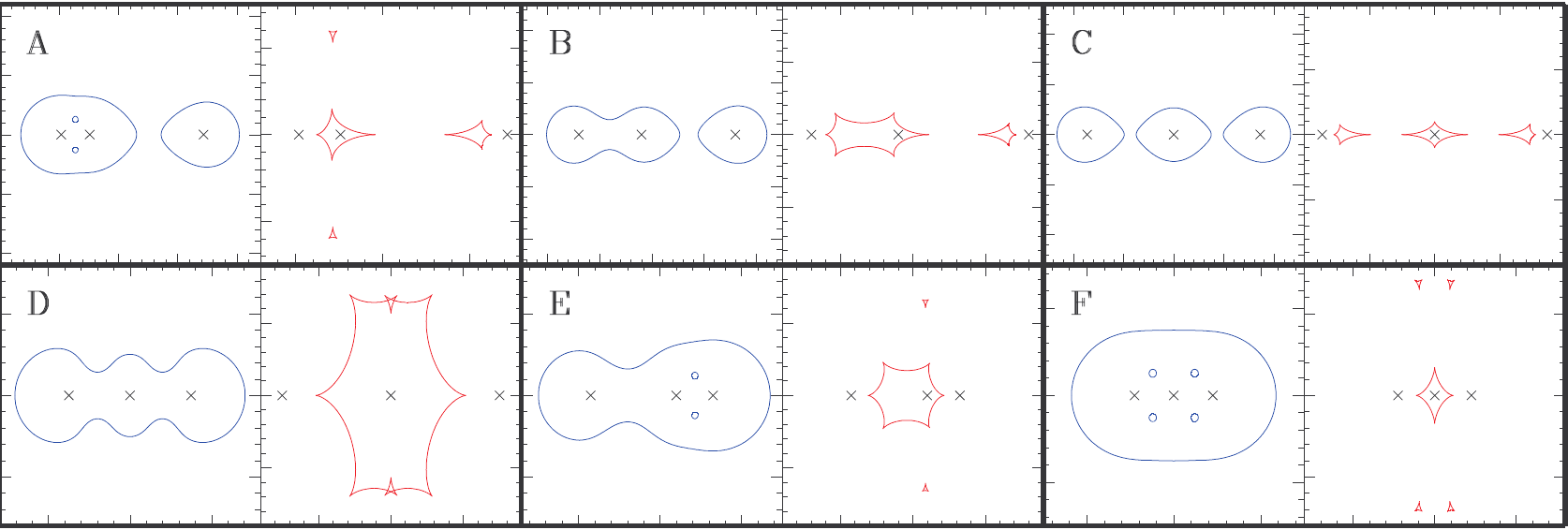}
\caption{LA model: gallery of topologies of critical curves (blue) and corresponding caustics (red) with lens positions marked by black crosses. Letters correspond to regions marked in Figure~\ref{fig:LA-parspace}. Caustic subregions and lens parameters $[p, s]$ of examples: A$_1$ $[0.2,1.2]$, B$_1$ $[0.4,1.5]$, C$_1$ $[0.5,1.725]$, D$_2$ $[0.5,0.75]$, E$_1$ $[0.7,0.75]$, F$_2$ $[0.5,0.45]$.}
\label{fig:LA-topologies}
\efi

\clearpage
\bfi
\hspace*{-2mm}
\includegraphics[scale=.38]{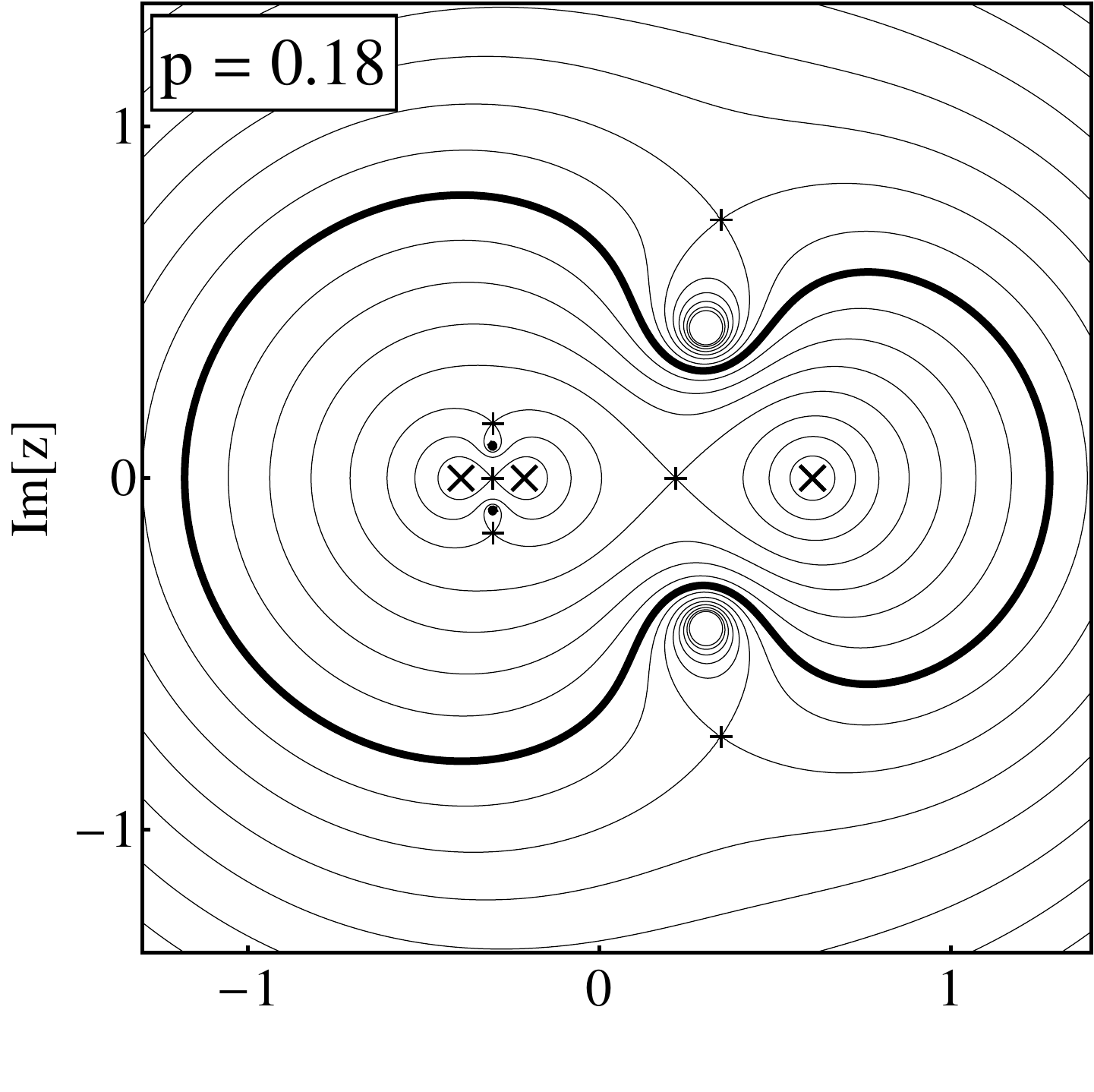}
\hspace{-5mm}
\includegraphics[scale=.38]{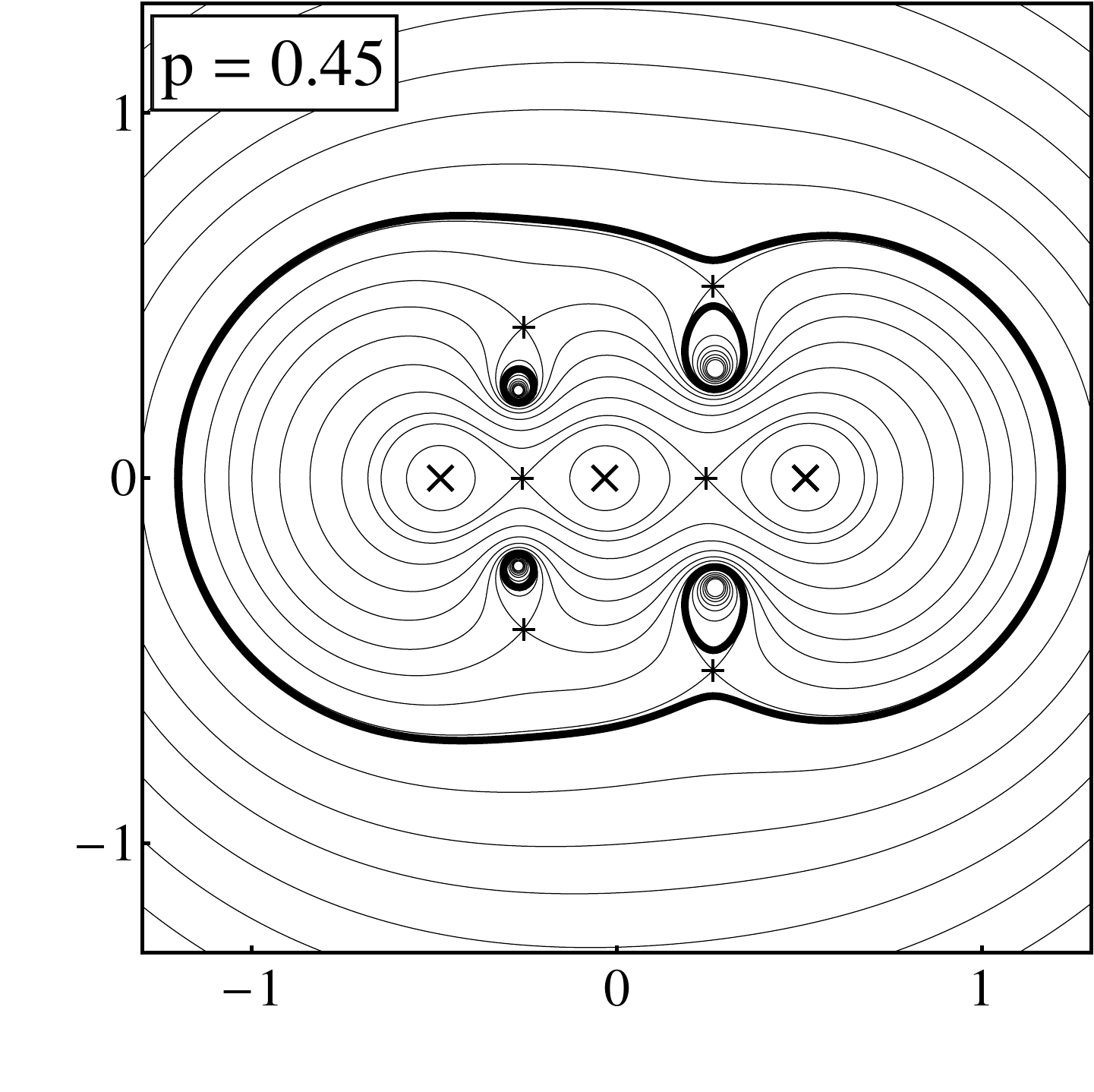}
\hspace{-5mm}
\includegraphics[scale=.38]{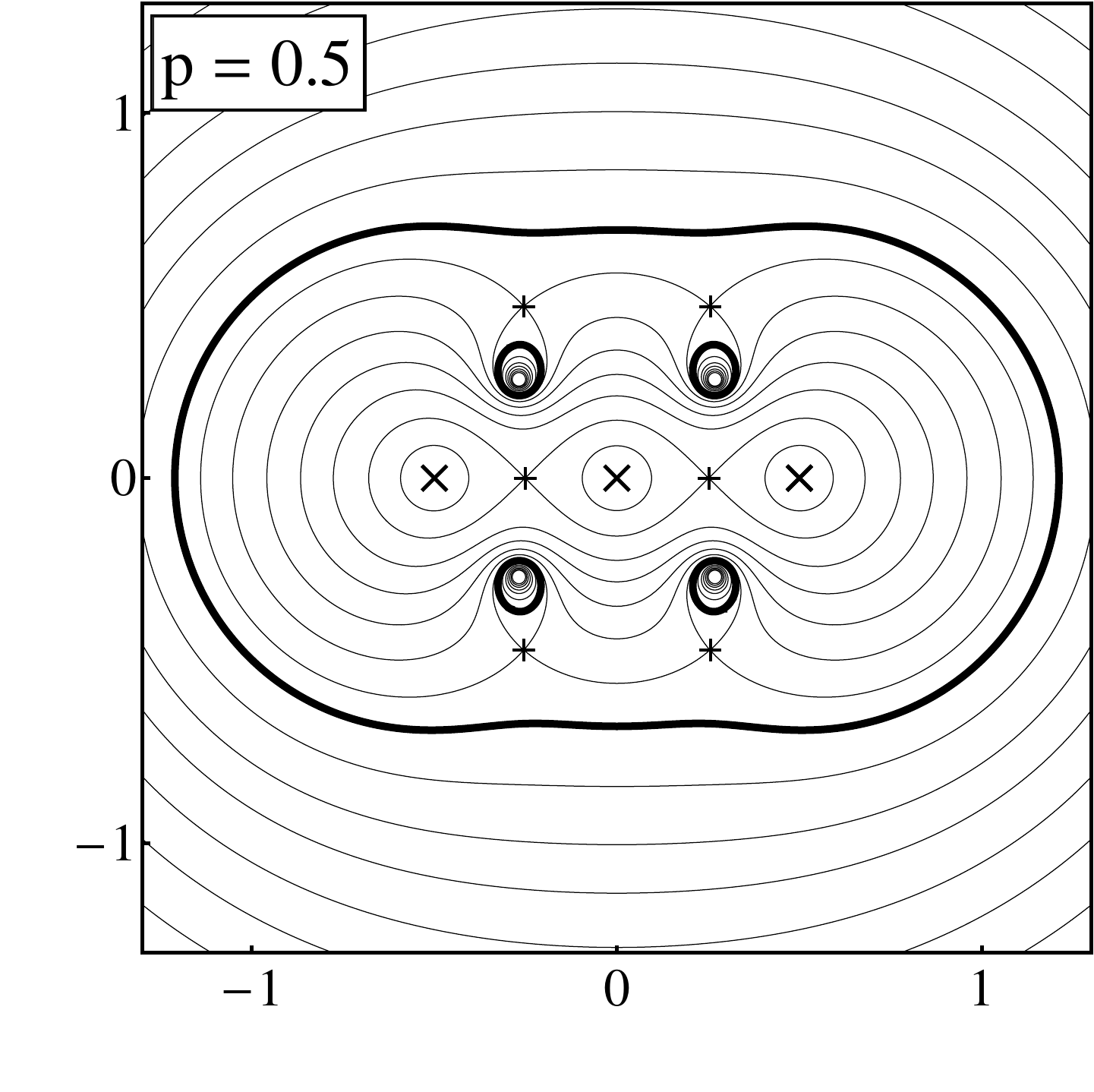}
\vspace*{-3mm}
\\
\hspace*{-1.6mm}
\includegraphics[scale=.38]{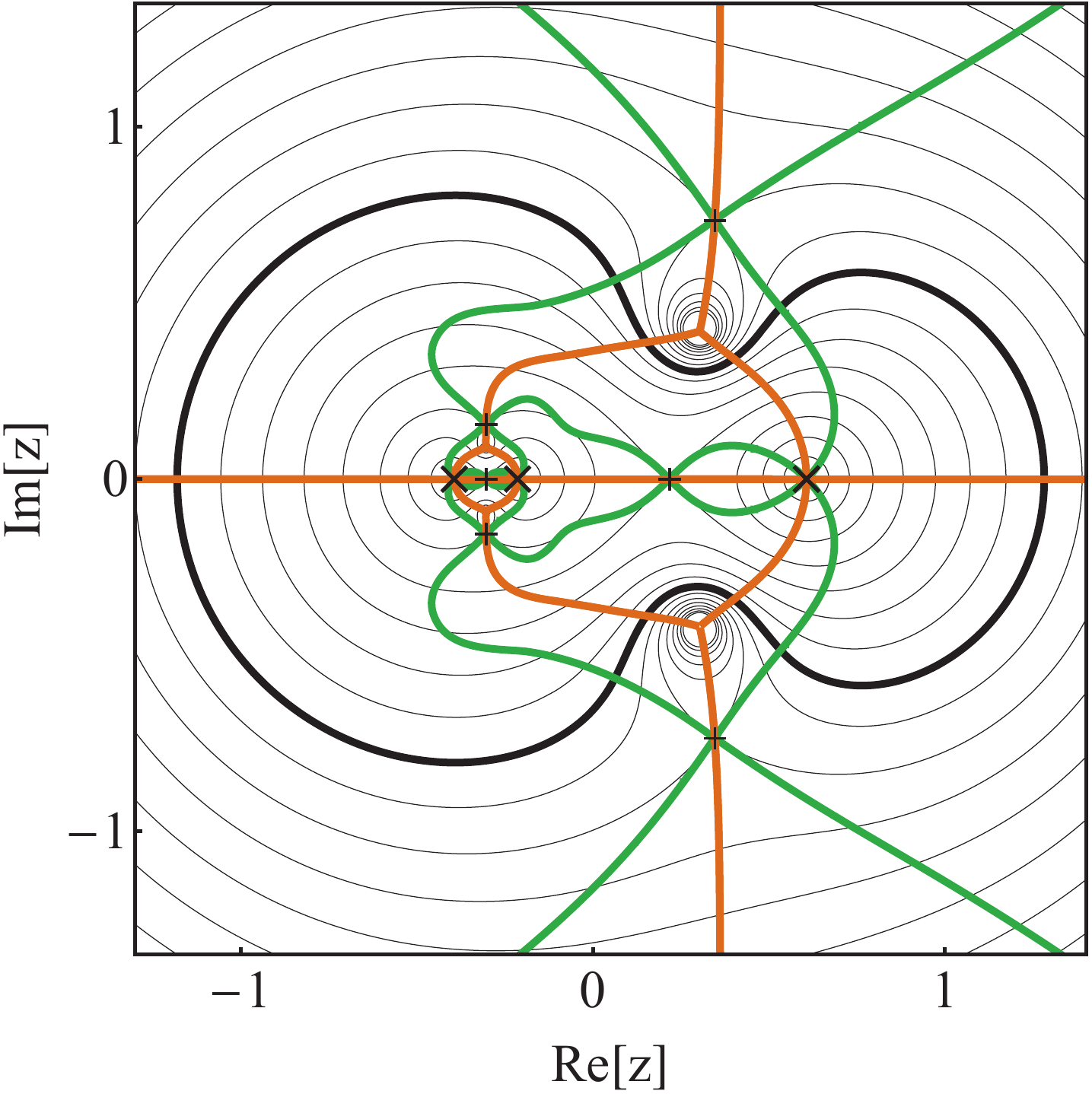}
\hspace{-1.4mm}
\includegraphics[scale=.38]{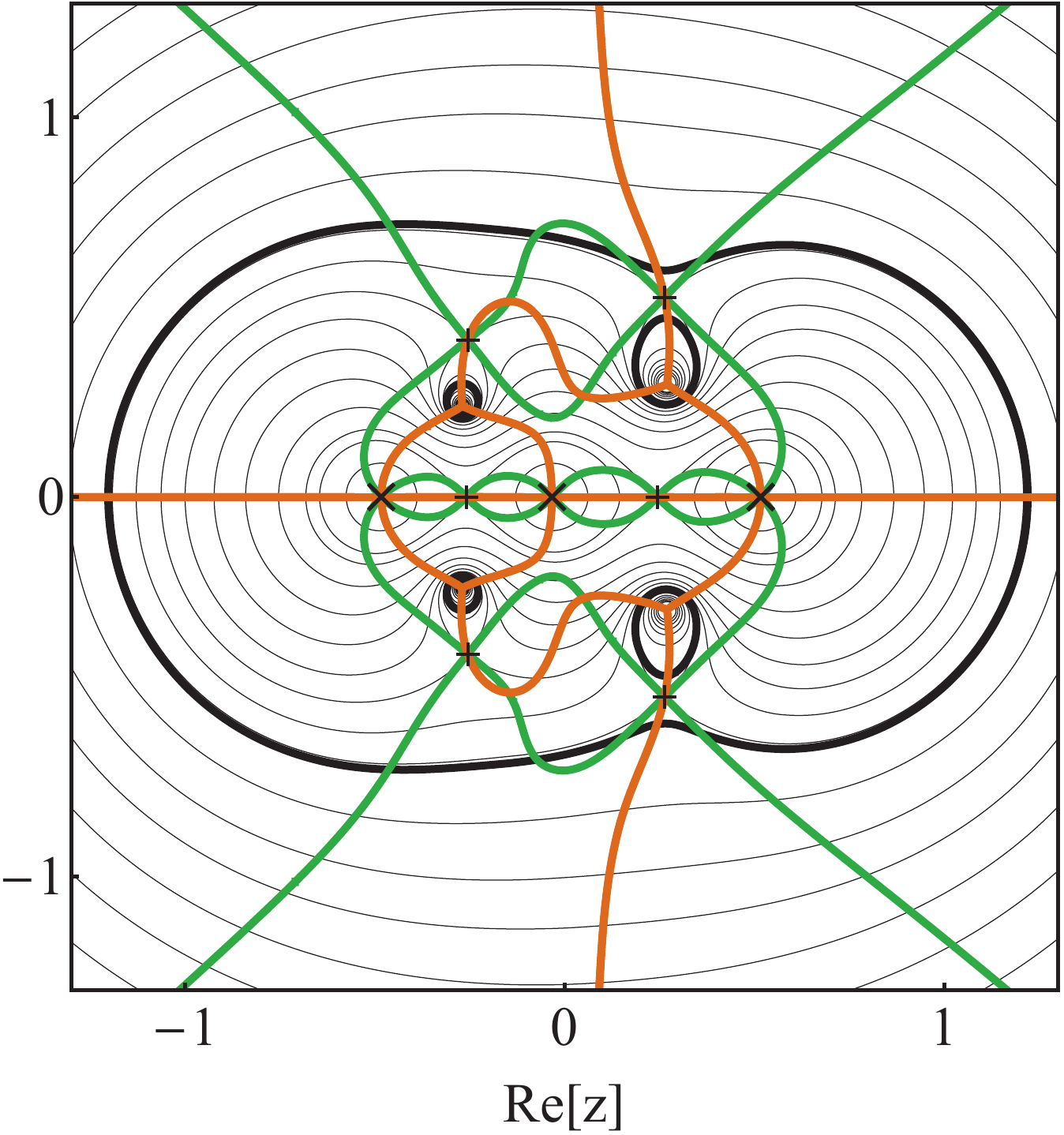}
\hspace{-1.4mm}
\includegraphics[scale=.38]{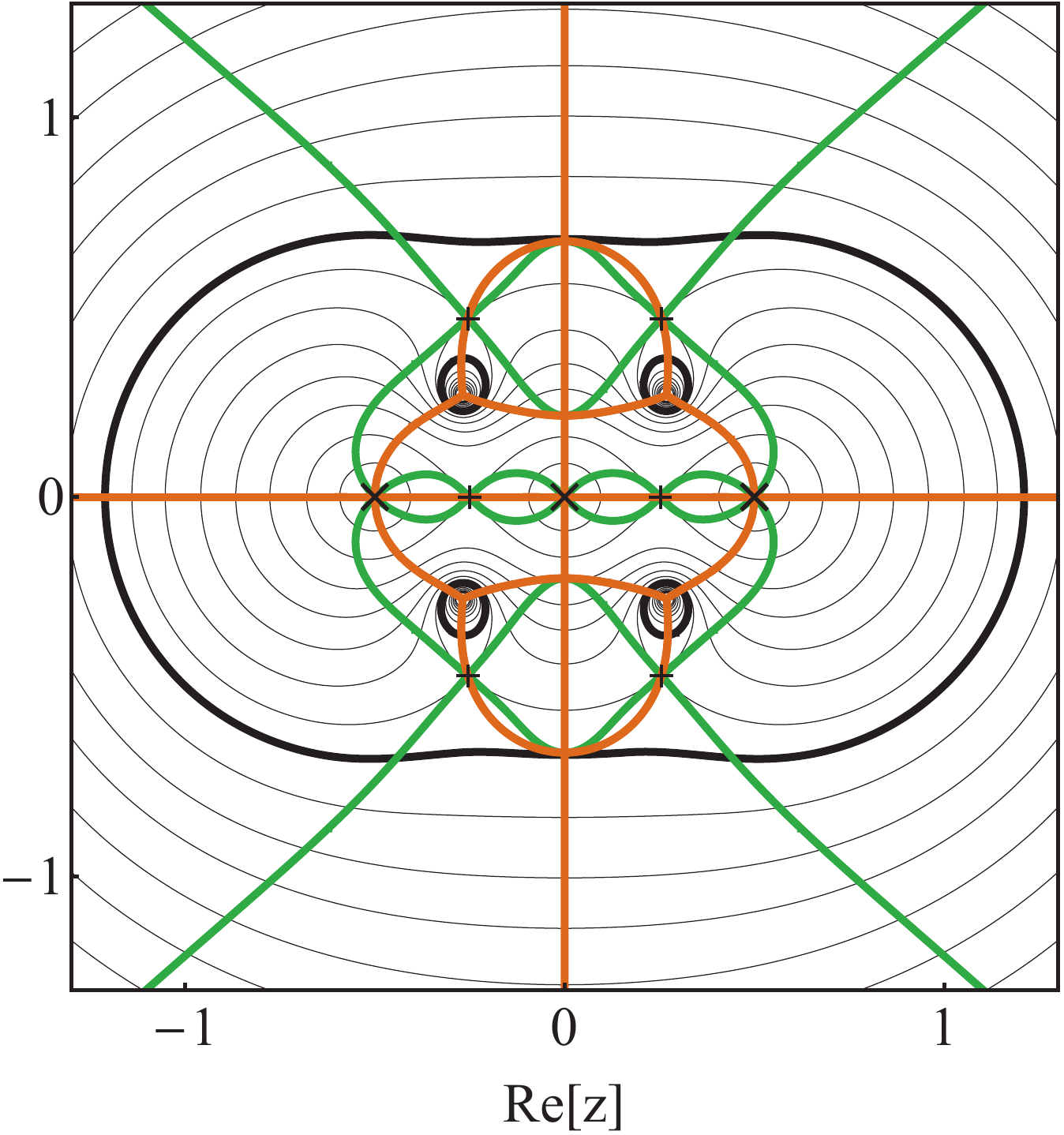}
\caption{LA model: Jacobian contour plots in regime of binary lens with distant companion at $p=0.18$ (left column), for the nearly symmetric $p=0.45$ (central column), and the symmetric case $p=0.5$ (right column). Lower row includes cusp curves (orange) and morph curves (green). Contour values differ from column to column. Notation as in Figure~\ref{fig:LS-contours}.}
\label{fig:LA-contours}
\efi

\clearpage
\bfi
\hspace{-2mm}
\includegraphics[scale=.9]{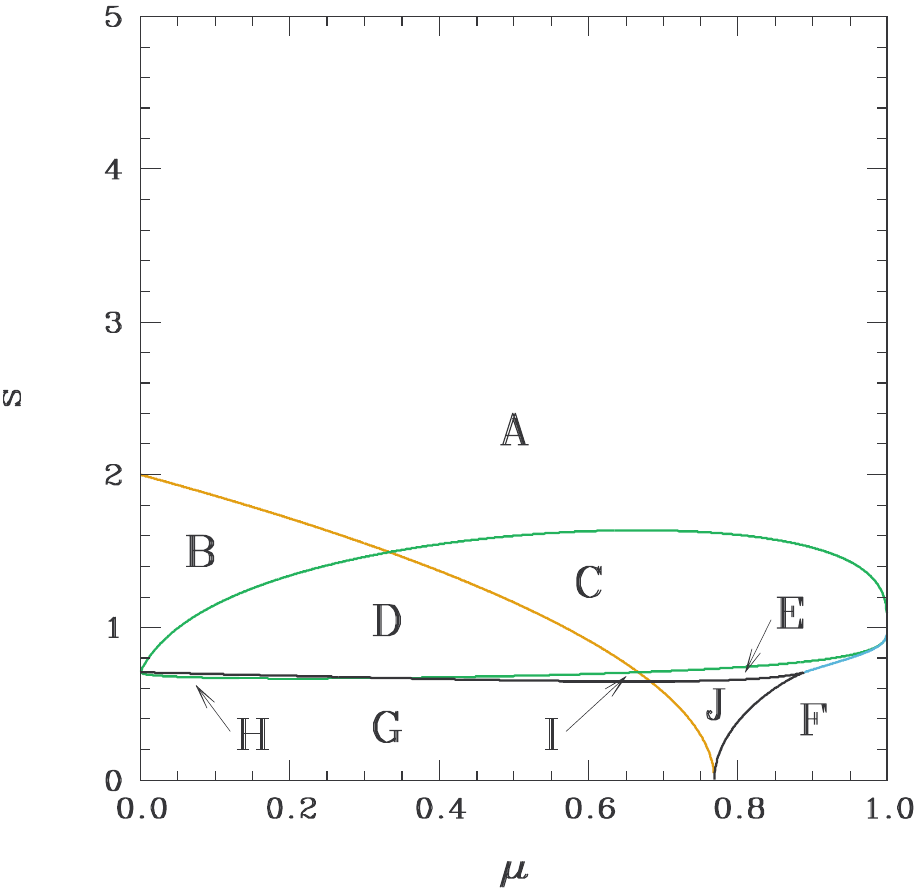}
\hspace{2mm}
\includegraphics[scale=.9]{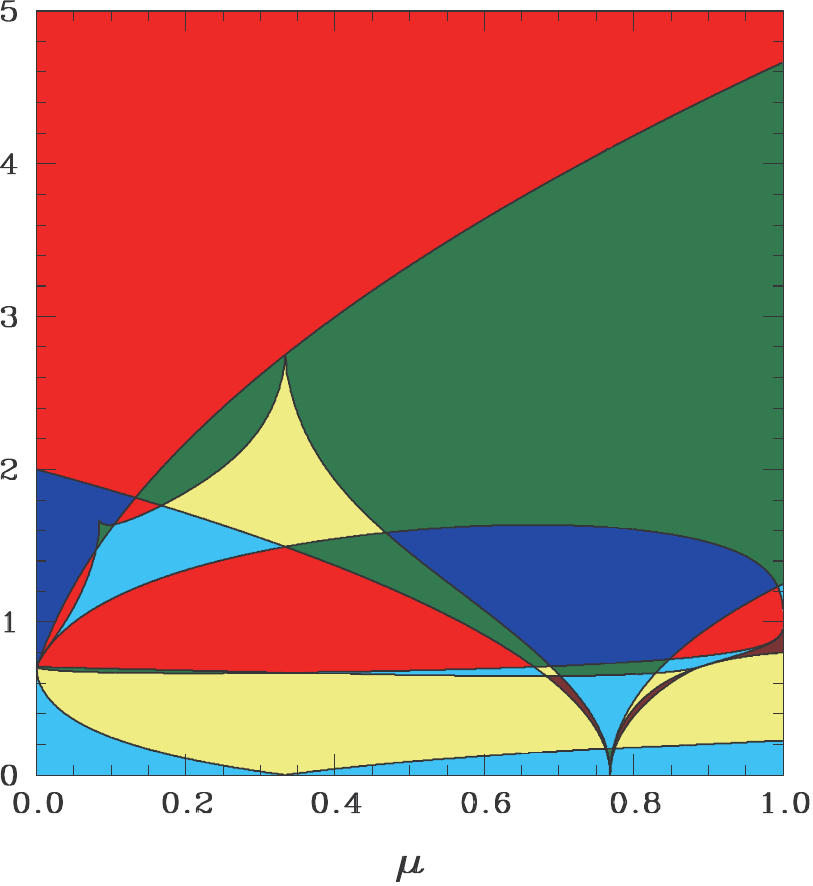}
\caption{TE model parameter-space division. Left panel: by critical-curve topology (marked by letters), with black boundary given by $p_{res}(1)=0$, orange by $p_{res}(-1)=0$, cyan by 3rd degree and green by 12th degree polynomials in $s^4$ (see \S~\ref{sec:equilateral_vertex}). Right panel: by total number of cusps on caustic [\,10 cusps -- blue; 12 -- red; 14 -- green; 16 -- cyan; 18 -- yellow; 20 -- brown]. Region near lower right corner of right panel is blown up in Figure~\ref{fig:TE-details}.}
\label{fig:TE-parspace}
\efi

\clearpage
\bfi
\includegraphics[width=16.5 cm]{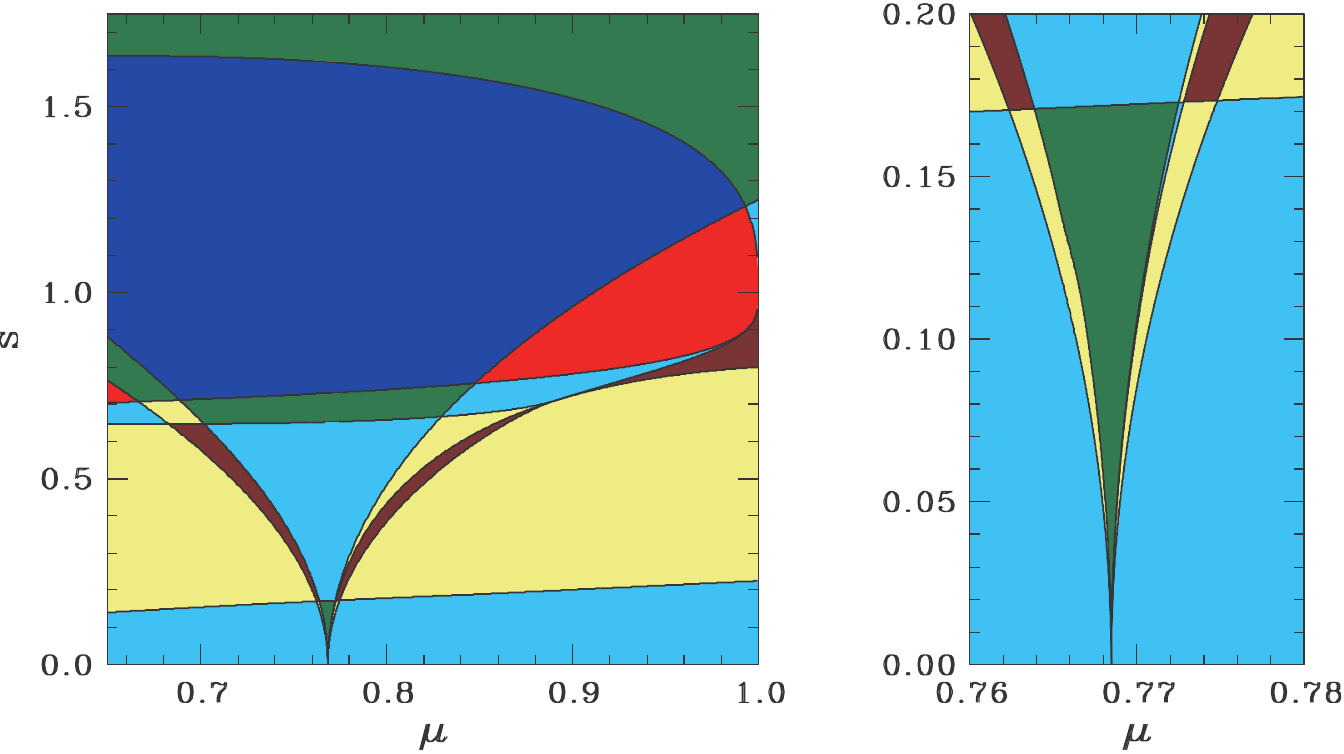}
\caption{TE model: parameter-space division by total cusp number. Left panel: detail from Figure~\ref{fig:TE-parspace}. Right panel: detail from left panel near close limit of double-maximum configuration.}
\label{fig:TE-details}
\efi

\clearpage
\bfi
\includegraphics[width=16.5 cm]{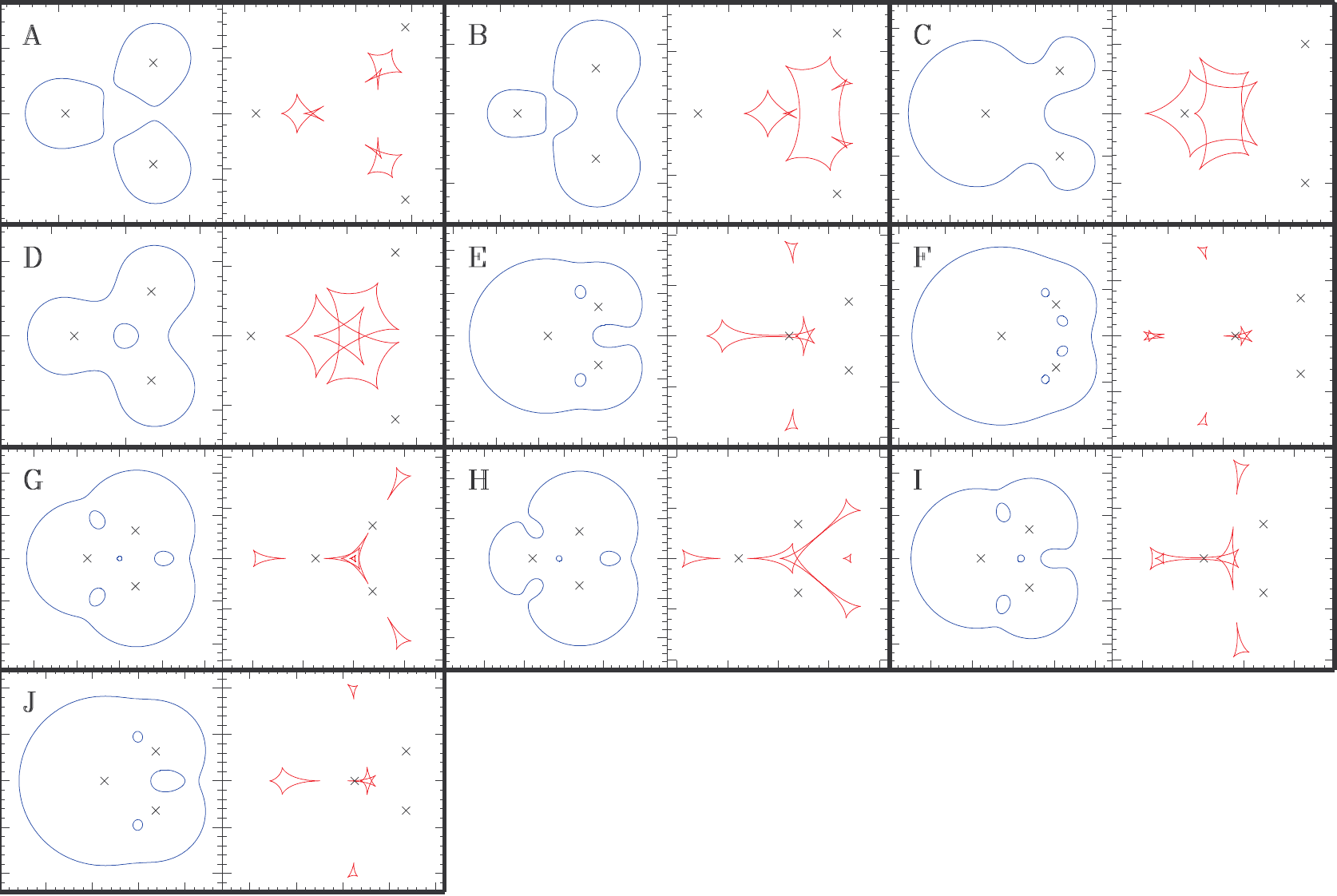}
\caption{TE model: gallery of topologies of critical curves (blue) and corresponding caustics (red) with lens positions marked by black crosses. Letters correspond to regions marked in Figure~\ref{fig:TE-parspace}. Caustic subregions and lens parameters $[\mu, s]$ of examples: A$_4$ $[1/3,1.55]$, B$_3$ $[0.15,1.3]$, C$_1$ $[0.7685,1.0]$, D$_1$ $[1/3,1.2]$, E$_2$ $[0.7685,0.68]$, F$_2$ $[0.9,0.68]$, G$_2$ $[1/3,0.65]$, H$_1$ $[0.15,0.68]$, I$_1$ $[0.6,0.68]$, J$_3$ $[0.7685,0.64]$.}
\label{fig:TE-topologies}
\efi

\clearpage
\bfi
\hspace*{-2mm}
\includegraphics[scale=.385]{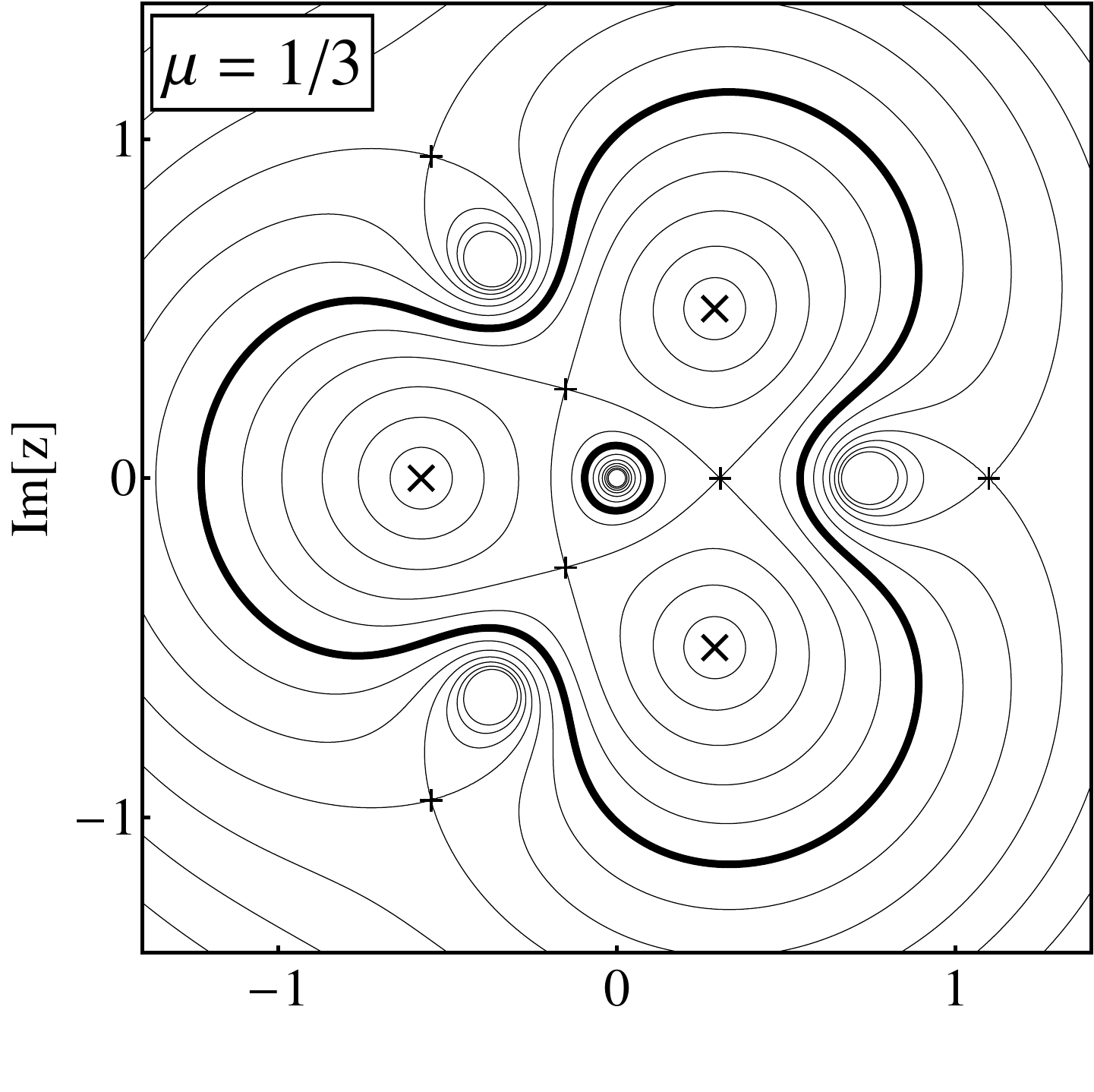}
\hspace{-5mm}
\includegraphics[scale=.385]{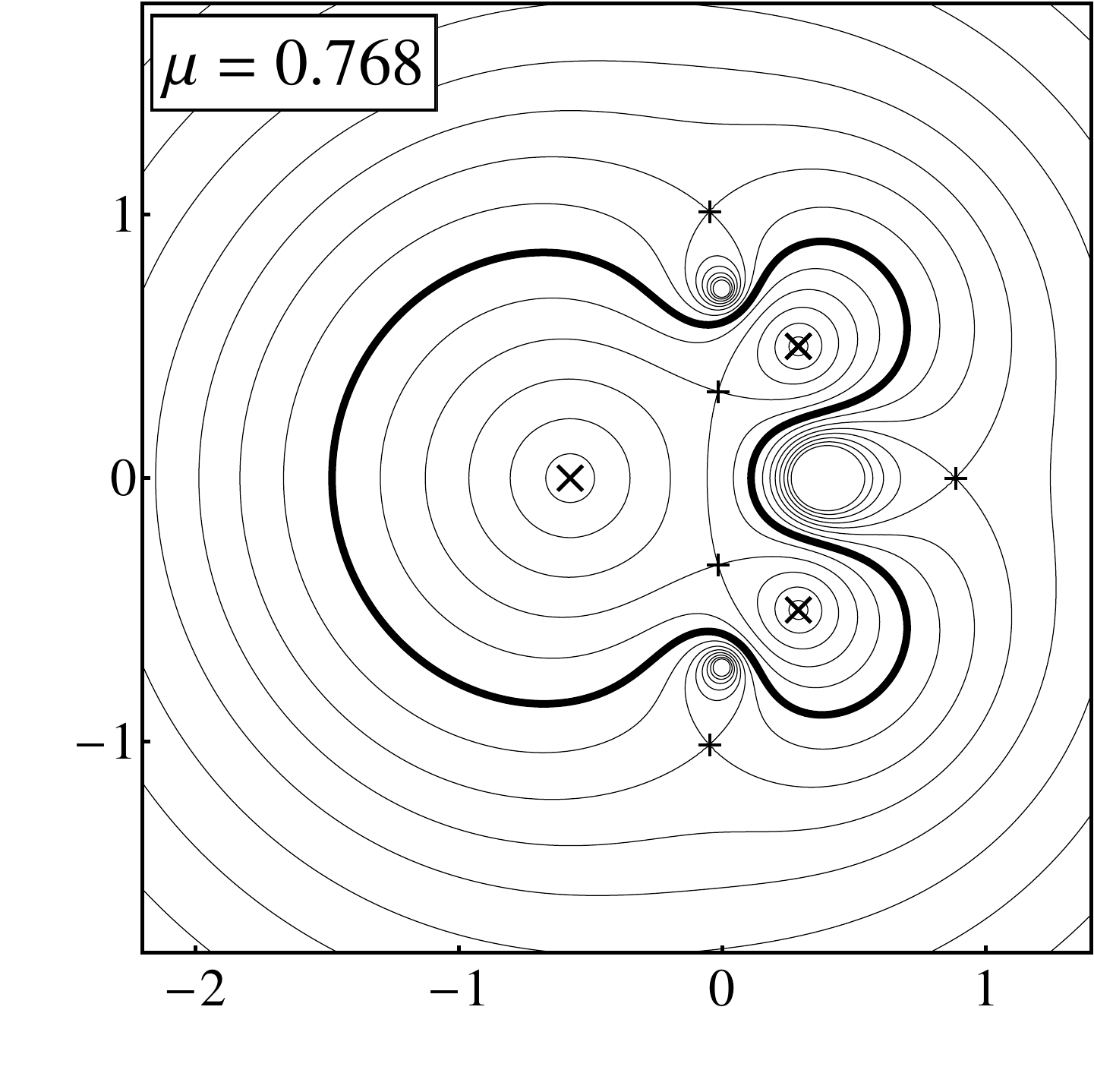}
\hspace{-5mm}
\includegraphics[scale=.385]{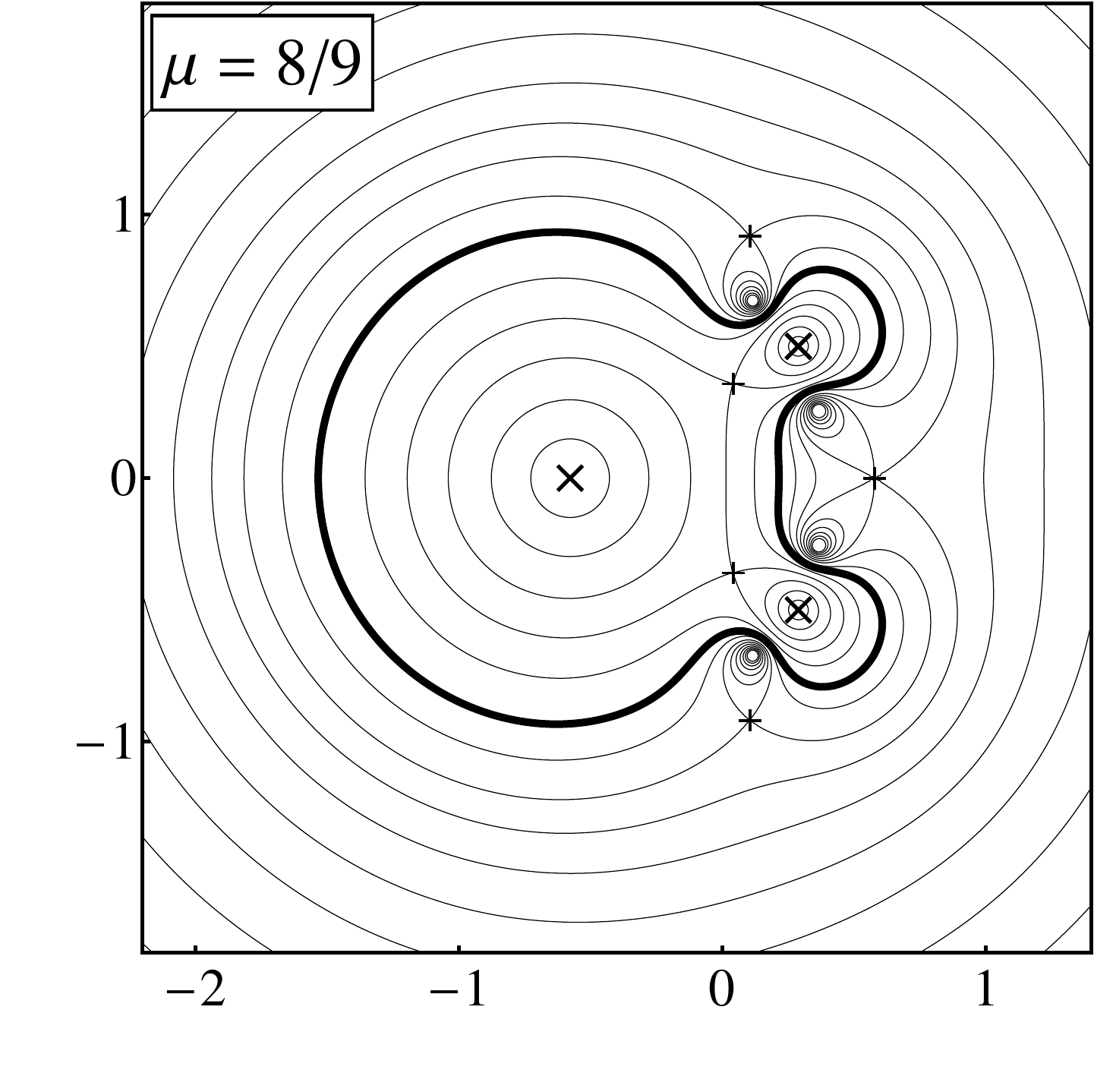}
\vspace*{-3mm}
\\
\hspace*{-1.6mm}
\includegraphics[scale=.385]{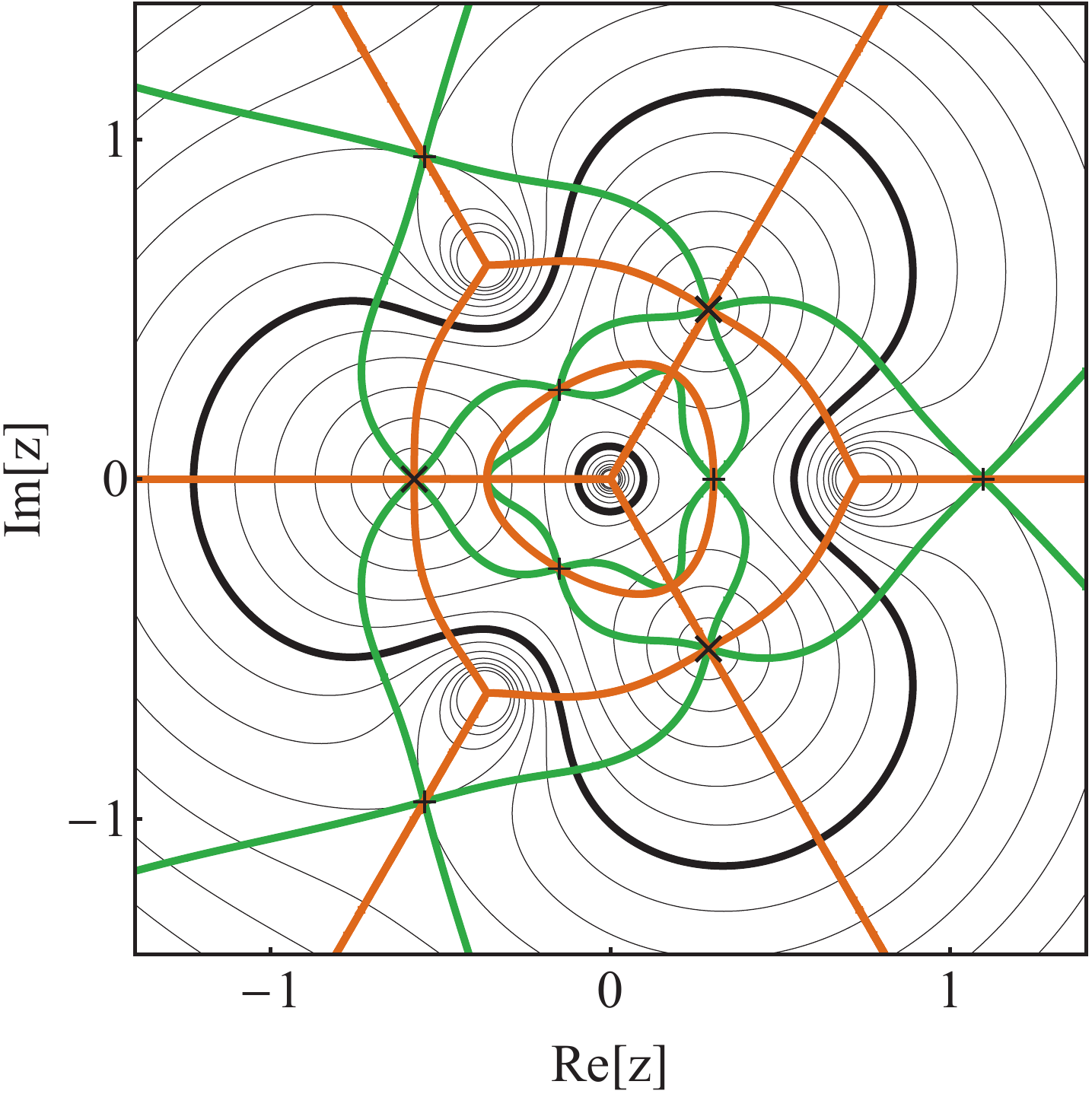}
\hspace{-1.4mm}
\includegraphics[scale=.385]{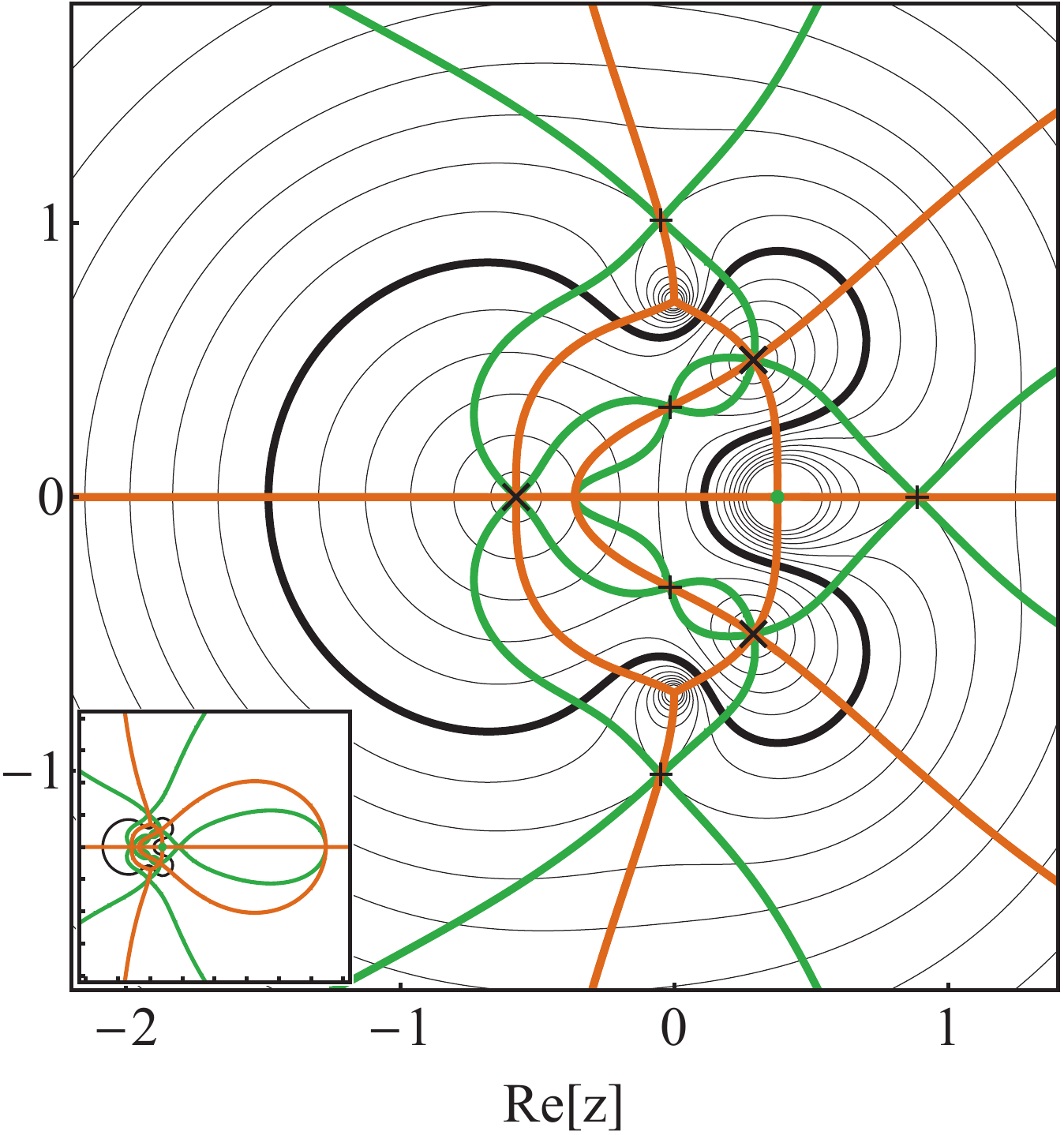}
\hspace{-1.4mm}
\includegraphics[scale=.385]{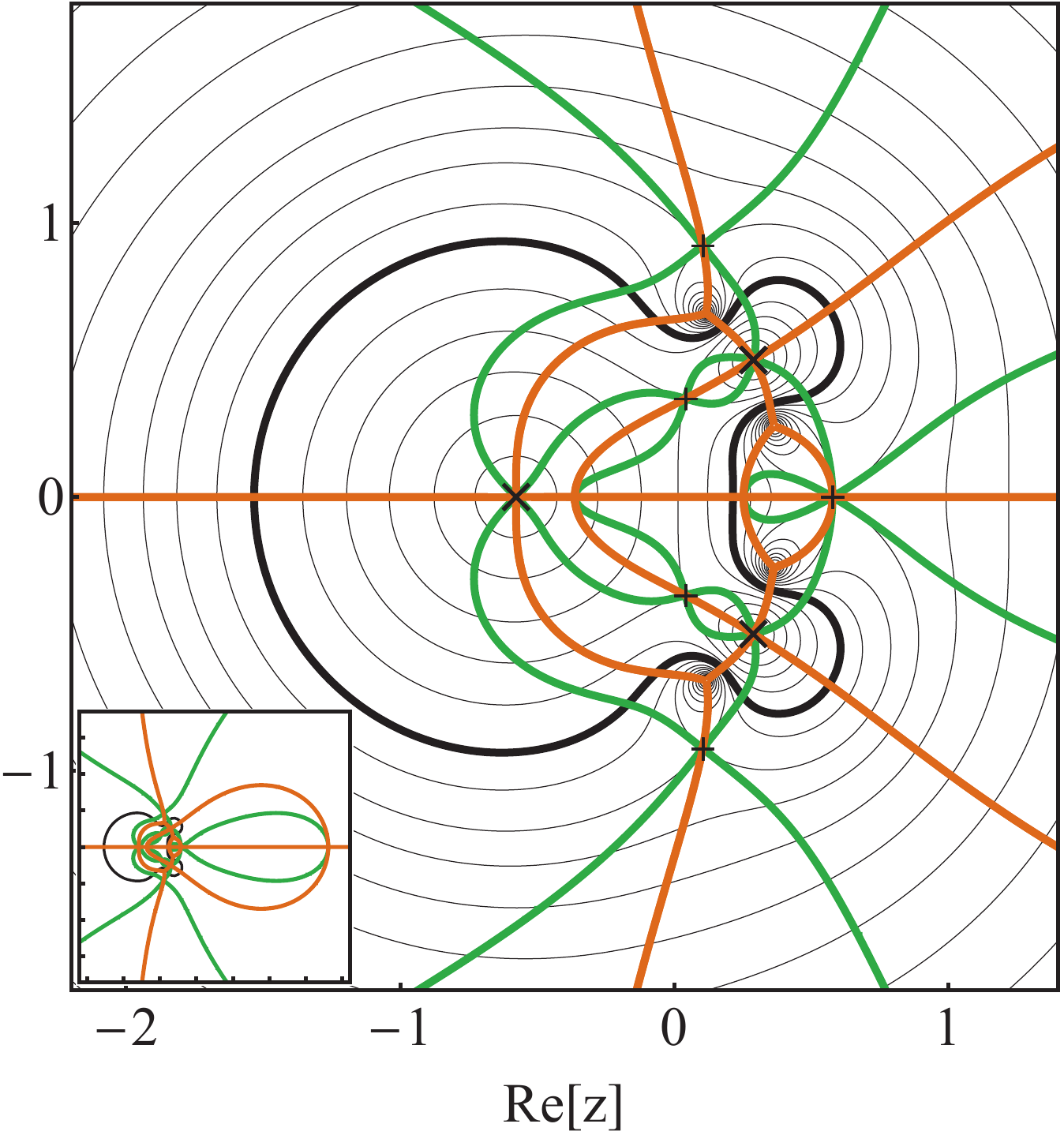}
\caption{TE model: Jacobian contour plots for equal masses $\mu=1/3$ (left column), for double maximum at $\mu\approx0.768$ (central column), and for monkey saddle at $\mu=8/9$ (right column). Lower row includes cusp curves (orange) and morph curves (green). Contour values differ from column to column. Insets in lower central and right panels: zoomed-out plots of critical, cusp, and morph curves. Notation as in Figure~\ref{fig:LS-contours}.}
\label{fig:TE-contours}
\efi

\clearpage
\bfi
\hspace{-2mm}
\includegraphics[scale=.9]{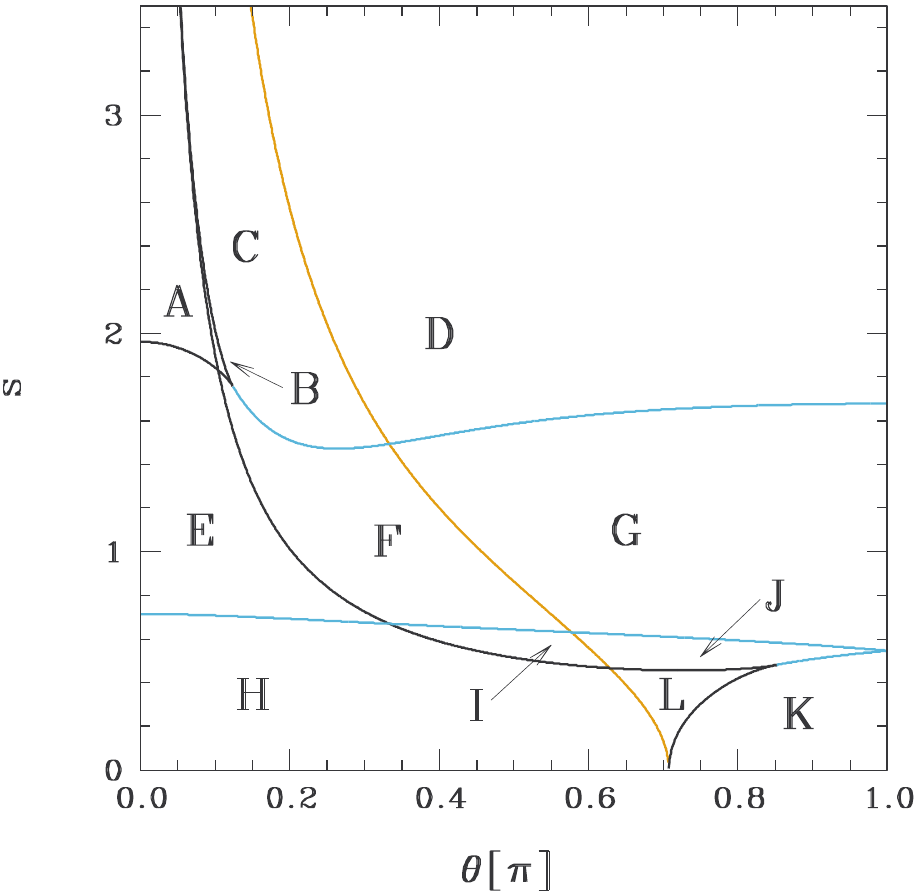}
\hspace{2mm}
\includegraphics[scale=.9]{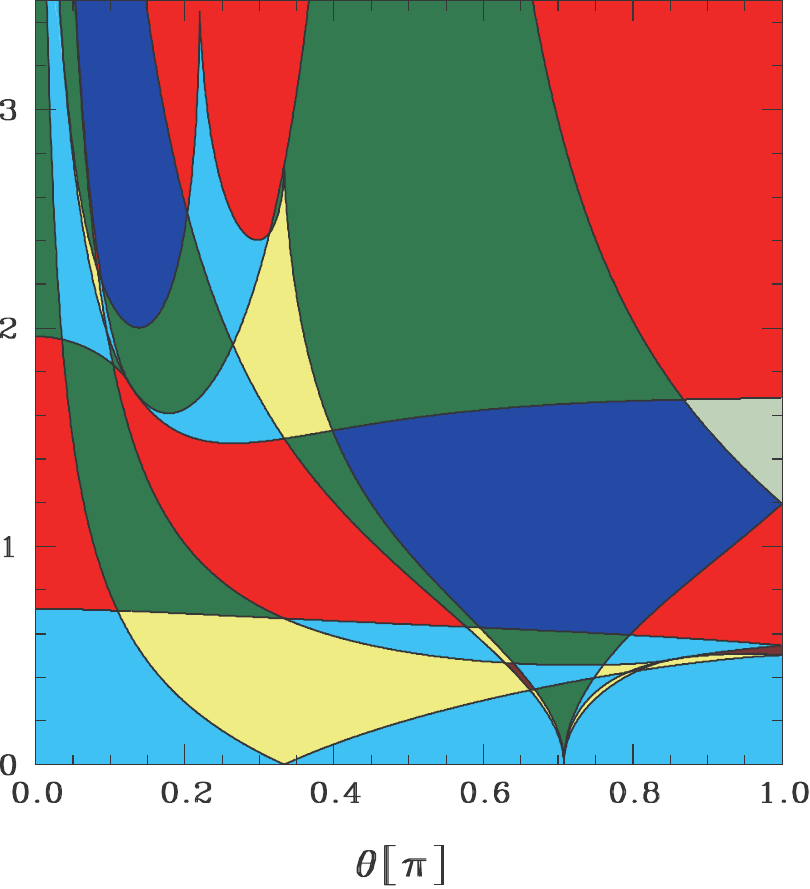}
\caption{TI model parameter-space division. Left panel: by critical-curve topology (marked by letters), with black boundary given by $p_{res}(1)=0$, orange by $p_{res}(-1)=0$, cyan by 15th degree polynomial in $s^4$ (see \S~\ref{sec:isosceles}). Right panel: by total number of cusps on caustic [\,8 cusps -- gray; 10 -- blue; 12 -- red; 14 -- green; 16 -- cyan; 18 -- yellow; 20 -- brown]. Regions near upper left and lower right corners of right panel are blown up in Figure~\ref{fig:TI-details}.}
\label{fig:TI-parspace}
\efi

\clearpage
\bfi
\includegraphics[width=16.5 cm]{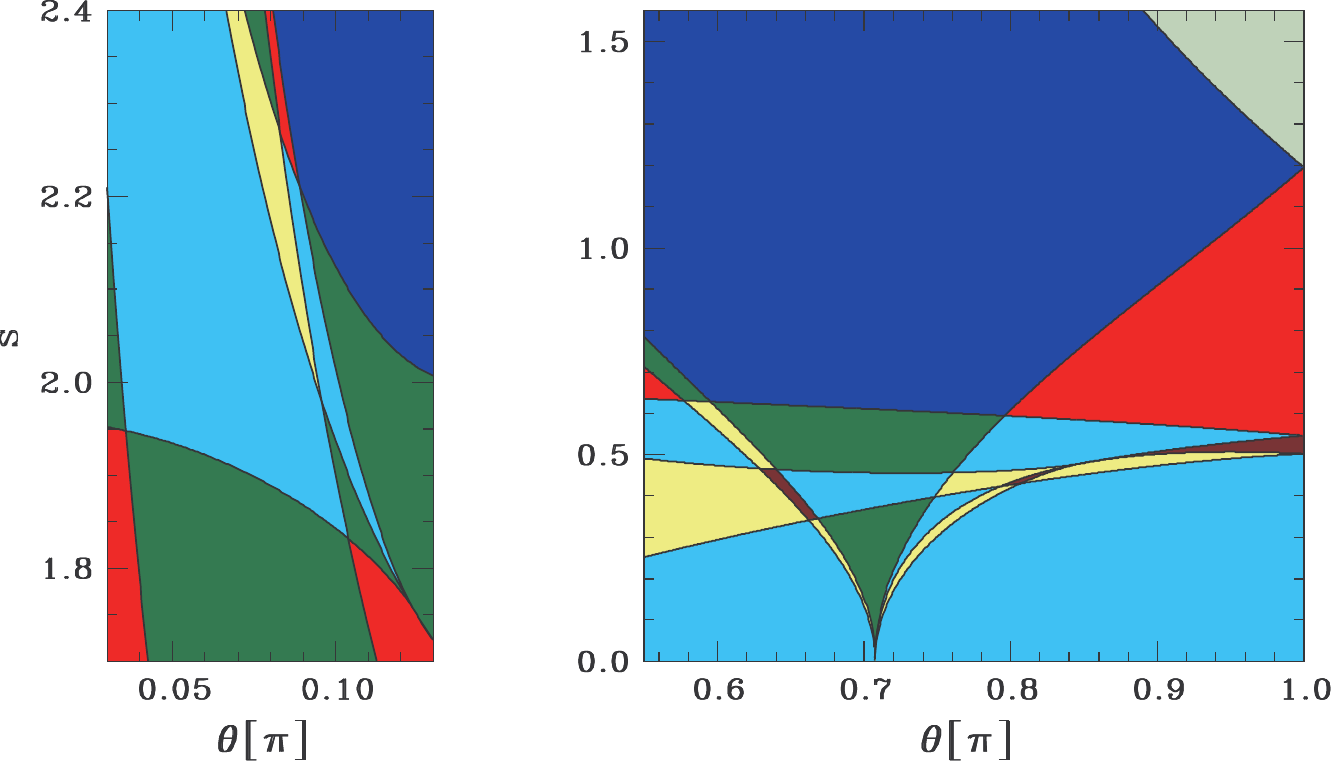}
\caption{TI model: parameter-space division by total cusp number. Left panel: detail from upper left part of Figure~\ref{fig:TI-parspace}. Right panel: detail of lower right corner of Figure~\ref{fig:TI-parspace}.}
\label{fig:TI-details}
\efi

\clearpage
\bfi
\includegraphics[width=16.5 cm]{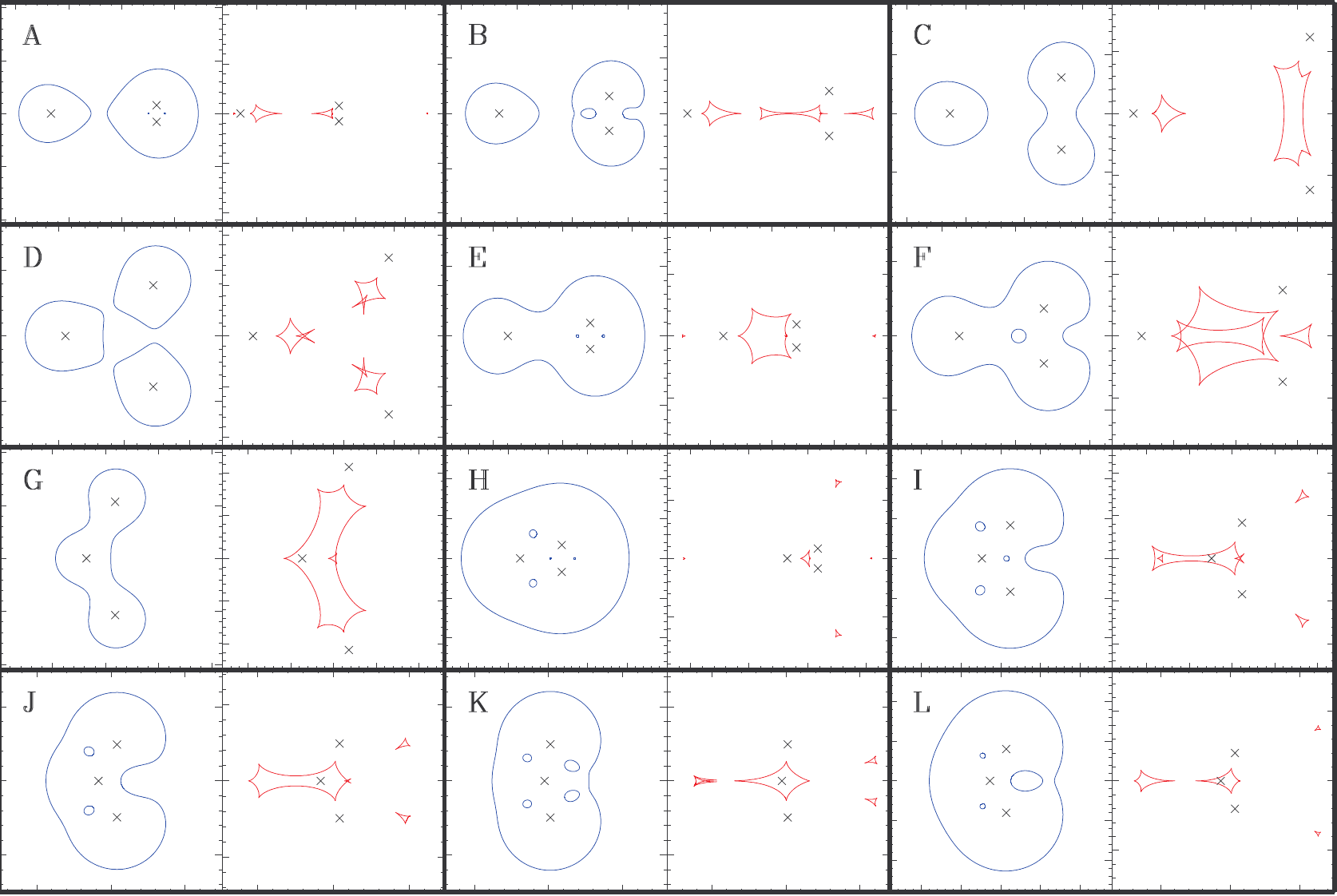}
\caption{TI model: gallery of topologies of critical curves (blue) and corresponding caustics (red) with lens positions marked by black crosses. Letters correspond to regions marked in Figure~\ref{fig:TI-parspace}. Caustic subregions and lens parameters $[\theta, s]$ of examples: A$_2$ $[0.05\,\pi,2]$, B$_2$ $[0.1\,\pi,2]$, C$_2$ $[0.2\,\pi,2]$, D$_6$ $[\pi/3,1.55]$, E$_2$ $[0.1\,\pi,1.2]$, F$_1$ $[0.2\,\pi,1.2]$, G$_2$ $[0.7\,\pi,1.2]$, H$_2$ $[0.2\,\pi,0.55]$, I$_1$ $[0.55\,\pi,0.55]$, J$_2$ $[0.7\,\pi,0.55]$, K$_2$ $[0.9\,\pi,0.5]$, L$_4$ $[0.7\,\pi,0.45]$.}
\label{fig:TI-topologies}
\efi

\clearpage
\bfi
\hspace*{-2mm}
\includegraphics[scale=.385]{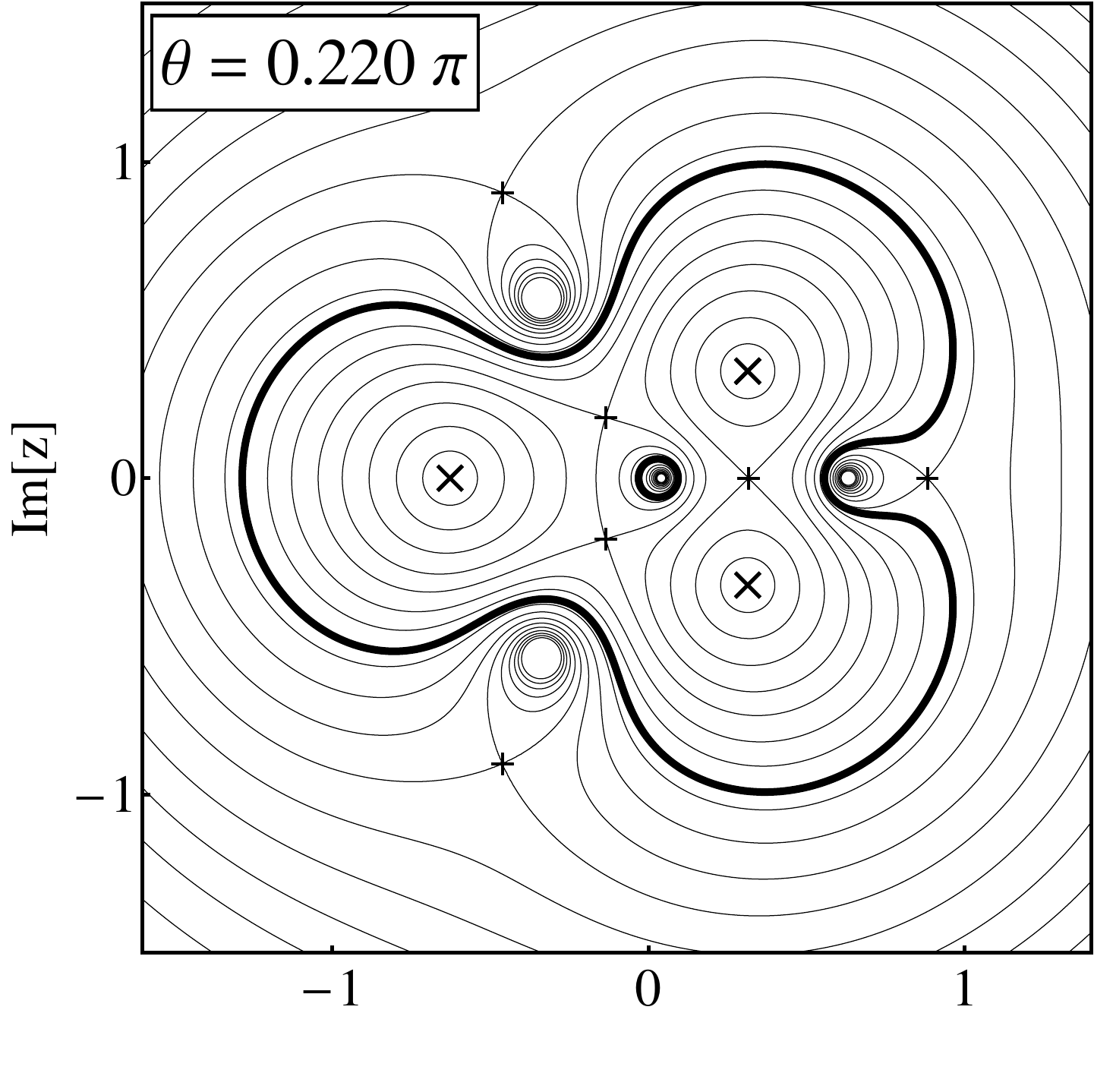}
\hspace{-5mm}
\includegraphics[scale=.385]{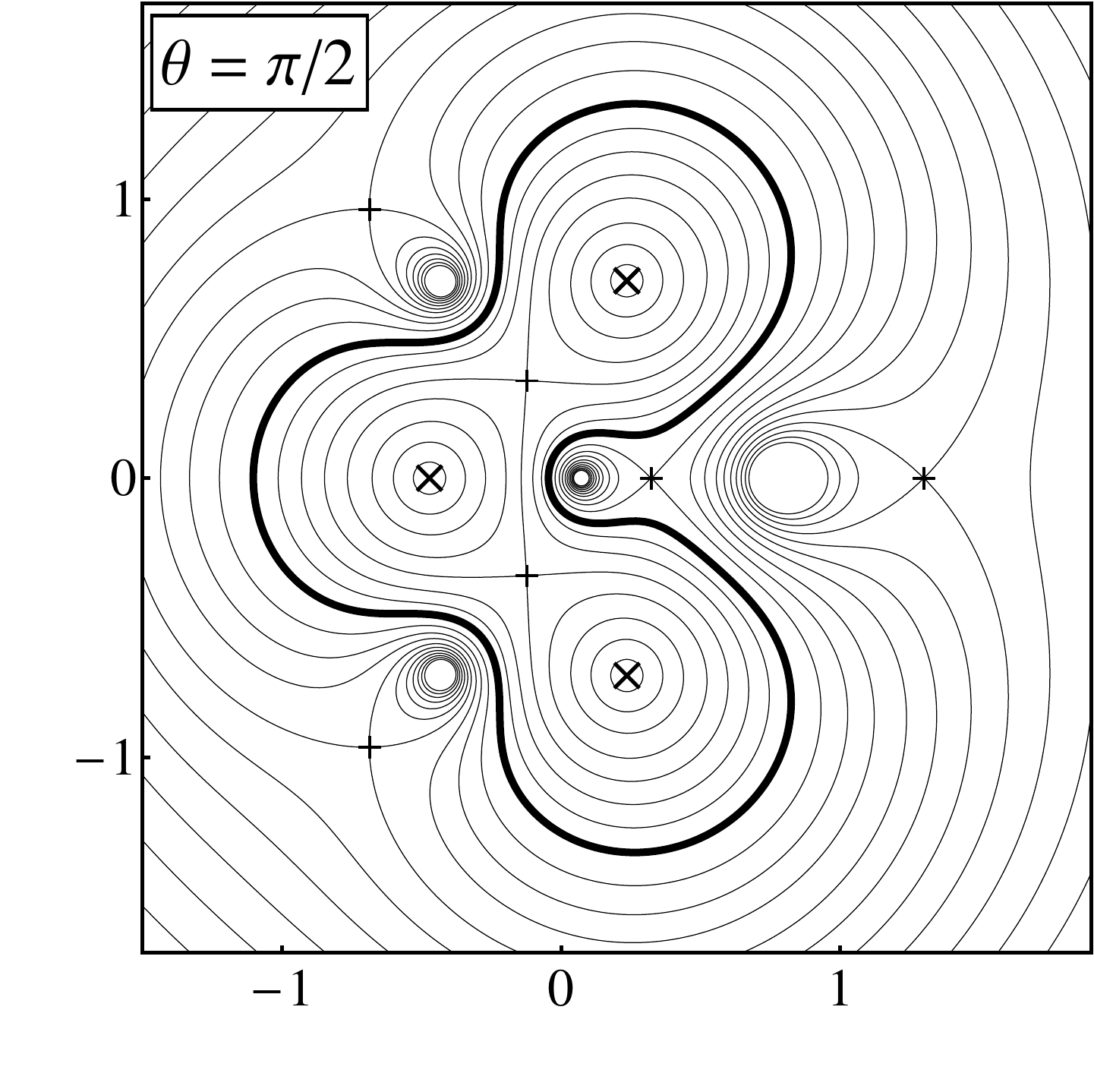}
\hspace{-5mm}
\includegraphics[scale=.385]{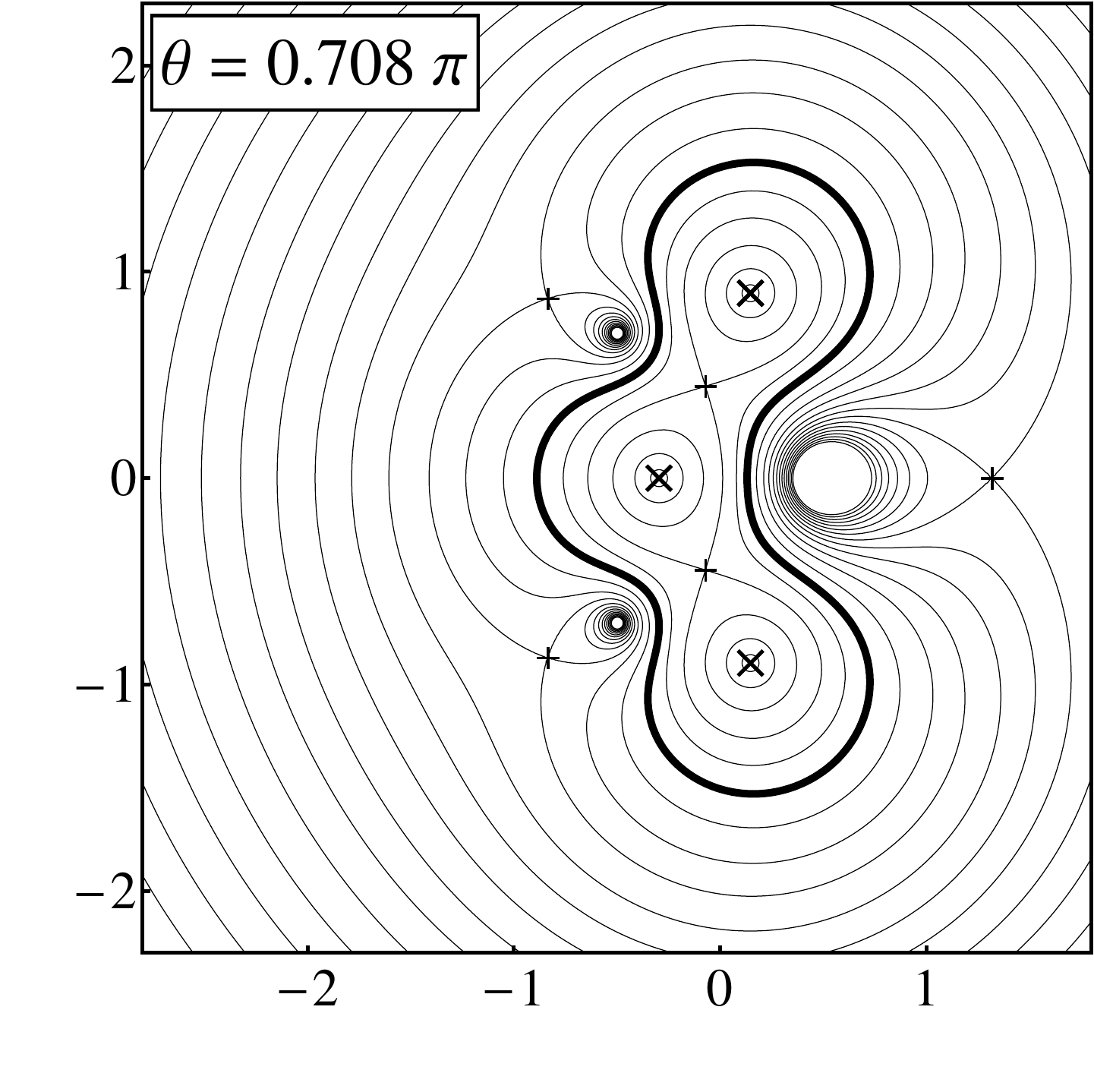}
\vspace*{-3mm}
\\
\hspace*{-1.6mm}
\includegraphics[scale=.385]{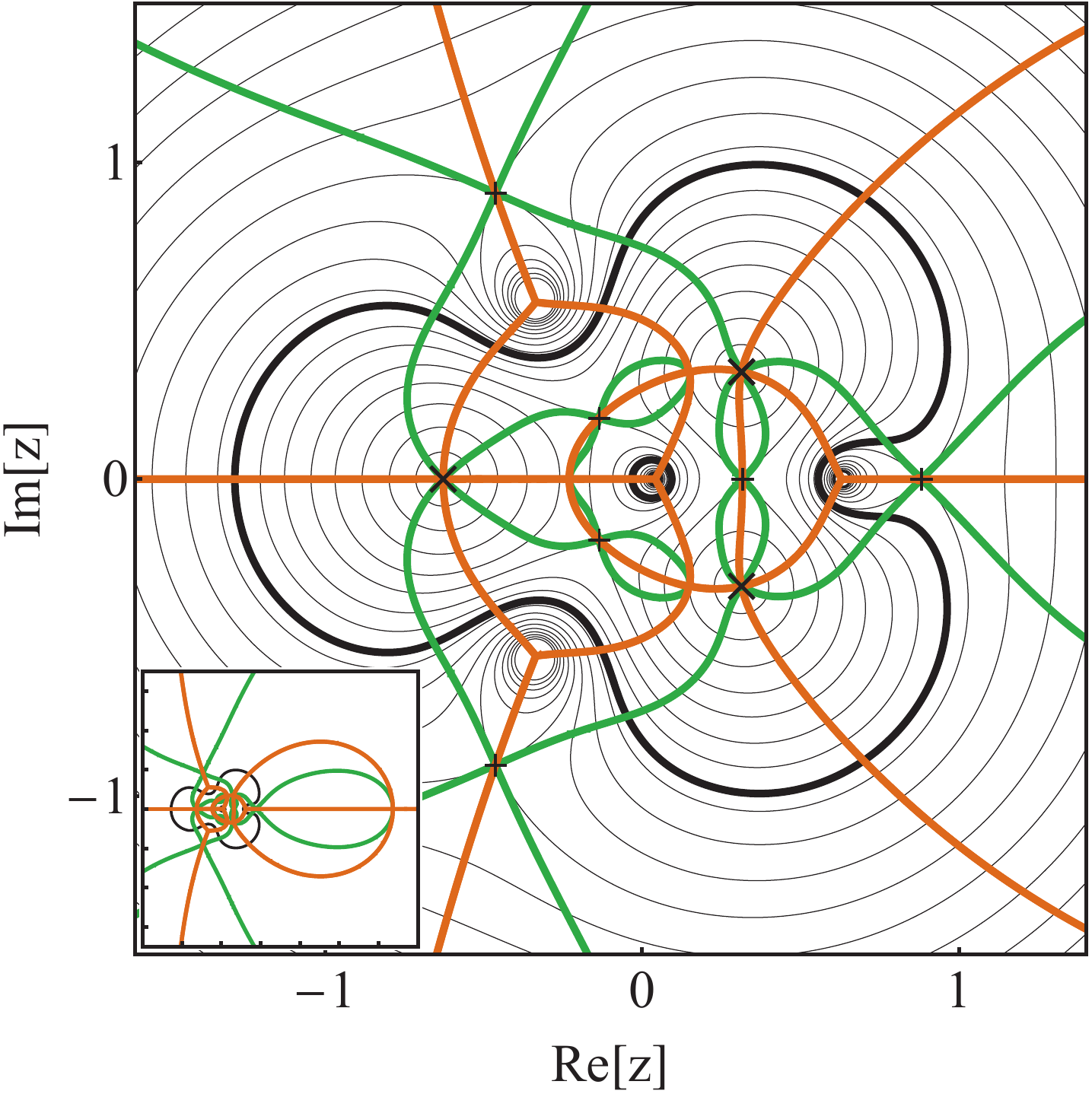}
\hspace{-1.4mm}
\includegraphics[scale=.385]{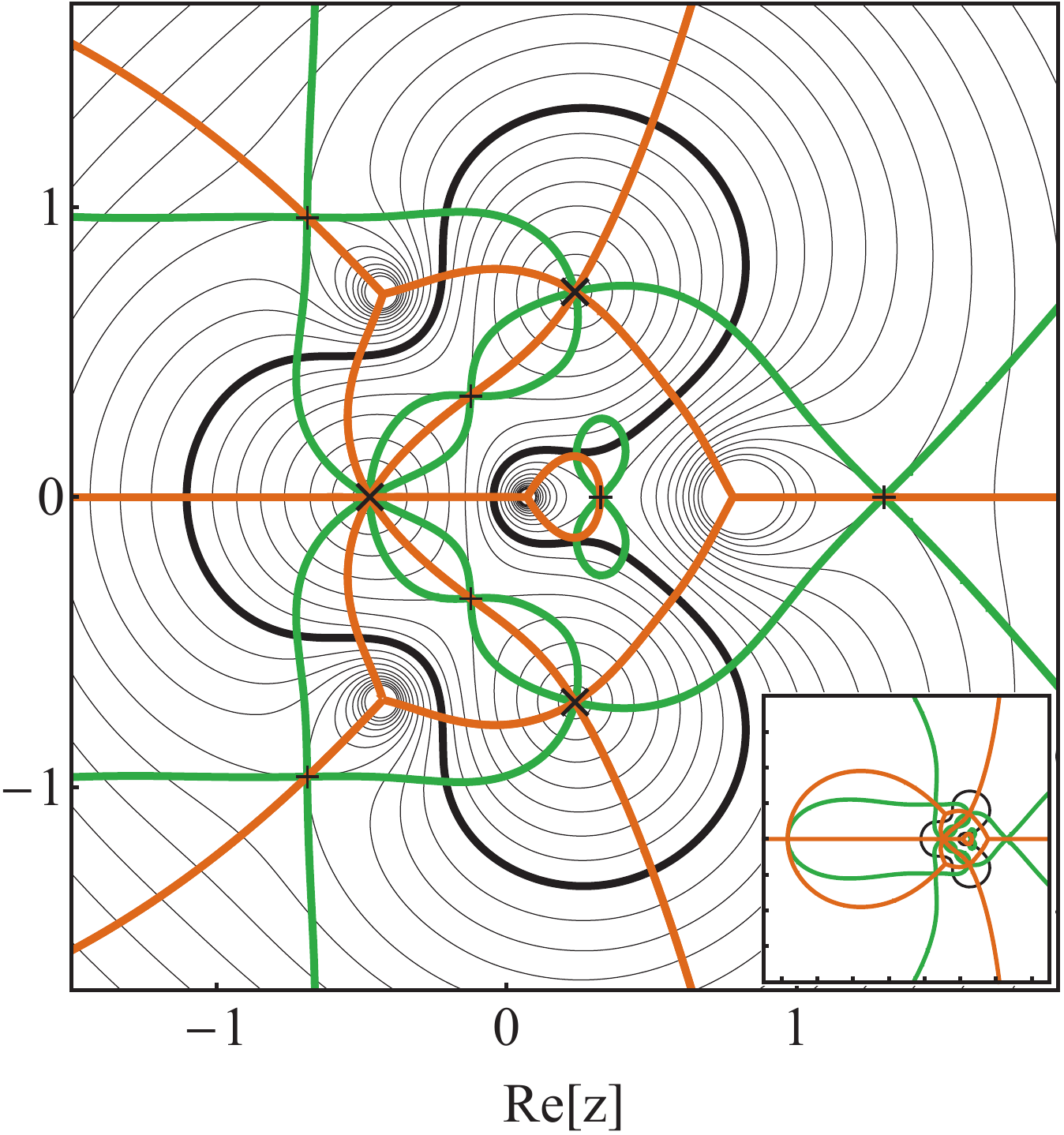}
\hspace{-1.4mm}
\includegraphics[scale=.385]{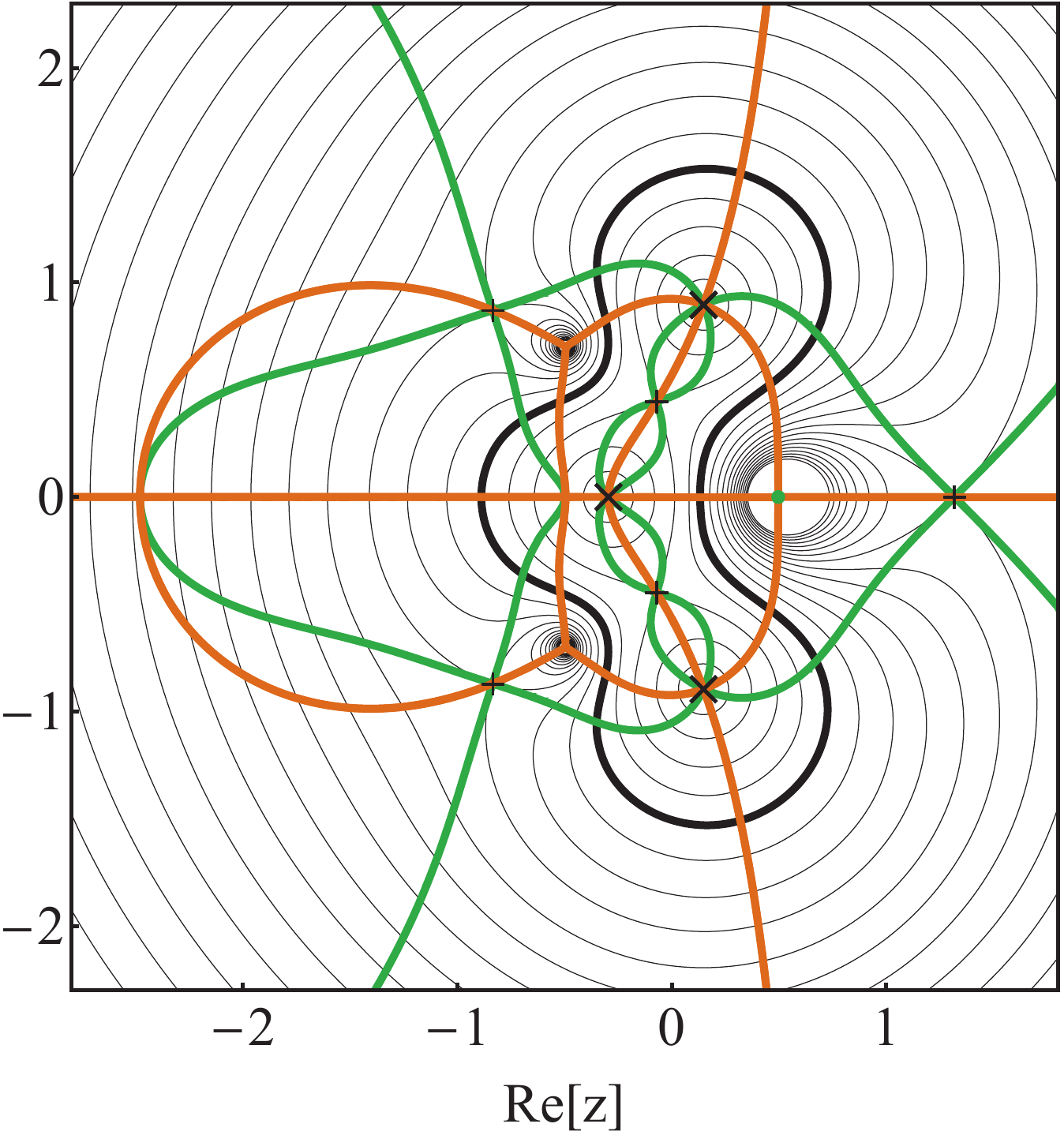}
\caption{TI model: Jacobian contour plots for off-axis butterflies at $\theta\approx0.220\,\pi$ (left column), for Chang-Refsdal violating $\theta=\pi/2$ (central column), and for double maximum at $\theta\approx0.708\,\pi$ (right column). Lower row includes cusp curves (orange) and morph curves (green). Contour values differ from column to column. Insets in lower panels: zoomed-out plots of critical, cusp, and morph curves. Notation as in Figure~\ref{fig:LS-contours}.}
\label{fig:TI-contours}
\efi

\clearpage
\begin{deluxetable}{lccccccccc}
\tabletypesize{\scriptsize}
\tablecaption{Critical-curve Topology Occurrence in Triple-lens Models}
\tablewidth{0pt}
\tablehead{
Model &\includegraphics[angle=90,height=0.3cm]{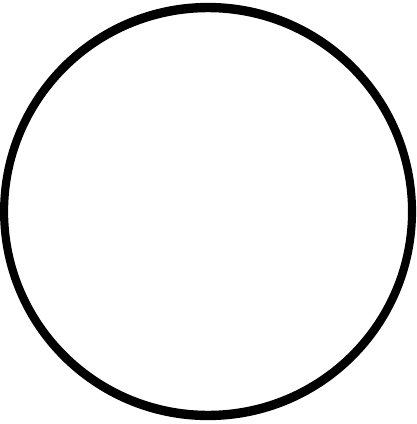} &\includegraphics[angle=90,height=0.3cm]{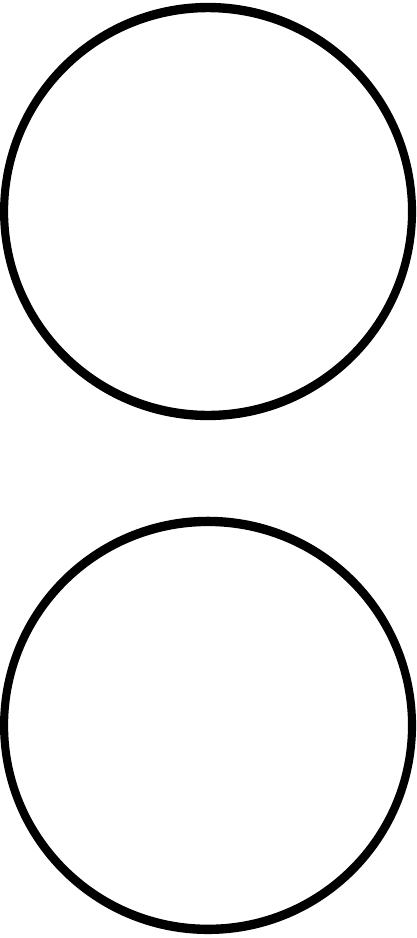}&\includegraphics[angle=90,height=0.3cm]{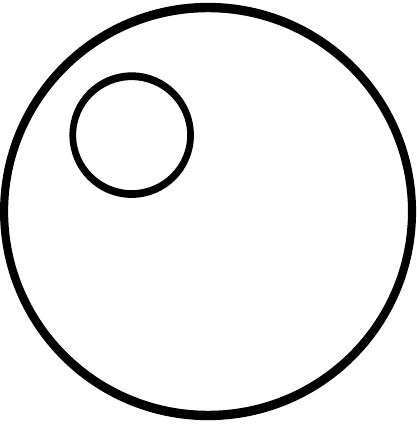} &\includegraphics[angle=90,height=0.3cm]{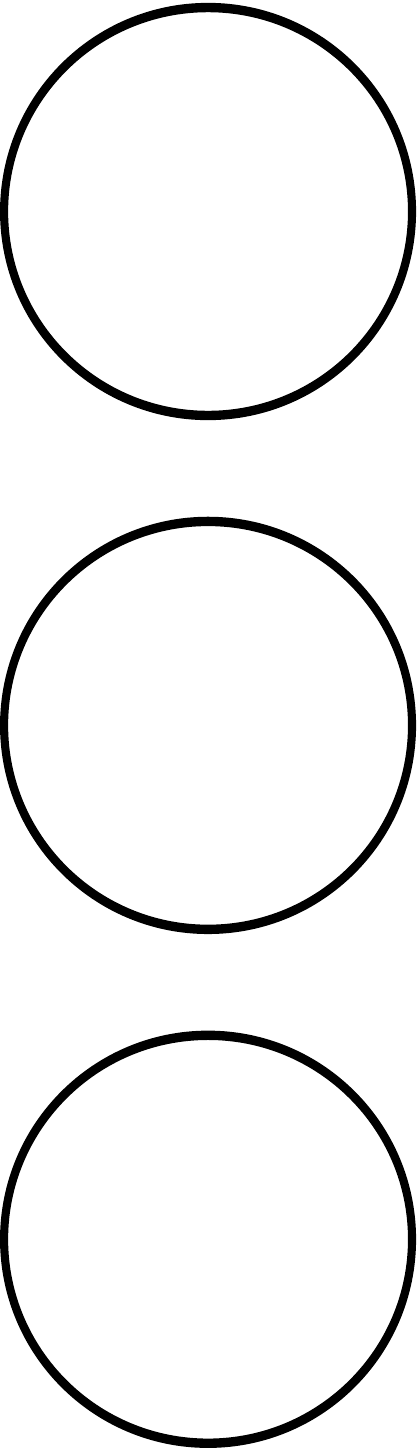}&\includegraphics[angle=90,height=0.3cm]{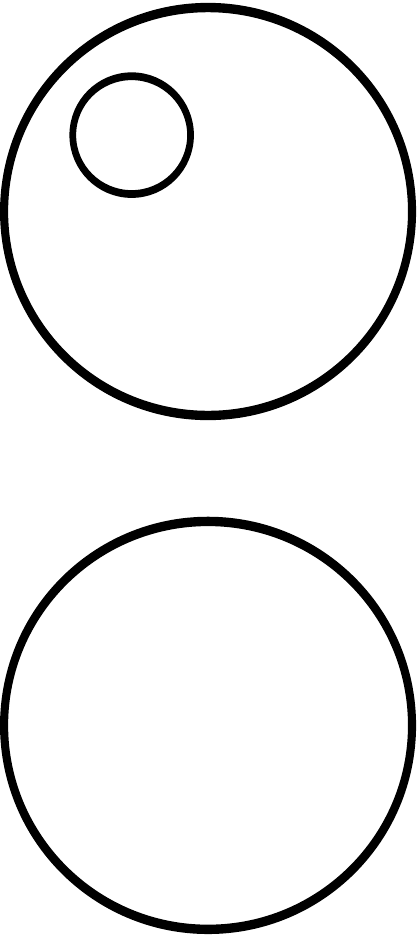} &\includegraphics[angle=90,height=0.3cm]{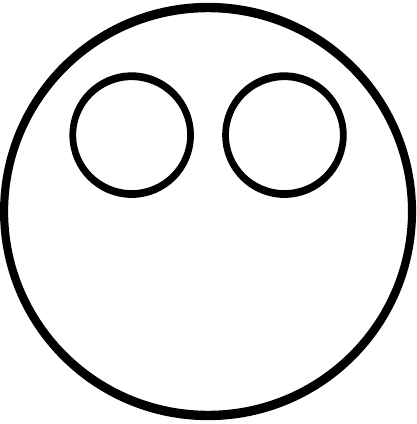}&\includegraphics[angle=90,height=0.3cm]{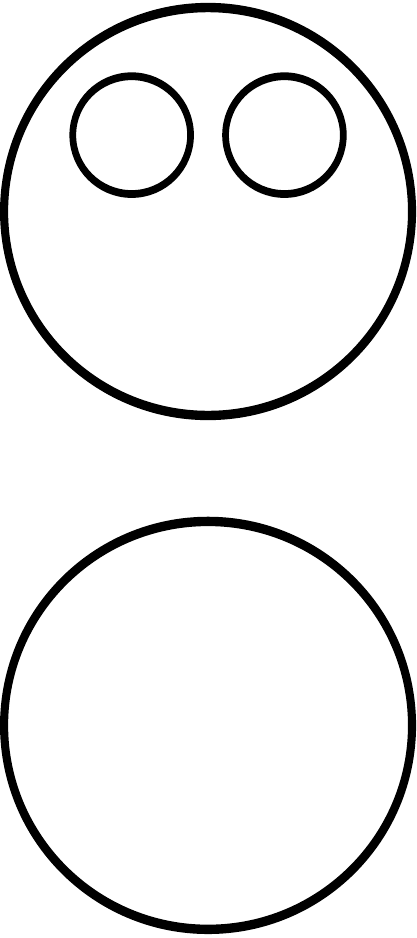} &\includegraphics[angle=90,height=0.3cm]{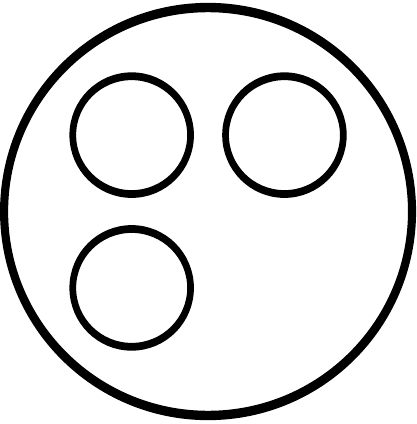}&\includegraphics[angle=90,height=0.3cm]{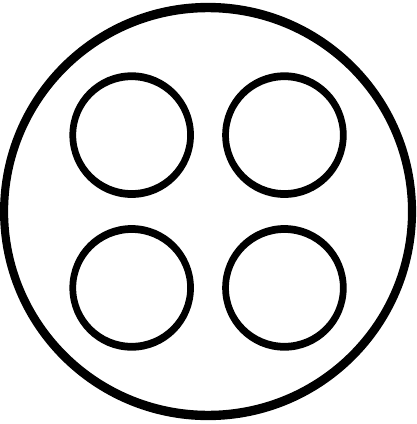}}
\startdata
Linear Symmetric (LS) &B&-&-&A&-&C,E&-&-&D,F    \\
Linear Asymmetric (LA) &D&B&-&C&-&E&A&-&F    \\
Triangular Equilateral (TE) &C&B&D&A&-&E,H&-&I,J&F,G    \\
Triangular Isosceles (TI) &G&C&F&D&B&E,J&A&I,L&H,K    \\
\enddata
\tablecomments{Topology regions of the model parameter spaces are marked by letters, as defined in Figure~\ref{fig:LS-parspace} for the LS model, Figure~\ref{fig:LA-parspace} for the LA model, Figure~\ref{fig:TE-parspace} for the TE model, and Figure~\ref{fig:TI-parspace} for the TI model.}
\label{tab:topologies}
\end{deluxetable}

\clearpage
\setlength{\tabcolsep}{2.5pt}
\begin{deluxetable}{lllll}
\tabletypesize{\footnotesize}
\rotate
\tablewidth{0pt}
\tablecaption{Caustic Structure Occurrence in Triple-lens Models}
\tablehead{Cusps & \multicolumn{4}{l}{Triple-lens Model Parameter-space Subregions [sample parameter combination]} \\
 Total/By Loop & LS [$\mu,s$] & LA [$p,s$] & TE [$\mu,s$] & TI [$\theta/\pi,s$] }
\startdata
8/8 & B$_1$[0.2,0.8] & D$_1$[0.45,1.2] & \nodata & G$_1$[0.95,1.5]   \\
\tableline
10/10 & \nodata & \nodata & C$_1$[0.6,1.35] & G$_2$[0.7,1.2]   \\
10/6+4 & \nodata & B$_1$[0.3,1.5] & B$_1$[0.05,1.5] & C$_1$[0.1,3]   \\
\tableline
12/12 & B$_2$[0.2,0.5] & D$_2$[0.45,0.7] & C$_3$[0.95,0.95] & G$_4$[0.95,0.8]   \\
12/9+3 & \nodata & \nodata & D$_1$[0.33,1.2] & F$_1$[0.33,1]   \\
12/6+6 & \nodata & \nodata & B$_2$[0.13,1.73],B$_5$[0.05,1.13] & \nodata   \\
12/6+3+3 & C$_1$[0.04,0.5] & E$_1$[0.35,0.65] & \nodata & E$_1$[0.05,1]   \\
12/5+4+3 & \nodata & \nodata & \nodata & B$_1$[0.08,2.38]   \\
12/4+4+4 & A$_1$[0.25,1.75],E$_3$[0.09,0.36] & C$_1$[0.45,1.95] & A$_1$[0.25,3] & D$_1$[0.3,3],D$_3$[0.9,3] \\
\tableline
14/14 & \nodata & \nodata & C$_2$[0.6,0.95] & G$_3$[0.4,1.3]   \\
14/10+4 & \nodata & \nodata & B$_4$[0.09,1.58] & C$_2$[0.2,2]   \\
14/8+3+3 & \nodata & \nodata & E$_2$[0.7,0.7],H$_1$[0.05,0.69] & E$_2$[0.15,1],J$_2$[0.7,0.55]   \\
14/7+4+3 & \nodata & \nodata & \nodata & B$_3$[0.11,1.83]   \\
14/6+4+4 & \nodata & \nodata & A$_2$[0.6,2],A$_3$[0.25,2.1] & D$_2$[0.5,2.5],D$_5$[0.32,2.5]   \\
14/4+4+3+3 & \nodata & A$_1$[0.1,1.5] & J$_7$[0.768,0.14] & A$_1$[0.01,2.1],A$_4$[0.05,3.1],L$_7$[0.7,0.34] \\
\tableline
\tablebreak
16/16 & B$_3$[0.04,0.6] & \nodata & \nodata & \nodata   \\
16/10+6 & \nodata & \nodata & B$_3$[0.25,1.45] & C$_3$[0.25,1.6]   \\
16/10+3+3 & \nodata & E$_2$[0.35,0.75] & E$_1$[0.85,0.7] & J$_3$[0.84,0.55]   \\
16/9+4+3 & \nodata & \nodata & \nodata & B$_2$[0.1,1.98]   \\
16/8+4+4 & A$_2$[0.95,1.45] & \nodata & A$_5$[0.995,1.1] & \nodata   \\
16/7+3+3+3 & \nodata & \nodata & I$_1$[0.6,0.67] & I$_1$[0.5,0.6] \\
16/6+6+4 & E$_1$[0.15,0.36] & \nodata & \nodata & D$_4$[0.25,2.3]   \\
16/6+4+3+3 & \nodata & \nodata & J$_3$[0.77,0.6],J$_6$[0.772,0.158] & A$_2$[0.05,2.1],L$_4$[0.7,0.4],L$_6$[0.76,0.39]   \\
16/4+3+3+3+3 & D$_2$[0.7,0.4],F$_1$[0.04,0.36] & F$_2$[0.5,0.45] & F$_3$[0.95,0.15],G$_1$[0.05,0.34],G$_3$[0.6,0.1] & H$_1$[0.1,0.5],H$_3$[0.65,0.2],K$_3$[0.85,0.25] \\
\tableline
18/12+3+3 & \nodata & \nodata & E$_3$[0.69,0.67] & J$_1$[0.62,0.52]   \\
18/8+4+3+3 & \nodata & \nodata & J$_5$[0.773,0.16],J$_8$[0.764,0.155] & A$_3$[0.08,2.24],L$_5$[0.77,0.37],L$_8$[0.67,0.33]   \\
18/6+6+6 & \nodata & \nodata & A$_4$[0.3,1.7] & D$_6$[0.33,2]   \\
18/6+6+3+3 & \nodata & \nodata & J$_1$[0.8,0.45] & L$_1$[0.78,0.44]   \\
18/6+3+3+3+3 & \nodata & \nodata & F$_2$[0.95,0.35],G$_2$[0.25,0.6] & H$_2$[0.33,0.5],K$_2$[0.88,0.48]   \\
\tableline
20/8+8+4 & E$_2$[0.15,0.34],E$_4$[0.09,0.29] & \nodata & \nodata & \nodata   \\
20/8+6+3+3 & \nodata & \nodata & J$_2$[0.8,0.4],J$_4$[0.7,0.6] & L$_2$[0.81,0.44],L$_3$[0.66,0.36]   \\
20/8+3+3+3+3 & D$_1$[0.7,0.65] & F$_1$[0.5,0.53] & F$_1$[0.99,0.85] & K$_1$[0.94,0.52]   \\
\enddata
\tablecomments{Caustic-structure subregions of the model parameter spaces are marked by letters and numerals. The sample parameter combinations in square brackets identify their position in Figures~\ref{fig:LS-parspace} and \ref{fig:LS-detail} for the LS model, Figure~\ref{fig:LA-parspace} for the LA model, Figures~\ref{fig:TE-parspace} and \ref{fig:TE-details} for the TE model, and Figures~\ref{fig:TI-parspace} and \ref{fig:TI-details} for the TI model.}
\label{tab:caustic_structures}
\end{deluxetable}
\setlength{\tabcolsep}{6pt}

\end{document}